\documentclass[preprint,3p,times]{elsarticle}
\journal{Computer Methods in Applied Mechanics and Engineering}

\usepackage{amssymb}
\usepackage{amsmath}
\usepackage{subcaption}
\usepackage{algorithm}
\usepackage{algpseudocode}

\usepackage{lineno}

\usepackage{xcolor}

\newtheorem{remark}{Remark}

\journal{Computer Methods in Applied Mechanics and Engineering}

\begin{document}

\begin{frontmatter}



\title{Sparse Narrow-Band Topology Optimization for Large-Scale Thermal-Fluid Applications}


\author[label1]{Vladislav Pimanov} 
\ead{vpimanov@ucsd.edu}
\author[label1]{Alexandre T. R. Guibert} 
\author[label3]{John-Paul Sabino} 
\author[label4]{Michael Stoia} 
\author[label1,label2]{H. Alicia Kim} 

\affiliation[label1]{organization={Department of Structural Engineering, University of California San Diego},
            addressline={9500 Gilman Drive}, 
            city={La Jolla},
            postcode={92122}, 
            state={CA},
            country={USA}}

\affiliation[label3]{organization={The Boeing Company},
            addressline={}, 
            city={Tukwila},
            postcode={98108}, 
            state={WA},
            country={USA}}

\affiliation[label4]{organization={The Boeing Company},
            addressline={}, 
            city={Huntington Beach},
            postcode={92647}, 
            state={CA},
            country={USA}}

\affiliation[label2]{organization={Program in Materials Science and Engineering, University of California San Diego},
            addressline={9500 Gilman Drive}, 
            city={La Jolla},
            postcode={92122}, 
            state={CA},
            country={USA}}

\begin{abstract}
We propose a fluid-based topology optimization methodology for convective heat-transfer problems that can manage an extensive number of design variables, enabling the fine geometric features required for the next generation of heat-exchangers design. Building on the classical Borrvall–Petersson formulation for the Stokes flow, we introduce an optimization algorithm that focuses computational effort on the fluid–solid interface, where it is most needed. To address the high cost of repeated forward and adjoint analyses {and to avoid leakage through nominally solid regions}, we exclude fictitious solid voxels from the analysis by imposing the no‑slip boundary conditions {in the vicinity of} the fluid-solid interface. {In contrast to the prior approaches, the fictitious solids are also excluded from the global optimization problem via reducing it} to a sequence of {local} narrow‑band subproblems with a variable design space. {The contribution of our method is that large-scale optimization can be solved efficiently by continuous simplex method while reliably obtaining binary designs without additional filtering or projection.}
We demonstrate efficiency of the method on multiple examples, including the optimization of a two-fluid heat exchanger at $Pe=10^4$ on a $370^3$ grid comprising $5\times10^7$ design variables using only a single desktop workstation. 
\end{abstract}


\begin{graphicalabstract}
\includegraphics[width=\textwidth]{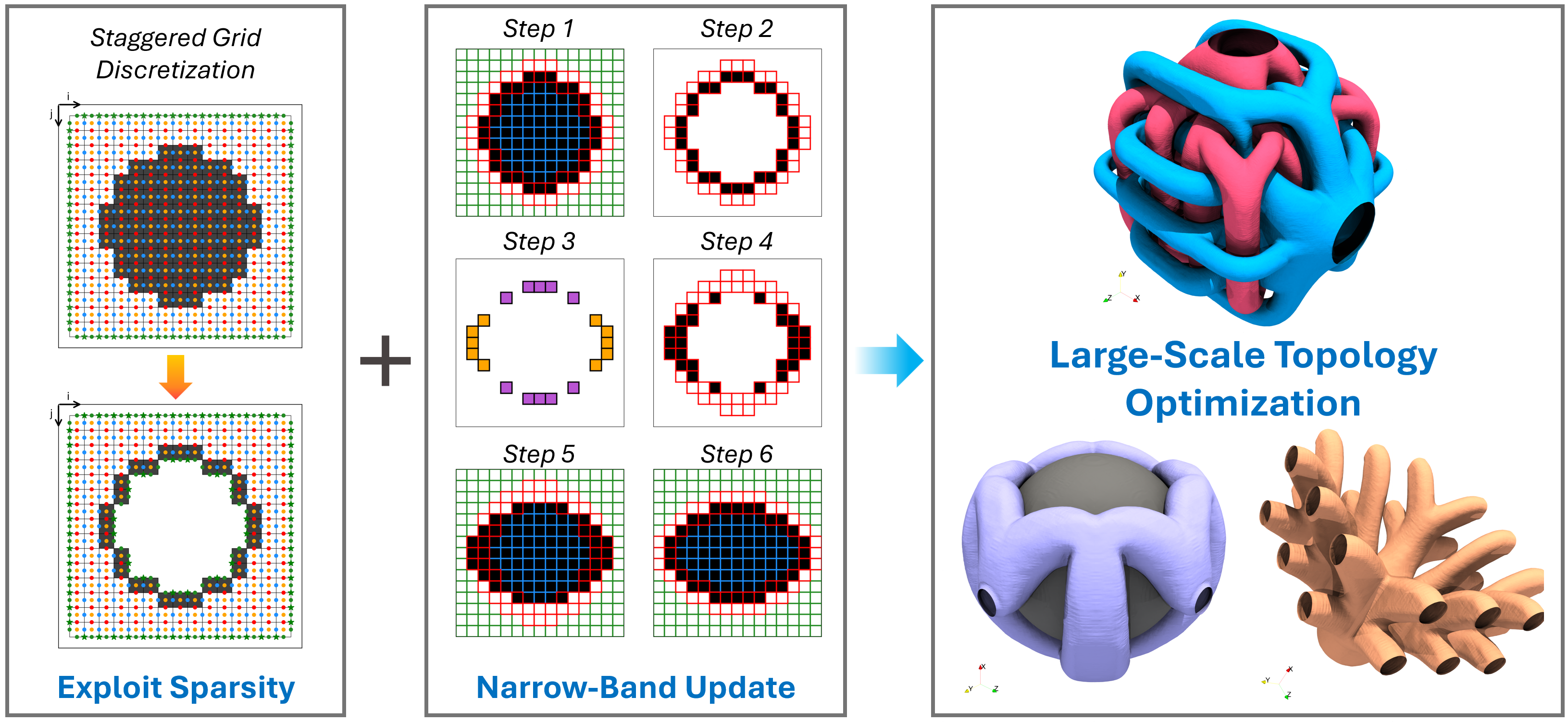}
\end{graphicalabstract}


\begin{keyword}

Multiphysics optimization \sep large scale \sep Stokes flow \sep convective heat transfer \sep heat exchanger




\end{keyword}

\end{frontmatter}

\section{Introduction}
Topology optimization has discovered non-intuitive configurations that achieve high performance under complex physical constraints. Fluid-based topology optimization for thermal management applications has been applied to diverse areas ranging from microelectronics \cite{zeng2019topology, zou2022topology} and aerospace \cite{kambampati2021level} to energy storage systems, including designs with phase change materials \cite{pizzolato2017design, pizzolato2017topology}. Early work in structural topology optimization \cite{bendsoe1988generating, bendsoe1989optimal, sigmund2013review} laid the foundation for extending these methods to fluid mechanics, where researchers have addressed both laminar and turbulent flows governed by the Stokes and Navier–Stokes equations \cite{borrvall2003, gersborg2005topology, dilgen2018topology, yoon2016topology}. The review by Alexandersen and Andreasen \cite{alexandersen2020review} summarizes advances in the field and highlights continued research needs on topics such as computational efficiency, manufacturing constraints, and multi-objective formulations. However, the prohibitive computational cost of large-scale, three-dimensional problems remains a challenge.

One of the earliest density-based methods for fluid problems was proposed by Borrvall and Petersson \cite{borrvall2003}, who suggested {material interpolation} via a Brinkman reaction term to distinguish between “fluid” and “solid” regions during topology optimization with Stokes flow. The flow is penalized in the solid domain by modeling it as a porous material with very low permeability. Gersborg-Hansen \textit{et al.} \cite{gersborg2005} later demonstrated how this approach can be extended to steady laminar Navier–Stokes flow at moderate Reynolds number. Olesen \textit{et al.} \cite{olesen2006} implemented a high-level Navier–Stokes topology optimization framework based on the fictitious porous media approach.

Alongside purely fluid problems, several works have studied topology optimization for conjugate heat transfer. Alexandersen \textit{et al.} \cite{alexandersen2016} explored large-scale topology optimization of three-dimensional heat sinks with natural convection, handling up to $65.5\times10^6$ design variables.
{Other representative studies on conjugate heat transfer that build on the Borrvall--Petersson framework include a convection-dominated formulation by Bruns \textit{et al.} \cite{bruns2007topology}, natural convection modeled with the Boussinesq approximation by Alexandersen \textit{et al.} \cite{alexandersen2014topology}, forced convection with turbulent extensions by Dilgen \textit{et al.} \cite{dilgen2018density}, and applications to gas-turbine cooling configurations by Pietropaoli \textit{et al.} \cite{pietropaoli2019three}.}

Another branch of research studies heat exchangers with multiple fluids. Notable examples include Høghøj \textit{et al.} \cite{hoghoj2020}, who applied a density-based optimization to handle two fluids within a single domain, and  Kobayashi \emph{et al.} \cite{kobayashi2021topology} also explored conjugate heat transfer in two-fluid thermal systems using a single design-variable field to represent the two fluids and the solid domain.
Fawaz \textit{et al.} \cite{fawaz2022topology}, reviewing and categorizing over two hundreds publications on topology optimization of heat exchangers, identified the substantial computational cost of repeatedly solving flow equations as one of the primary barriers to industrial adoption. They noted that the resulting topologies often become highly complex, adding challenges to robust convergence of the flow solver.

Exhaustive reviews indicate that density-based methods {relying on Borrvall–Petersson material interpolation} are the common approach for the fluid-based topology optimization, accounting for more than 80\% of the published research in this field \cite{alexandersen2020review,dbouk2017review,fawaz2022topology}. The advantages of {this} approach include the simplicity of structured-grid discretizations, eliminating the need for discontinuity treatments at the interface, and the availability of well-established techniques for computing sensitivities. Nonetheless, several challenges remain. First, the Brinkman term only approximates the no-slip boundary condition at the fluid–solid interface. {
 Consequently, for faster flows, the porous-medium surrogate may exploit “flow seepage” through nominally solid regions, creating numerically permeable obstacles that would be physically impermeable. As demonstrated and discussed extensively by Kreissl and Maute \cite{kreissl2012levelset} in a level-set context, and observed to a lesser extent in Lin \emph{et al.} \cite{lin2015topology}, such leakage paths can materially degrade the performance and may violate imposed mass-flow constraints. } Consequently, the penalization factor must be large enough to ensure negligible flow velocity within the solid domain. However, it is known that the convergence of iterative solvers for the Stokes–Brinkman equations is highly sensitive to the penalization parameter \cite{alexandersen2020review}. Although various preconditioners and numerical methods have been proposed \cite{iliev2013numerical, vassilevski2013block, efendiev2012robust, iliev2011variational, pimanov2022workflow}, a robust numerical solution strategy for heterogeneous permeability coefficients with large jumps has yet to be established; the numerical solution of Stokes–Brinkman equations continues to be an active area of research. Second, density-based methods typically pose an optimization problem over the entire design domain, which means the velocity and pressure fields are also present in the solid region. In three-dimensional settings, this can quickly lead to a prohibitively large number of degrees of freedom. Even a moderate $400^3$ grid results in $64\times10^6$ design variables, with roughly half a billion total unknowns (including three velocity components, pressure, and temperature), inevitably requiring distributed-memory computing resources. In addition, density-based optimization methods often employ parameter continuation (or homotopy) schemes \cite{pereira2016fluid, alexandersen2023detailed} for improved convergence, further increasing the computational cost due to the growth in the number of iterations.
To overcome these difficulties, a number of strategies have been proposed. One well‑known approach is to use unstructured body‑fitted meshes with a level‑set representation of the geometry (see, e.g., \cite{feppon2021body}). {Body‑fitted discretizations naturally allow solving the governing equations only in the fluid domain}; however, frequent remeshing at every design modification becomes a serious bottleneck in three‑dimensional applications. {Another family of methods that only solves for the fluid regions relies on surface‑capturing discretizations}.
{Examples of surface‑capturing discretizations applied to topology optimization include XFEM \cite{kreissl2012levelset, jenkins2015level, hoghoj2025density, noel2023xfem}, cutFEM \cite{villanueva2017cutfem}, the moments method \cite{kambampati2023cad}, and immersed‑boundary techniques \cite{jenkins2016immersed, kubo2021level}. In topology optimization settings, these discretizations are commonly paired with a level‑set geometry description: The interface is represented implicitly and then discretized by cutting a uniform background grid.}
{In immersed‑boundary methods, in order to impose boundary conditions directly at the interface, a localized forcing (e.g., Brinkman‑type penalization or distributed Lagrange multipliers) is introduced rather than using a heuristic material interpolation. While these methods can yield an improved accuracy near the boundaries, they typically require specialized quadrature for cut elements or ghost‑cell/extended stencils.}
{In level‑set–based topology optimization with surface‑capturing discretizations, one typically computes an interface velocity that minimizes an objective functional while respecting the constraints and advances a Hamilton–Jacobi level‑set equation to update the geometry; the associated indicator (density) field is implied by the new interface. In our approach, we do not determine the interface position explicitly. Instead, we update the indicator field directly, restricting changes to a narrow band around the current interface. Under this restriction, we assume the updated field corresponds to a small interface displacement, i.e., the new interface lies within the band.}

To reduce the computational cost and alleviate the memory bottleneck of uniform‑grid methods, Kambampati \emph{et al.} \cite{kambampati2021geometry} proposed representing designs using an efficient VDB data structure \cite{museth2013vdb}. The VDB representation is based on sparse grids, enabling memory and work to be allocated only where needed with negligible overhead. An illustrative example is level‑set propagation, where memory and computation are concentrated within a narrow band around the evolving interface, avoiding unnecessary work across the entire domain \cite{adalsteinsson1995fast}. We adopt this narrow‑band concept in our optimization algorithm, described in detail in Section~\ref{sec:narrow_band}. Specifically, during topology optimization, design variables are dynamically activated or deactivated so that only a small subset remains active at each step. Consequently, the global problem is reduced to a sequence of subproblems solved iteratively for these active subsets using a Sequential Linear Programming (SLP) strategy.
This methodology is similar to the Topology Optimization of Binary Structures (TOBS) algorithm \cite{sivapuram2018topology, souza2021topology, picelli2022topology}. However, to maintain the validity of linear approximations, TOBS introduces an additional constraint limiting design changes between successive iterations, whereas our approach explicitly confines the active set to a narrow band around the fluid–solid interface. {Moreover, in contrast to the TOBS method that relies on integer optimizers, we employ continuous‑variable simplex linear programming updates within the SLP framework enabling good scaling properties.}

The narrow-band optimization methodology allows us to leverage the sparsity advantage: Since design updates are restricted to the predefined active subset around the fluid-solid interface, solid voxels that are known to remain solid throughout optimization can be excluded from the analysis, offering significant computational acceleration.  In fluid topology optimization, the fluid volume fraction may be low, in which case a significant portion of computational effort is spent unnecessarily on fictitious solid regions. This sparsity concept was advocated in fluid-based level-set topology optimization by Challis and Guest \cite{challis_guest}, who discussed eliminating fictitious solid voxels and imposing no-slip boundary conditions directly on the fluid-solid interface. By discarding degrees of freedom associated with these solids, they enabled solving large-scale three-dimensional problems using a basic iterative method without any special preconditioning. In the current study, we extend this idea to conjugate heat transfer problems and further accelerate it with an efficient numerical method.  
{Most prior topology‑optimization approaches that solve only in the fluid domain rely on explicit interface descriptions (typically level‑set methods). A drawback of such level‑set‑based approaches is that they require the derivation of shape sensitivities, which is considered more cumbersome than in the Brinkman penalization approach. Excluding fictitious voxels has been shown to be effective in structural density‑based topology optimization (see, e.g., \cite{bruns2003element}), but transferring this idea to a Brinkman framework for fluids is not straightforward: The key challenge in using adaptive, design‑dependent boundary conditions is to ensure that introducing boundary conditions does not overly restrict the design space \cite{behrou2019adaptive,theulings2023towards}. Behrou \emph{et al.} \cite{behrou2019adaptive} proposed a density‑based optimization method for incompressible Navier--Stokes flow that utilizes adaptive no‑slip boundary conditions. The authors exclude fictitious solid voxels using a simple threshold. However, they remove them from the finite element analysis but not from the optimization problem. As a result, they acknowledge that a discontinuity into an otherwise continuous topology optimization problem is introduced, and that the designs are likely to depend on the chosen threshold magnitude (see the discussion in Section~3.1 \cite{behrou2019adaptive}). In our proposed method, we alleviate this difficulty by removing solids also from the optimization, via considering a sequence of optimization subproblems with a variable design space instead of a single optimization problem. In \cite{behrou2019adaptive}, the authors use a Heaviside projection to obtain binary designs.  A distinctive feature of our method is that, despite using a continuous optimizer (simplex) within the standard Brinkman penalization approach, we obtain binary designs without any filtering or projection. }
{Another density-based method that reconstructs explicit interface from densities and uses cut-cell method
for computing sensitivities was proposed by Galanos \textit{et al.} \cite{galanos2024cut,galanos2025continuous}. However, the Helmholtz filter \cite{lazarov2011filters} is utilized in their work which results in a diffuse-interface.} 

Our key contributions can be summarized as follows: {\emph{(i)} We propose a narrow-band topology optimization algorithm that focuses on the evolving fluid–solid interface; the proposed algorithm allows for robust removal of fictitious solids in the classical Borrvall-Petersson formulation \emph{(ii)} Despite using a continuous optimizer (simplex) within the classical Brinkman framework, we obtain binary designs without any filtering or projection.} \emph{(iii)} We provide a detailed implementation, including sensitivity analysis and discretization schemes for both fluid flow and heat transfer. We demonstrate the method on a suite of examples, including a pure-fluid problem, a conjugate heat-transfer case, and a two-fluid heat exchanger with a minimum-thickness constraint.

\subsection*{Notation}
In this work, we consider domains $\Omega \subset \mathbb{R}^d$ with $d=2,3$. In Sections \ref{sec:gov_eqns}-\ref{sec:discrete}, we describe problem formulation, optimization method, and discretization for $d=2$ for clarity. However, the numerical experiments in Section \ref{sec:numerical} are given for $d=3$.
The following notation is used:
\begin{itemize}
    \item Continuous fields are written using lowercase letters. For example, $u(\mathbf{x})$ denotes the velocity field, $p(\mathbf{x})$ the pressure, $\theta(\mathbf{x})$ the temperature, and $\gamma(\mathbf{x})$ the design or density field, all defined on $\Omega$.
    \item Uppercase letters denote discrete analogs of the corresponding continuous fields. For example $U\in\mathbb{R}^{N_U}$,$P\in\mathbb{R}^{N_P}$,$\Theta\in\mathbb{R}^{N_\Theta}$, and $\Gamma\in\mathbb{R}^{N_\Gamma}$.
    \item Bold uppercase letters, such as $\mathbf{A}$, $\mathbf{B}$, $\mathbf{C}$, $\mathbf{L}_U$, and $\mathbf{L}_\Theta$, denote matrices that arise from discretizing the PDEs.
\end{itemize}

\section{Governing equations}\label{sec:gov_eqns}

\begin{figure}[htbp]
    \centering
        \includegraphics[width=0.4\textwidth]{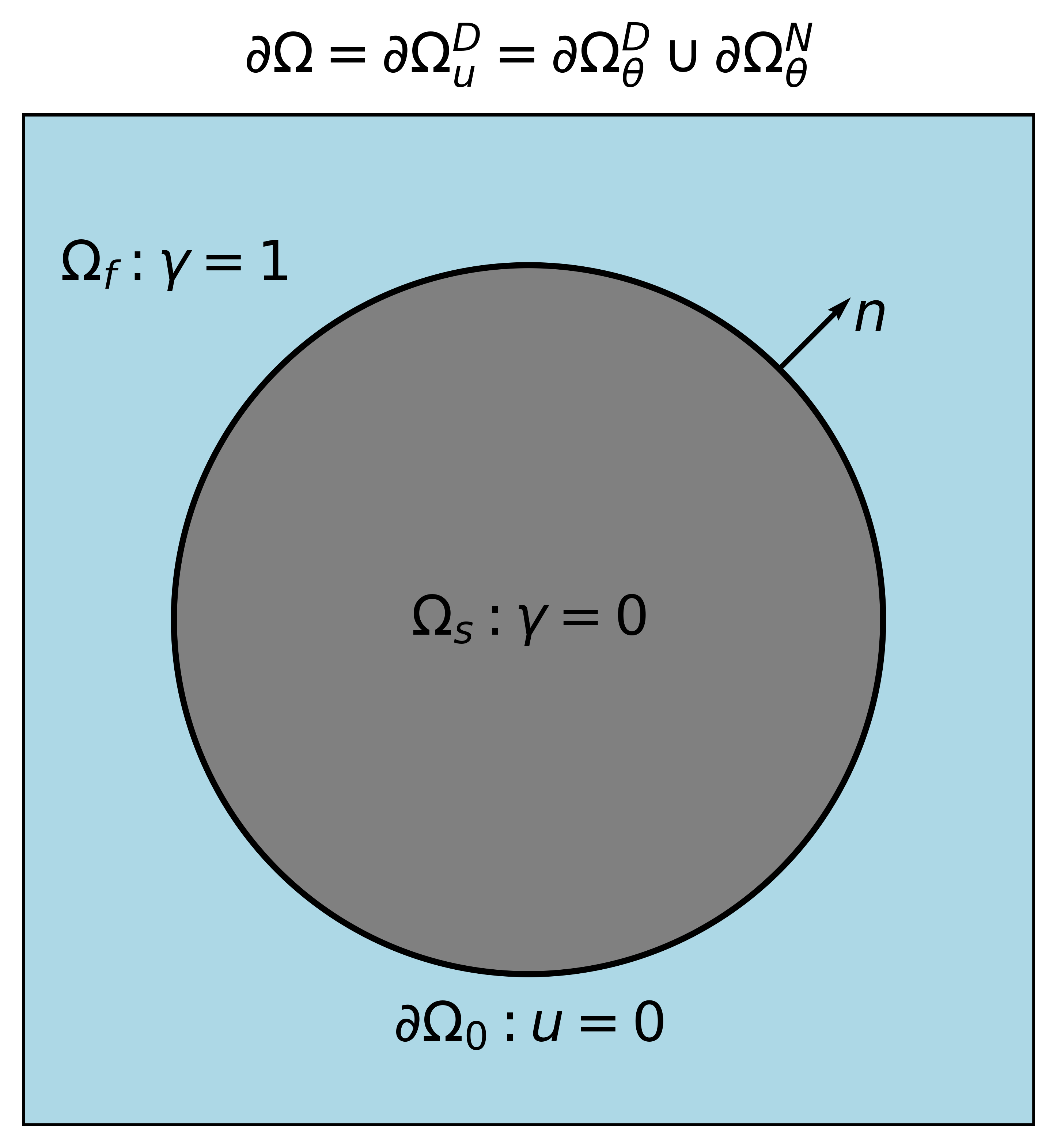}
    \caption{Computational domain and boundary conditions.}
    \label{fig:problem_setup}
\end{figure}

We consider a two-dimensional domain 
\(\Omega = (0,1)\times(0,1)\) represented in Fig.~\ref{fig:problem_setup}. This domain is decomposed into a fluid region \(\Omega_f\) and a solid region \(\Omega_s\), such that $\Omega = \Omega_f\cup\Omega_s$. Following the single domain approach, we introduce a spatially varying {density} field, denoted \(\gamma(\mathbf{x})\), such that \(\gamma=1\) in \(\Omega_f\), \(\gamma=0\) in \(\Omega_s\), and the fluid--solid interface, denoted \({\partial\Omega}_0\), is defined implicitly by the distribution of $\gamma$.

\paragraph{Flow equation}
Following the classical approach of Borrvall and Petersson \cite{borrvall2003}, we describe fluid flow throughout the entire domain \(\Omega\) via introducing the Brinkman penalization term.
Let \(u(\mathbf{x})\) be the velocity field, and \(p(\mathbf{x})\) the pressure field.
The Stokes-Brinkman equations are given as follows:
\begin{equation}
\label{eq:stokes_brinkman}
\begin{aligned}
-\nu\,\Delta {u} \;+\;\alpha(\gamma)\,{u} \;+\;\nabla p &= 0 
&& \text{in } \Omega,\\
\nabla \cdot {u} &= 0 
&& \text{in } \Omega, \\
{u} &= {u}_D 
&& \text{on } {\partial\Omega}_{u}^D,
\end{aligned}
\end{equation}
where \(\nu\) is the kinematic viscosity, and:
\begin{equation}
\label{brinkman_interpolation}
\alpha(\gamma) \;=\;\alpha_{\max}\,\frac{(1-\gamma)}{1+q_a\,\gamma}.
\end{equation}
The parameter $\alpha_{\max} \gg 1$ denotes the inverse permeability in solids, and the parameter $q_a \geq 0$ controls the convexity of the interpolation scheme.
The Dirichlet boundary condition is imposed on the entire boundary $\partial\Omega = {\partial\Omega}_{{u}}^D$, where ${u}_D$ describes the inflow/outflow velocity, or is equal to zero for the no-slip boundary condition. In addition, the penalization term $\alpha(\gamma)$ effectively replaces the no-slip boundary condition on the fluid-solid interface \({\partial\Omega}_0\) by forcing \({u}\) to be small in solid regions. 

\paragraph{Heat equation}
Let \(\theta(\mathbf{x})\) be the temperature field. The heat propagation in $\Omega$ is described by the steady convection--diffusion equation, given as follows:
\begin{equation}
\label{eq:heat_equation}
\begin{aligned}
 -\,k\,\Delta \theta + \nabla \cdot\bigl({u}(\gamma)\,\theta\bigr) &= q 
&& \text{in } \Omega, \\
\theta &= \theta_D 
&& \text{on } {\partial\Omega}_\theta^D, \\
-\,k\,\frac{\partial \theta}{\partial n} &= 0 
&& \text{on } {\partial\Omega}_\theta^N,
\end{aligned}
\end{equation}
where \(k\) is the thermal conductivity, \(q\) is the volumetric heat source, and ${u}(\gamma)$ is the velocity solution of the Stokes-Brinkman problem \eqref{eq:stokes_brinkman}. The entire boundary $\partial\Omega = {\partial\Omega}_\theta^D \cup {\partial\Omega}_\theta^N$ is decomposed into two parts ${\partial\Omega}_\theta^D$ and ${\partial\Omega}_\theta^N$ where either the Dirichlet ($\theta = \theta_D$) or the homogeneous Neumann ($-k\frac{\partial \theta}{\partial n} = 0$) boundary conditions are prescribed. Note that the convective term does not completely vanish in the solid region since a small velocity still exists there in the single-domain approach.

In what follows, the velocity $u$, pressure $p$ and temperature \(\theta\) are referred to as {state variables} and the density \(\gamma\) is referred to as {design variable}.

\section{Narrow-Band Topology Optimization}\label{sec:narrow_band}

\begin{figure}[htbp]
    \centering
    \begin{subfigure}[b]{0.25\textwidth}
        \includegraphics[width=\textwidth]{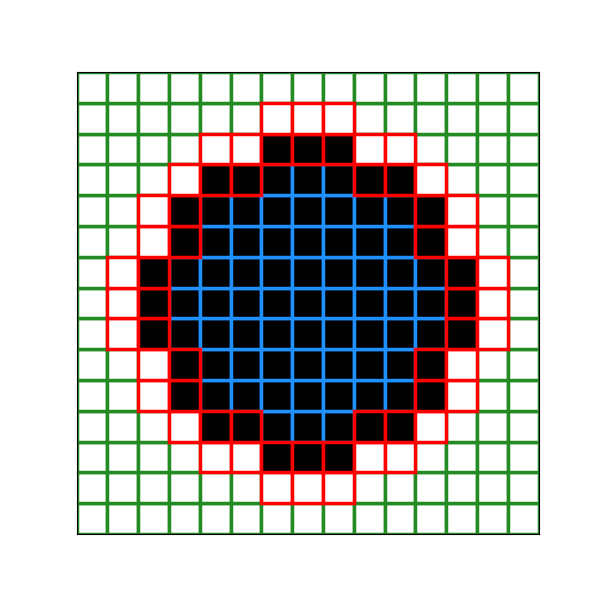}
        \caption{}
        \label{fig:steps1}
    \end{subfigure}
    \begin{subfigure}[b]{0.25\textwidth}
        \includegraphics[width=\textwidth]{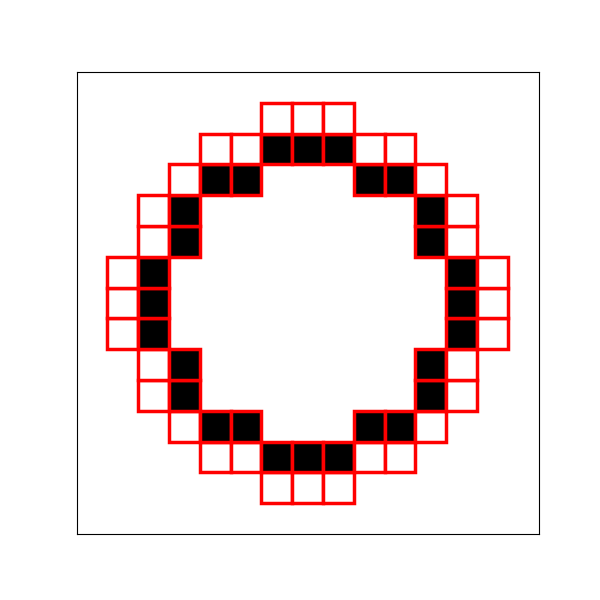}
        \caption{}
        \label{fig:steps2}
    \end{subfigure}
    \begin{subfigure}[b]{0.25\textwidth}
        \includegraphics[width=\textwidth]{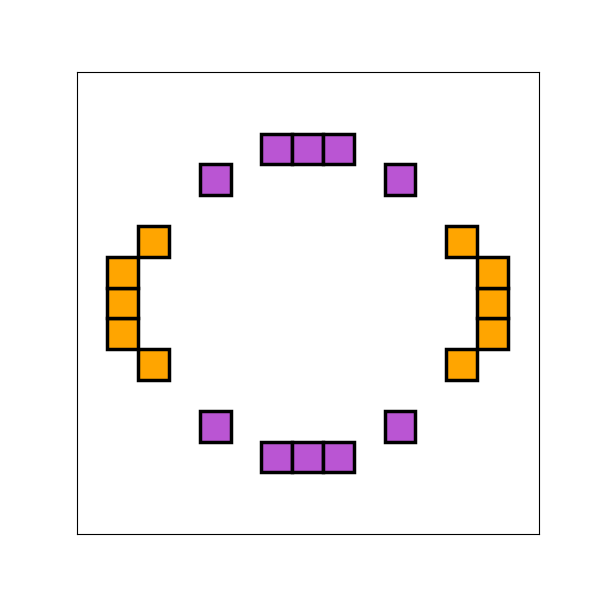}
        \caption{}
        \label{fig:steps3}
    \end{subfigure}
    \hfill
    \begin{subfigure}[b]{0.25\textwidth}
        \includegraphics[width=\textwidth]{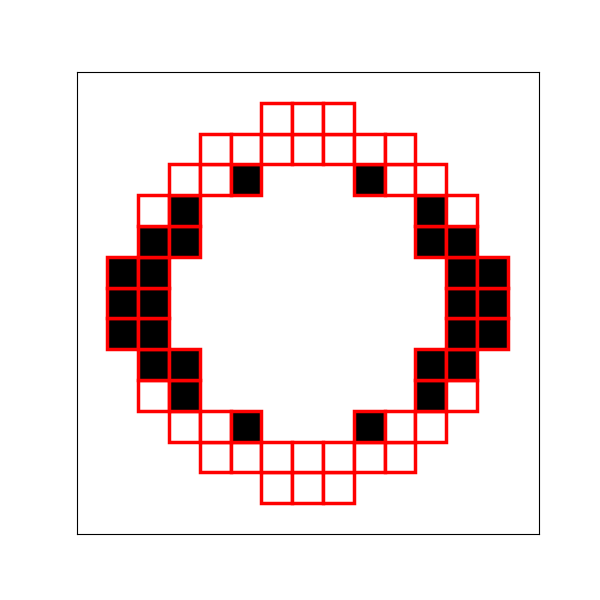}
        \caption{}
        \label{fig:steps4}
    \end{subfigure}
    \begin{subfigure}[b]{0.25\textwidth}
        \includegraphics[width=\textwidth]{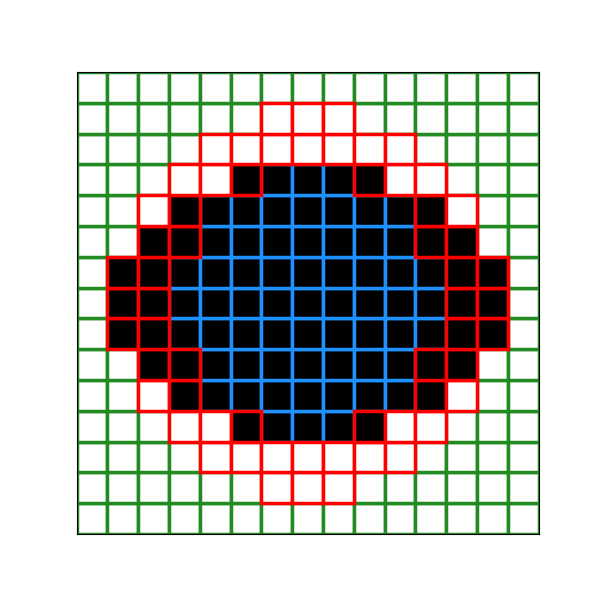}
        \caption{}
        \label{fig:steps5}
    \end{subfigure}
    \begin{subfigure}[b]{0.25\textwidth}
        \includegraphics[width=\textwidth]{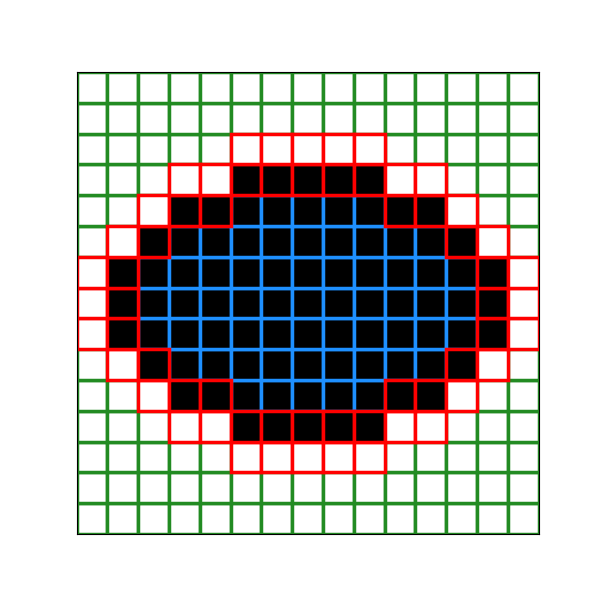}
        \caption{}
        \label{fig:steps6}
    \end{subfigure}
    \caption{%
Illustration of one iteration  $\Gamma^0 \to \Gamma^1$ of the proposed narrow band optimization algorithm.
{(a)} {Classification} of the current design~$\Gamma_0$ according to \eqref{eq:voxel_types}.
{(b)} {Projection onto the active set}, where the local optimization subproblem is solved.
{(c)} Solution of the optimization subproblem \eqref{eq:subprob_pert}: yellow for removed fluid, purple for added fluid.
{(d)} {Updated active set} with conserved volume.
{(e)} {Recovering of the frozen solid/fluid regions}.
{(f)} {Classification of the updated design~$\Gamma^1$ to proceed to the next iteration}.
}
    \label{fig:steps}
\end{figure}

We discretize the continuous problem as illustrated in Fig.~\ref{fig:steps}, where the design is shown on a regular background grid of size $N_\Gamma = n^2$, 
where \(N_\Gamma\) is the total number of cells, \(n\) is the number of cells in one direction. We define the vector of design variables as follows:
\begin{equation}
\Gamma \in \mathbb{R}^{N_\Gamma},
\end{equation}
where the discrete density \(\Gamma_{i,j} \in [0,1]\) for $i,j \in \{1,\ldots,n\}$ represents the volume fraction of fluid in the corresponding cell.
We aim to solve the following optimization problem:
\begin{equation}
\begin{aligned}
\min_{{\Gamma}\,\in\,\mathbb{R}^{N_\Gamma}}
\quad & \Phi({\Gamma})
\\
\text{subject to} 
\quad 
& \Gamma_j \in \{0,1\} \quad \forall j \in \{1,\ldots,N_\Gamma\},
\\
& \sum_{j=1}^{N_\Gamma} \Gamma_j = V_0,
\end{aligned}
\label{eq:orig}
\end{equation}
where \(\Phi(\Gamma)\) is a nonlinear objective functional and \(V_0\) is the desired volume constraint. The optimization problem \eqref{eq:orig} is formulated as a binary problem, i.e. each component \(\Gamma_j\) indicates whether cell \(j\) is solid or fluid.

Instead of directly solving \eqref{eq:orig} in the entire design space \(\{0,1\}^{N_\Gamma}\), inspired by the sparse grid concept we propose an iterative procedure that updates only an {active} subset of design variables at each step. Namely, we construct a sequence of designs:
\begin{equation}
{\Gamma}^0,\; {\Gamma}^1,\; \ldots,\; {\Gamma}^M,\;\ldots
\end{equation}
where \({\Gamma}^0\) is an initial design, and the next design is obtained by solving an optimization subproblem localized to a narrow band around the current fluid-solid interface where the computational effort is most needed. 
In what follows, we describe one iteration of the proposed optimization algorithm $\Gamma^0 \to \Gamma^1$.

\paragraph{Design Classification}
We partition the current design \({\Gamma}^0\) into two subsets:
\begin{equation}
\left\{
\begin{aligned}
&{\Gamma}^0_{a} \in \mathbb{R}^{N_a} &&: \text{narrow band of active cells}, \\
&{\Gamma}^0_{na} \in \mathbb{R}^{N_{na}} &&: \text{nonactive cells},
\end{aligned}
\right.
\label{eq:voxel_types}
\end{equation}
with the subset sizes satisfying \(N_{a} + N_{na} = N_\Gamma\). This partition is demonstrated in Fig. \ref{fig:steps1} where:
\begin{itemize}
    \item Active cells are colored in red, which are in the narrow band of cells that have at least one different neighbor;
    \item Nonactive cells include isolated solids colored in blue, i.e. solid cells that do not have fluid neighbors, and isolated fluids colored in green, i.e. fluid cells that do not have solid neighbors.
\end{itemize}
Note, in 2D we consider neighbors to be defined in the sense of 5-point stencil, i.e. for a cell $\Gamma_{i,j}$ there are four neighbors: $\Gamma_{i+1,j}, \Gamma_{i-1,j}, \Gamma_{i,j+1}, \Gamma_{i,j-1}$. Alternatively, 9-point stencil can be considered. Also, the presented algorithm can be easily extended to a larger narrow band width. 
For the ease of notation, we facilitate the above decomposition \eqref{eq:voxel_types} by the {permutation matrix} \(\mathbf{P}\in\mathbb{R}^{N_\Gamma\times N_\Gamma}\), given as follows:
\begin{equation}\label{eq:projection_matrix}
\mathbf{P} 
\;=\;
\begin{bmatrix}
\mathbf{P}_{a\phantom{n}}\\[4pt]
\mathbf{P}_{na}\\[4pt]
\end{bmatrix}, \quad \text{such that }
\mathbf{P}\,{\Gamma}_0
\;=\;
\begin{bmatrix}
\mathbf{P}_{a\phantom{n}}{\Gamma}^0\\
\mathbf{P}_{na}{\Gamma}^0
\end{bmatrix}
\;=\;
\begin{bmatrix}
{\Gamma}^0_{a\phantom{n}}\\
{\Gamma}^0_{na}
\end{bmatrix},
\end{equation}
\text{with }
$\mathbf{P}_{a}\in\mathbb{R}^{N_{a}\times N_\Gamma},\;
\mathbf{P}_{na}\in\mathbb{R}^{N_{na}\times N_\Gamma}$.
For the inverse permutation, we have: 
\begin{equation}
\mathbf{P}^\top
\begin{bmatrix}
{\Gamma}^0_{a\phantom{n}}\\
{\Gamma}^0_{na}\\
\end{bmatrix} 
\;=\;
\begin{bmatrix}
\mathbf{P}_{a}^\top &
\mathbf{P}_{na}^\top
\end{bmatrix} 
\begin{bmatrix}
{\Gamma}^0_{a}\\
{\Gamma}^0_{na}\\
\end{bmatrix} 
\;=\;
\mathbf{P}_{a}^\top\,{\Gamma}^0_{a\phantom{n}}
\;+\;
\mathbf{P}_{na}^\top\,{\Gamma}^0_{na}
\;=\;
{\Gamma}_0,
\end{equation}
where the operators $\mathbf{P}_{a}^\top$ and $\mathbf{P}_{na}^\top$ extend with zeros the nonactive and active cells, respectively.
So, for the permutation matrix, the following property is satisfied: 
\begin{equation}
\mathbf{P}^\top\,\mathbf{P} \;=\; \mathbf{P}\,\mathbf{P}^\top \;=\; \mathbf{I} \in\mathbb{R}^{N_\Gamma\times N_\Gamma}.
\end{equation}

\paragraph{Design update}
After classifying the current design $\Gamma^0$ according to \eqref{eq:voxel_types}, we look for an updated design in the following form:
\begin{equation}\label{gamma1_sc}
    {\Gamma}_{1}
    \;=\;
    \mathbf{P}_{na}^\top\,{\Gamma}^0_{na}
    \;+\;
    \mathbf{P}_{a}^\top\,\tilde{{\Gamma}}^0_{a}.
\end{equation}
In words, we freeze the nonactive cells $\Gamma^0_{na}$ so that they do not change during the update step. The frozen cells are eliminated in Fig. \ref{fig:steps2}. An updated narrow band of active cells, denoted \(\tilde{{\Gamma}}^0_{a}\), is found by solving the following local optimization subproblem: 
  \begin{equation}
  \begin{aligned}
  \min_{\tilde{{\Gamma}}^0_{a} \,\in\, \mathbb{R}^{N_a}}
  & \quad
  \Phi(
    \Gamma^1
  )
  \\
\text{subject to} 
\quad 
& (\tilde{{\Gamma}}^0_{a})_j \in \{0,1\} \quad \forall j \in \{1,\ldots,N_a\},
\\
& \sum_{j=1}^{N_{a}} (\tilde{{\Gamma}}^0_{a})_j = \sum_{j=1}^{N_{a}} ({{\Gamma}}^0_{a})_j.
\end{aligned}
  \label{eq:subprob}
\end{equation}
Thus, we aim to redistribute the material across the interface to improve the objective value: the solid active layer allows the interface to move in one direction while the fluid active layer allows it to move in another direction.
The last equation in \eqref{eq:subprob} guarantees that the volume constraint is satisfied for the updated design ${\Gamma}^{1}$ as soon as it is satisfied for the current design ${\Gamma}^{0}$.


\paragraph{Solving local subproblems using Sequential Linear Programming} 
We solve the local subproblem \eqref{eq:subprob} using the {Sequential Linear Programming} (SLP) approach \cite{dunning2015introducing}. 
In order to apply SLP, we first reformulate the optimization subproblem in the perturbation form. Namely, we look for a new design in the form:
\begin{equation}
    {\Gamma}^{1}
    \;=\;
    {\Gamma}^{0}
    \;+\;    \mathbf{P}_{a\phantom{n}}^\top\,\Delta{{\Gamma}}^0_{a\phantom{n}}.
\end{equation}
The optimization problem \eqref{eq:subprob} is then reformulated as follows:
\begin{equation}
  \begin{aligned}
  \min_{\Delta{{\Gamma}}^0_{a} \,\in\, \mathbb{R}^{N_{a}}}
  & \quad
  \Phi(
    {{\Gamma}}^{1}
    )
  \\
\text{subject to} 
\quad 
& (\Delta{{\Gamma}}^0_{a\phantom{n}})_j \in \{-({{\Gamma}}^{0}_{a})_j,1-({{\Gamma}}^0_{a})_j\} \quad \forall j \in \{1,\ldots,N_{a}\},
\\
& \sum_{j=1}^{N_{a}} (\Delta{{\Gamma}}^0_{a})_j = 0.
\end{aligned}
  \label{eq:subprob_pert}
\end{equation}
In SLP, a sequence of approximate solutions is generated:
\begin{equation}\label{eq:inner_iterations}
    \Gamma^{0,0}, \; \Gamma^{0,1}, \;\ldots,\; \Gamma^{0,m},
\end{equation}
where \(m\) is the number of SLP iterations, the initial design is \(\Gamma^{0,0} = \Gamma^0\), and subsequent iterates are given by:
\begin{equation}
    \Gamma^{0,k+1} = \Gamma^{0,k} + \mathbf{P}_{a\phantom{n}}^\top \Delta\Gamma^{0,k}_{a\phantom{n}},
    \quad
    k=0,\dots,m-1,
    \quad
    \Gamma^{1} = \Gamma^{0,m}.
\end{equation}
At each iteration, the update \(\Delta\Gamma_{0,k}^a\) is found by solving the following linear subproblem:
\begin{equation}
  \begin{aligned}
    \min_{\Delta{{\Gamma}}^{0,k}_{a} \in \mathbb{R}^{N_a}}
    & \quad
    \Bigl(\mathbf{P}_{a}\,\frac{d \Phi({\Gamma}^{0,k})}{d {\Gamma}} , \Delta{{\Gamma}}^{0,k}_{a} \Bigr)
    \\
    \text{subject to} 
    & \quad (\Delta{{\Gamma}}^{0,k}_{a\phantom{n}})_j \in \{-({{\Gamma}}^{0,k}_{a\phantom{n}})_j,\,1-({{\Gamma}}^{0,k}_{a\phantom{n}})_j\},
    \quad j = \{1,\ldots,N_{a}\},
    \\
    & \quad \sum_{j=1}^{N_{a\phantom{n}}} (\Delta{{\Gamma}}^{0,k}_{a\phantom{n}})_j = 0,
  \end{aligned}
  \label{eq:subprob_pert_linear}
\end{equation}
where the first-order Taylor expansion of the nonlinear objective \(\Phi(\Gamma)\) around the current iterate \({\Gamma}^{0,k}\) was used. The computation of the corresponding sensitivity vector $\frac{d \Phi(\Gamma_0)}{d\Gamma} \in \mathbb{R}^{N_\Gamma}$ is described in Section~\ref{sec:sens}.

The proposed optimization method is related to the SLP method for level-set topology optimization proposed by Dunning and Kim \cite{dunning2015introducing}, customized for fluid-based optimization. In comparison to the level-set optimization, our method operates directly on the density level and does not require an auxiliary evolution equation such as the Hamilton–Jacobi equation. Thus, the proposed narrow-band optimization algorithm can be seamlessly integrated into standard density-based frameworks for large-scale thermal-fluid topology optimization. The overlap and similarities between level set and density-based optimization were also pointed out by Andreasen \textit{et al.} \cite{andreasen2020level}.

{
\begin{remark}
    The fixed‑volume constraint used in \eqref{eq:subprob}–\eqref{eq:subprob_pert_linear} (i.e., $\sum_{j\in a} (\Delta\Gamma^{0,k}_a)_j = 0$) is not a severe limitation of the method. Varying volume during optimization can be achieved, e.g., by an additional outer loop where the target volume is adjusted by shrinking/expanding the interface.
\end{remark}
}





\section{Sparse grid discretization}
\label{sec:discrete}

The Stokes--Brinkman problem \eqref{eq:stokes_brinkman} is discretized using the finite difference method on staggered grids, also known as Marker-and-Cell (MAC) scheme \cite{harlow1965numerical,lebedev1964difference,griebel1998numerical}. 
{We adapt the Stokes–Brinkman solver of Pimanov \emph{et al.} \cite{pimanov2022workflow,pimanov2024efficient} to fluid‑based topology optimization. We prefer the staggered finite difference discretization over the finite element method (FEM) because (i) equal‑order velocity–pressure FEM pairs violate the discrete inf–sup condition and thus require stabilization  \cite{brezzi2012mixed,tezduyar1991stabilized}, whereas the MAC scheme satisfies a discrete inf–sup condition by construction \cite{rui2017stability}; and (ii) for Brinkman flow with high‑contrast permeability, standard Stokes‑stable pairs (e.g., Taylor–Hood) are not robust and typically need additional stabilization or specialized preconditioners \cite{hannukainen2011computations,efendiev2012robust,vassilevski2013block}.}

\begin{figure}[htbp]
    \centering
    \includegraphics[width=0.5\textwidth]{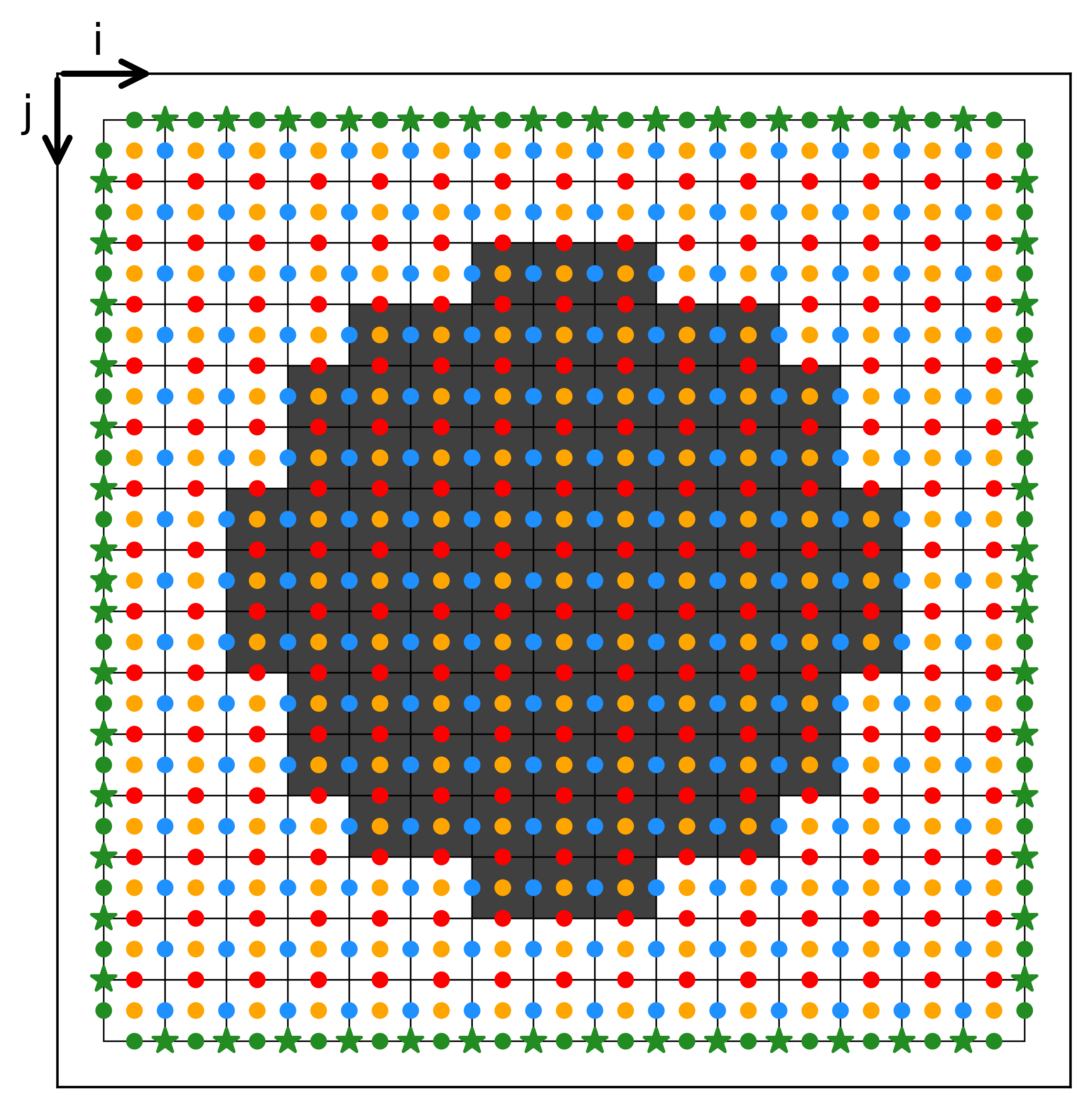}
    \caption{Staggered grid DoFs arrangement: blue and red for velocity components, orange for pressure, temperature, and density. Boundary conditions are applied at green circles and stars.}
    \label{fig:staggered}
\end{figure}

In the staggered grid arrangement, the pressure grid coincides with the background grid while the components of the velocity $U$, denoted $U^x$ and $U^y$, are defined on the cell faces. For the heat equation \eqref{eq:heat_equation}, we use the centered finite-difference scheme for discretizing the Laplacian operator $\Delta \theta$ and the upwind scheme tailored for a high–Péclet number regime \cite{shyy1985study} for discretizing the convective term $\nabla \cdot\bigl({u}(\gamma)\,\theta\bigr)$. The degrees of freedom (DoFs) are located as follows (see Fig. \ref{fig:staggered}):
\begin{itemize}
    \item \(P_{i,j}\), \(\Theta_{i,j}\), \(\Gamma_{i,j}\) are defined at the cell centers (orange);
    \item \(U^x_{i+\frac{1}{2},j}\) is defined at the vertical edges (blue);
    \item \(U^y_{i,j+\frac{1}{2}}\) is defined at the horizontal edges (red).
\end{itemize}
For the interior DoFs, the discretization of the state equations \eqref{eq:stokes_brinkman} and \eqref{eq:heat_equation} proceeds as follows.

\noindent
At the points \(U^x_{i+\frac{1}{2},j}\), the discrete \(x\)-momentum equation is given by,
\begin{equation}
\label{discrete_momentum_x}
\begin{aligned}
    -\nu\,\frac{U^x_{i+\frac{3}{2},j} - 2U^x_{i+\frac{1}{2},j} + U^x_{i-\frac{1}{2},j}}{h^2} 
    - \nu\,\frac{U^x_{i+\frac{1}{2},j+1} - 2U^x_{i+\frac{1}{2},j} + U^x_{i+\frac{1}{2},j-1}}{h^2} \\
    + \frac{\alpha(\Gamma_{i+1,j}) + \alpha(\Gamma_{i,j})}{2}\,U^x_{i+\frac{1}{2},j}
    + \frac{P_{i+1,j} - P_{i,j}}{h} = 0,
\end{aligned}
\end{equation}
 where $h = 1/n$ is the cell size.
At the points \(U^y_{i,j+\frac{1}{2}}\), the discrete \(y\)-momentum equation is written as,
\begin{equation}
\label{discrete_momentum_y}
\begin{aligned}
    -\nu\,\frac{U^y_{i+1,j+\frac{1}{2}} - 2U^y_{i,j+\frac{1}{2}} + U^y_{i-1,j+\frac{1}{2}}}{h^2} 
    - \nu\,\frac{U^y_{i,j+\frac{3}{2}} - 2U^y_{i,j+\frac{1}{2}} + U^y_{i,j-\frac{1}{2}}}{h^2} \\
    + \frac{\alpha(\Gamma_{i,j+1}) + \alpha(\Gamma_{i,j})}{2}\,U^y_{i,j+\frac{1}{2}}
    + \frac{P_{i,j+1} - P_{i,j}}{h} = 0.
\end{aligned}
\end{equation}
The discrete incompressibility condition is written at the cell centers, {where \(P_{i,j}\) is defined}, as follows:
\begin{equation}
\label{discrete_incompressibility}
    \frac{U^x_{i+\frac{1}{2},j} - U^x_{i-\frac{1}{2},j}}{h} +
    \frac{U^y_{i,j+\frac{1}{2}} - U^y_{i,j-\frac{1}{2}}}{h} = 0.
\end{equation}

\noindent
The discrete heat equation \eqref{eq:heat_equation} is written at the cell centers for $\Theta_{i,j}$ as follows:
\begin{equation}
\label{heat_discrete}
    -k\,\frac{\Theta_{i+1,j} - 2\Theta_{i,j} + \Theta_{i-1,j}}{h^2} 
    - k\,\frac{\Theta_{i,j+1} - 2\Theta_{i,j} + \Theta_{i,j-1}}{h^2} + \left[ \frac{\partial (U^x\Theta)}{\partial x} \right]_{i,j}  + \left[ \frac{\partial (U^y\Theta)}{\partial y} \right]_{i,j} = Q_{i,j},
\end{equation}
where the convective terms are given by,
\begin{equation}
\begin{aligned}
\left[ \frac{\partial (U^x\Theta)}{\partial x} \right]_{i,j} = & \,
\frac{1}{h} \Big(
    U^x_{i+\frac{1}{2},j} \frac{\Theta_{i+1,j} + \Theta_{i,j}}{2} 
     - U^x_{i-\frac{1}{2},j} \frac{\Theta_{i,j} + \Theta_{i-1,j}}{2}
\Big) \\
& - \frac{1}{h} \Big(
    |U^x_{i+\frac{1}{2},j}| \frac{\Theta_{i+1,j} - \Theta_{i,j}}{2} 
     - |U^x_{i-\frac{1}{2},j}| \frac{\Theta_{i,j} - \Theta_{i-1,j}}{2}
\Big),
\end{aligned}
\end{equation}
and,
\begin{equation}
\begin{aligned}
\left[ \frac{\partial (U^y\Theta)}{\partial y} \right]_{i,j} = & \,
\frac{1}{h} \Big(
    U^y_{i,j+\frac{1}{2}} \frac{\Theta_{i,j+1} + \Theta_{i,j}}{2} 
     - U^y_{i,j-\frac{1}{2}} \frac{\Theta_{i,j} + \Theta_{i,j-1}}{2}
\Big) \\
& - \frac{1}{h} \Big(
    |U^y_{i,j+\frac{1}{2}}| \frac{\Theta_{i,j+1} - \Theta_{i,j}}{2} 
     - |U^y_{i,j-\frac{1}{2}}| \frac{\Theta_{i,j} - \Theta_{i,j-1}}{2}
\Big).
\end{aligned}
\end{equation}

\paragraph{System Assembly}
We consider the vectors of state variables,
\begin{equation}
U \in \mathbb{R}^{N_U}, \quad P \in \mathbb{R}^{N_P}, \quad \Theta \in \mathbb{R}^{N_\Theta},
\end{equation}
with,
\begin{equation}
N_U = 2n(n-1), \quad N_P = N_\Theta = n^2 = N_\Gamma,
\end{equation}
representing the number of DoFs for the velocity, pressure and temperature, respectively.

By combining equations \eqref{discrete_momentum_x}--\eqref{discrete_incompressibility} and imposing the Dirichlet boundary conditions on \({\partial\Omega}_{{u}}^D\), the discrete system can be assembled into the following block form:
\begin{equation}
\label{stokes_brinkman_system}
\begin{bmatrix}
-\nu\mathbf{L}_U + \mathbf{A}(\Gamma) & \mathbf{B}^\top \\
\mathbf{B} & 
\end{bmatrix}
\begin{bmatrix}
{U} \\
P
\end{bmatrix}
=
\begin{bmatrix}
{F_U} \\
F_P
\end{bmatrix},
\end{equation}
where \(\mathbf{L}_U \in \mathbb{R}^{N_U\times N_U}\) denotes the discrete velocity Laplacian, \(\mathbf{B} \in \mathbb{R}^{N_U\times N_P}\) the discrete divergence operator (with its transpose corresponding to the discrete pressure gradient), and the nonzero right-hand side vectors \({F_U} \in \mathbb{R}^{N_U}\) and \(F_P \in \mathbb{R}^{N_P}\) arise from the imposed Dirichlet boundary conditions. In our implementation, the boundary values for the normal velocity component are directly given at the cell faces (denoted by green circles in Fig.~\ref{fig:staggered}), while those for the tangential velocity component are incorporated via linear interpolation from the respective boundary points (denoted by green stars in Fig.~\ref{fig:staggered}). 
The dependence of the state variables on the design variables is reflected in the operator $\mathbf{A}(\Gamma):\mathbb{R}^{N_\Gamma}\to\mathbb{R}^{N_U\times N_U}$, where the matrix $\mathbf{A}$ corresponds to the discretized Brinkman term $\alpha(\Gamma)$ for any $\Gamma$.

By combining equation \eqref{heat_discrete} with the Dirichlet or Neumann boundary conditions on \({\partial\Omega}_{T}^D\) or \({\partial\Omega}_{T}^N\), the discrete system can be assembled into the following form:
\begin{equation}
\label{heat_system}
    (-k\mathbf{L}_\Theta + \mathbf{C}({U}(\Gamma))) \Theta = F_\Theta,
\end{equation}
where $\mathbf{L}_\Theta \in \mathbb{R}^{N_\Theta \times N_\Theta}$ is the discrete temperature Laplacian.  The dependency of the state variable $\Theta$ on the design variable is reflected by the operator $\mathbf{C}(U(\Gamma)):\mathbb{R}^{N_\Gamma} \to \mathbb{R}^{N_U} \to \mathbb{R}^{N_\Theta\times N_\Theta}$, where $\mathbf{C} \in \mathbb{R}^{N_\Theta \times N_\Theta}$ is the convection matrix assembled using the velocity solution ${U}$ obtained via solving the system \eqref{stokes_brinkman_system} for any design $\Gamma$.
The values for the Neumann boundary condition are directly given at the cell faces (green circles in Fig.~\ref{fig:staggered}), while those for the Dirichlet boundary condition are incorporated via linear interpolation from the same boundary points.
In what follows, we denote \eqref{stokes_brinkman_system} and \eqref{heat_system} as the state equations. 

\paragraph{Excluding fully isolated solid cells} 
One key feature of the proposed narrow-band optimization algorithm is that it allows us to exclude the solid cells  known to remain solid from the analysis during optimization. \begin{figure}[htbp]
    \centering
    \includegraphics[width=0.5\textwidth]{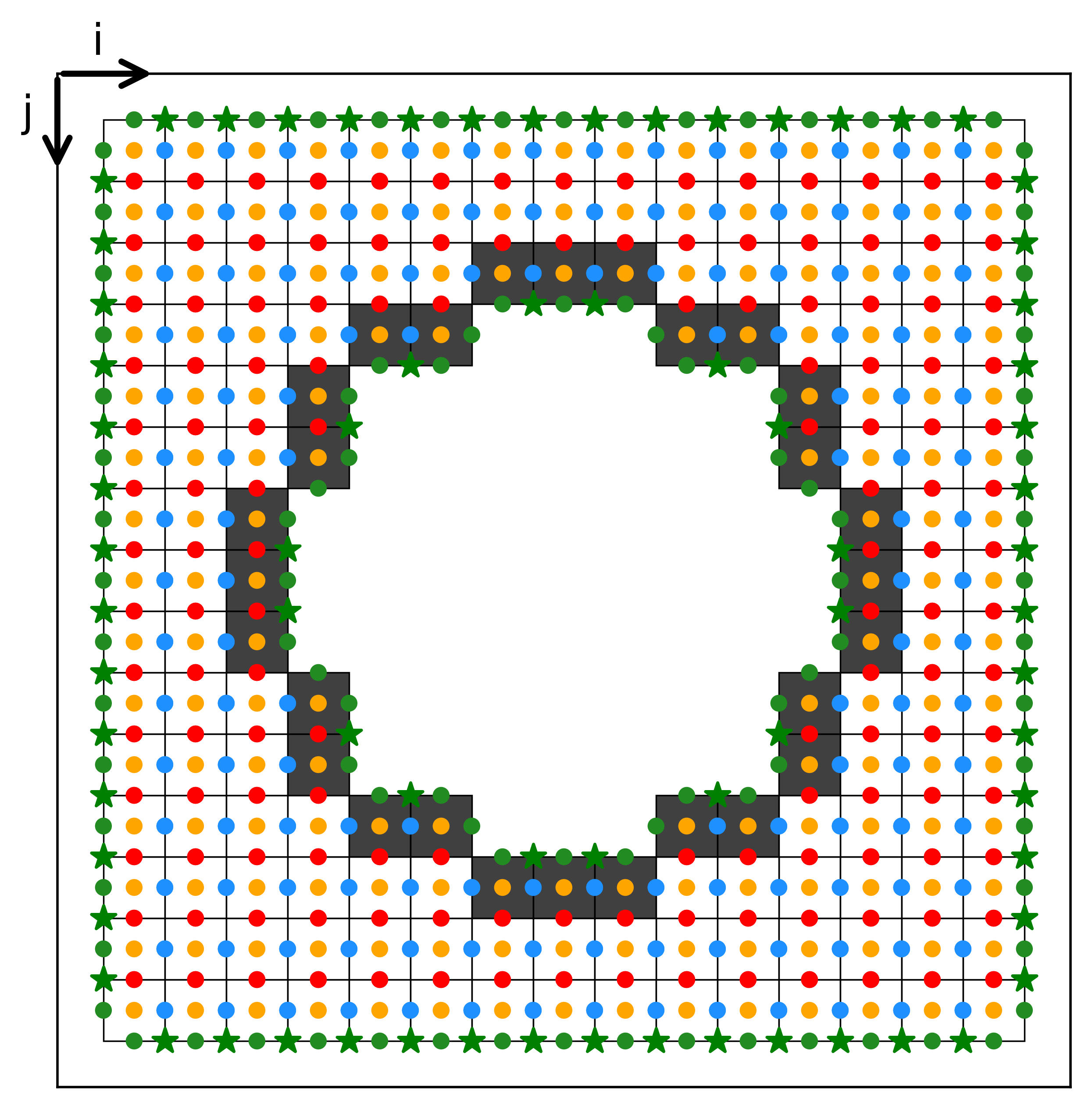}
    \caption{Excluding fully isolated solid cells and imposing the no-slip boundary conditions on the fluid-solid interface. One layer of Brinkman cells allows the interface to move inward during optimization.}
    \label{fig:staggered_removed}
\end{figure}
Similarly to \eqref{eq:voxel_types}, we consider the following partition of the current design \({\Gamma}^0\) into two subsets (see Fig. \ref{fig:steps1}):
\begin{equation}
\left\{
\begin{aligned}
&{\Gamma}^0_{a+if} \in \mathbb{R}^{N_{a}+N_{if}} &&: \text{narrow band of active cells (red) and isolated fluids (green)}, \\
&{\Gamma}^0_{is} \in \mathbb{R}^{N_{is}} &&: \text{isolated solids (blue)},
\end{aligned}
\right.
\label{eq:voxel_types_2}
\end{equation}
where \(N_{if}\) and \(N_{is}\) denote the number of fully isolated fluids and fully isolated solids, respectively, such that \(N_{if} + N_{is} = N_{na}\).


To improve computational efficiency, we solve the state equation \eqref{stokes_brinkman_system} only in the reduced domain \(\Gamma^0_{a+if}\), excluding the isolated solid cells \(\Gamma^0_{is}\). In Fig. \ref{fig:staggered_removed}, we demonstrate the resulting domain \(\Gamma^0_{a+if}\), where only one layer of solid cells is retained to allow the boundary to move inward during the optimization process.
Formally, we solve the following system of a smaller size:
\begin{equation}
\label{stokes_brinkman_system_ex}
\begin{bmatrix}
-\nu\overline{\mathbf{L}}_U + \overline{\mathbf{A}}(\Gamma) & \overline{\mathbf{B}}^\top \\
\overline{\mathbf{B}} & 
\end{bmatrix}
\begin{bmatrix}
\overline{{U}} \\
\overline{P}
\end{bmatrix}
=
\begin{bmatrix}
{\overline{F}_U} \\
\overline{F}_P
\end{bmatrix},
\end{equation}
with $\overline{{U}},{\overline{F}_U} \in \mathbb{R}^{N_{\overline{U}}}$ and $\overline{P},{\overline{F}_P} \in \mathbb{R}^{N_{\overline{P}}}$. Thus, the number of pressure DoFs coincides with the number of remaining cells, $N_{\overline{P}} = N_{a}+N_{if}$, and the corresponding velocity DoFs are denoted by red and blue circles in Fig. \ref{fig:staggered_removed}. In addition, we impose zero Dirichlet boundary condition on the stair-like fluid-solid interface in between of $\Gamma^0_{a+if}$ and $\Gamma^0_{is}$. Similarly to the outer boundary, the boundary values for the normal velocity component are directly given at the cell faces (green circles on the inner boundary in Fig.~\ref{fig:staggered_removed}), while those for the tangential velocity component are incorporated via linear interpolation from the respective boundary points (green stars on the inner boundary in Fig.~\ref{fig:staggered_removed}).
This removal of voxels enables the use of a sparse data structure. We store the 3D locations of active voxels in a linear list (1D array) and maintain an additional 3D array that maps each voxel to its corresponding 1D index. This approach allows for efficient iteration over the 1D list and quick determination of neighboring voxels using the auxiliary 3D array, eliminating the need to store a full adjacency graph.

\subsection{Numerical method and preconditioning}\label{sec:cguzawa}
\paragraph{Flow equation}
To efficiently solve the Stokes-Brinkman system, we follow the pressure Schur complement approach: We reduce the coupled system~\eqref{stokes_brinkman_system_ex} to an equivalent pressure‑only problem:
\begin{equation}\label{schur_system}
  \mathbf{S} \overline{P} \;=\; \overline{F}_S,
\end{equation}
where the Schur complement matrix $\mathbf{S}$ and the right-hand side $\overline{F}_S$ are defined as follows:
\begin{equation}\label{schur_rhs}
  \mathbf{S}
  \;=\;
  \overline{\mathbf{B}}\,\bigl(-\nu\overline{\mathbf{L}}_{U}+\overline{\mathbf{A}}(\Gamma)\bigr)^{-1}\overline{\mathbf{B}}^{\top},
  \qquad
  \overline{F}_S
  \;=\;
  \overline{\mathbf{B}}\,\bigl(-\nu\overline{\mathbf{L}}_{U}+\overline{\mathbf{A}}(\Gamma)\bigr)^{-1}\overline{F}_{U}
  - \overline{F}_{P}.
\end{equation}
After the pressure $\overline{P}$ is computed, the velocity field is recovered as follows:
\begin{equation}\label{velocity_schur}
  \bigl(-\nu\overline{\mathbf{L}}_{U}+\overline{\mathbf{A}}(\Gamma)\bigr)\,\overline{U}
  \;=\;
  \overline{F}_{U}-\overline{\mathbf{B}}^{\top}\overline{P}.
\end{equation}
Because the matrix $\mathbf{S}$ is symmetric positive definite, we solve~\eqref{schur_system} with the conjugate-gradient (CG) method, which can be viewed as a Krylov acceleration of the classical Uzawa algorithm.  
Each application of~$\mathbf{S}$ requires the action of the inverse of the perturbed velocity Laplacian \(-\nu\overline{\mathbf{L}}_{U}+\overline{\mathbf{A}}(\Gamma)\). Forming this inverse explicitly would produce a dense matrix, so we approximate the operation by an inner CG solve preconditioned with the classical algebraic multigrid (AMG) method. AMG has proved robust for the irregular sparsity patterns that arise in complex geometries ~\cite{ruge1987algebraic}—such as those produced by topology optimization.

For the Stokes equations, this numerical method—which we refer to as CG-Uzawa—was first proposed in~\cite{verfurth1984combined}, and its superlinear convergence at the operator level for smooth domains was later demonstrated by Axelsson \textit{et~al.}~\cite{axelsson2021krylov}. In the context of porous-media flow, the CG-Uzawa method was implemented and analyzed by Pimanov \textit{et~al.}~\cite{pimanov2024efficient}. To the best of the authors’ knowledge, the present work is the first application of the method to the Stokes–Brinkman equations.

\paragraph{Heat equation}
The discretized heat convection–diffusion equation \eqref{heat_system} and its adjoint are solved using the flexible GMRes method, preconditioned with AMG.

    

\subsection{Computing Sensitivities}\label{sec:sens}
Throughout an optimization process, it is required a repeated computation of the total derivative of the objective functional \(\Phi(\Gamma)\) with respect to design variable $\Gamma$, which is defined as follows:
\begin{equation}\label{eq:total_derivative}
\Phi(\Gamma^0+\Delta\Gamma^0)=\Phi(\Gamma^0)+\left(\frac{d\Phi(\Gamma^0)}{d\Gamma},\,\Delta\Gamma^0\right)+o(\|\Delta\Gamma^0\|).
\end{equation}
In general, we consider linear functionals with respect to the state variables $U$, $P$ in the following form: 
\begin{equation}
\label{objective_functional_up}
    \Phi_{(U,P)}(\Gamma) = ({G_U},{U}(\Gamma)) + (G_P,P(\Gamma)),
\end{equation}
where ${G_U} \in \mathbb{R}^{N_U}$, ${G_P} \in \mathbb{R}^{N_P}$ and $(\cdot,\cdot)$ denotes the vector dot product.
Similarly, linear functionals with respect to the state variable $\Theta$ are considered in the form:
\begin{equation}
\label{objective_functional_T}
    \Phi_\Theta(\Gamma) = (G_\Theta,\Theta(\Gamma)),
\end{equation}
where ${G_\Theta} \in \mathbb{R}^{N_\Theta}$. For instance, when optimizing a heat exchanger, \(\Phi_{(U,P)}(\Gamma)\) represents the pressure drop between the flow inlet and outlet, while \(\Phi_\Theta(\Gamma)\) represents the convective heat exchange between the cold and hot phases.
In our numerical experiments, we consider a weighted objective functional:
\begin{equation}
\label{objective_functional}
    \Phi(\Gamma) = \omega\Phi_{(U,P)}(\Gamma) + (1-\omega) \Phi_\Theta(\Gamma),
\end{equation}
where $0 < \omega \leq 1$ is the weight. Thus, an evaluation of $\Phi(\Gamma)$ for any $\Gamma$ requires solving the state equations \eqref{stokes_brinkman_system} and \eqref{heat_system}.
For notational simplicity, we redefine,
\begin{equation}\label{eq:definitions}
X = \begin{bmatrix} U \\ P \end{bmatrix},\quad 
G_X = \begin{bmatrix} G_U \\ G_P \end{bmatrix},\quad 
F_X = \begin{bmatrix} F_U \\ F_P \end{bmatrix} \in \mathbb{R}^{N_X},\quad N_X = N_U+N_P.
\end{equation}
Then, the state equation \eqref{stokes_brinkman_system} can be written as follows:
\begin{equation}\label{eq:state_equation}
\mathbf{M}_X(\Gamma)X = F_X,
\end{equation}
and the corresponding objective \eqref{objective_functional_up} becomes:
\begin{equation}\label{eq:objective_x}
\Phi_X(\Gamma) = \bigl(G_X,\,X(\Gamma)\bigr) = \Bigl(G_X,\,\mathbf{M}_X^{-1}(\Gamma)F_X\Bigr).
\end{equation}
Differentiating \eqref{eq:objective_x} with respect to \(\Gamma_i, \forall i \in \{1,\ldots,N_\Gamma\}\) and using the identity,
\begin{equation}\label{eq:derivative_inverse}
\frac{d\mathbf{M}_X^{-1}}{d\Gamma_i} = -\mathbf{M}_X^{-T}\frac{\partial\mathbf{M}_X}{\partial\Gamma_i}\mathbf{M}_X^{-1},
\end{equation}
we obtain,
\begin{equation}\label{eq:derivative_phi}
\frac{d\Phi_X(\Gamma_0)}{d\Gamma_i} = \Bigl(G_X,\,\frac{dX(\Gamma_0)}{d\Gamma_i}\Bigr)
= -\Bigl(X^*,\,\frac{\partial\mathbf{M}_X(\Gamma_0)}{\partial\Gamma_i}X\Bigr),
\end{equation}
where \(X^*\) is found by solving the adjoint problem,
$
X^* = \mathbf{M}_X^{-\top}G_X = \mathbf{M}_X^{-1}G_X.
$
Since only the $(0,0)$-th block in \(\mathbf{M}_X(\Gamma)\) depends on $\Gamma$, we have,
\begin{equation}\label{eq:structure_M}
\frac{\partial\mathbf{M}_X(\Gamma_0)}{\partial\Gamma_i} =
\begin{bmatrix}
\frac{\partial \mathbf{A}(\Gamma_0)}{\partial \Gamma_i} & \mathbf{0} \\
\mathbf{0} & \mathbf{0}
\end{bmatrix}.
\end{equation}
Thus, \eqref{eq:derivative_phi} simplifies to,
\begin{equation}\label{eq:sensitivity_u}
\frac{d\Phi_X(\Gamma_0)}{d\Gamma_i} = -\Bigl(U^*,\,\frac{\partial \mathbf{A}(\Gamma_0)}{\partial \Gamma_i}U\Bigr).
\end{equation}
Next, consider the heat equation \eqref{heat_system}, redefined as follows,
\begin{equation}\label{eq:heat_state}
\mathbf{M}_\Theta(X(\Gamma))\Theta = F_\Theta,
\end{equation}
with the corresponding objective \eqref{objective_functional_T} written as,
\begin{equation}\label{eq:objective_theta}
\Phi_\Theta(\Gamma) = \bigl(G_\Theta,\,\Theta(\Gamma)\bigr)
= \Bigl(G_\Theta,\,\mathbf{M}_\Theta^{-1}(X(\Gamma))F_\Theta\Bigr).
\end{equation}
Then, differentiating \eqref{eq:objective_theta} by the chain rule, we have,
\begin{equation}\label{eq:chain_rule}
\frac{d\Phi_\Theta(X(\Gamma^0))}{d\Gamma_i} = \left(\frac{d\Phi_\Theta(X^0)}{dX},\,\frac{dX(\Gamma^0)}{d\Gamma_i}\right), \text{ with } X^0=X(\Gamma^0).
\end{equation}
 Next, using \eqref{eq:derivative_phi} with
\(
G_X = \dfrac{d\Phi_\Theta(X^0)}{dX}
\),
 equation \eqref{eq:chain_rule} can be written as,
\begin{equation}\label{eq:derivative_phi_theta}
\frac{d\Phi_\Theta(X(\Gamma^0))}{d\Gamma_i} = -\Bigl(X^{**},\,\frac{\partial\mathbf{M}_X(\Gamma_0)}{\partial\Gamma_i}X\Bigr),
\end{equation}
where $X^{**}$ is found by solving the adjoint problem,
$
X^{**} = \mathbf{M}_X^{-1}\dfrac{d\Phi_\Theta(X^0)}{dX}.
$
Finally, differentiating \(\Phi_\Theta\) with respect to \(X_j, \forall j \in \{1,\ldots,N_X\}\) gives,
\begin{equation}\label{eq:phi_theta_derivative}
\frac{d\Phi_\Theta(X^0)}{dX_j} = -\Bigl(\Theta^*,\,\frac{\partial \mathbf{M}_\Theta(X^0)}{\partial X_j}\Theta\Bigr),
\end{equation}
where $\Theta^*$ is found by solving the heat adjoint,
$\Theta^* = \mathbf{M}_\Theta^{-\top}G_\Theta$.
Note, that the heat adjoint operator for the considered upwind discretization scheme satisfies,
\begin{equation}\label{eq:adjoint_operator}
\mathbf{M}_\Theta^\top = -k\mathbf{L}_\Theta + \mathbf{C}^\top(U(\Gamma)) = -k\mathbf{L}_\Theta - \mathbf{C}(U(\Gamma)),
\text{ and }
\frac{\partial \mathbf{M}_\Theta(X^0)}{\partial X_j} = \frac{\partial \mathbf{C}(X^0)}{\partial X_j}.
\end{equation}
It is noted, that due to the nondifferentiability of the modulus at \(U_i=0\) (see \eqref{heat_discrete}), for differentiating $\mathbf{C}(U)$ we use one-sided derivatives based on \(\operatorname{sign}(U_i)\) (alternatively, a smoothing approach or an \(\varepsilon\)-neighborhood can be employed).
In general, three solutions (one forward and two adjoint) are required for the Stokes–Brinkman problem \eqref{stokes_brinkman_system} and two solutions (one forward and one adjoint) are required for the heat equation \eqref{heat_system}. The number of forward Stokes-Brinkman solves is reduced to two (one forward and one adjoint) in the self-adjoint case $F_U = G_U$ since $\mathbf{M}_U^\top = \mathbf{M}_U$.

\section{Numerical Experiments}
\label{sec:numerical}
\paragraph{Physical units and parameters}

We begin by specifying the physical units that are {omitted hereafter} and understood implicitly.  
Throughout this section we employ SI units: Velocity \(u\,\bigl[\text{m}\,\text{s}^{-1}\bigr]\),  
pressure \(p\,\bigl[\text{Pa}\bigr]\),  
kinematic viscosity \(\nu\,\bigl[\text{m}^{2}\,\text{s}^{-1}\bigr]\),  
temperature \(\theta\,\bigl[\text{K}\bigr]\),  
thermal conductivity \(k\,\bigl[\text{W}\,\text{m}^{-1}\,\text{K}^{-1}\bigr]\),  
and volumetric heat source \(q\,\bigl[\text{W}\,\text{m}^{-3}\bigr]\).

Several parameters remain fixed in all numerical experiments.  
The design domain has a characteristic length of \(1\,\text{m}\), i.e.\ \(\Omega = (0,1)^{3}\,\text{m}^{3}\),  
and the viscosity is set to \(\nu = 1\,\text{m}^{2}\,\text{s}^{-1}\).  
We also assume a fluid density of \(\rho = 1\,\text{kg}\,\text{m}^{-3}\) and a specific heat capacity of \(c_{p} = 1\,\text{J}\,\text{K}^{-1}\,\text{kg}^{-1}\).

For the material–interpolation law in Eq.~\eqref{brinkman_interpolation}, which supplies the inverse-permeability term in the Stokes–Brinkman equation~\eqref{eq:stokes_brinkman}, we choose  
\(\alpha_{\text{max}} = 10^{6}\) and \(q_{a} = 10\).

\paragraph{Third‑party software}

The optimization subproblems are solved with GLOP, the linear programming component of Google OR‑Tools\,\cite{ortools_routing}, which implements a revised primal–dual simplex algorithm optimized for sparse matrices.  
Linear systems involving the perturbed velocity Laplacian, \(-\nu\mathbf{L}_{U} + \mathbf{A}(\Gamma)\), and the heat convection–diffusion operator, \(-k\mathbf{L}_{\Theta} + \mathbf{C}\bigl(U(\Gamma)\bigr)\), together with its adjoint, are solved using the algebraic multigrid library SAMG \,\cite{stuben2017algebraic} with the default settings.

Solver tolerances are kept fixed throughout: for the CG–Uzawa method we set the outer tolerance to \(\varepsilon_{S} = 10^{-3}\), whereas the algebraic multigrid solver uses a relative residual tolerance of \(10^{-6}\).

\paragraph{Computational environment} 
All numerical results presented in this work were produced on a workstation with an Intel(R) Xeon(R) w7-2495X CPU (24 cores, 48 threads) and 512\,GB of memory. To avoid saturating the memory bandwidth, the number of active threads was limited to 12.

\subsection{Example 1: Optimization of a fluid manifold with multiple outlets}
\label{subsec:manifold}
\begin{figure}[htbp]
    \centering
    \begin{subfigure}[b]{0.35\textwidth}
        \includegraphics[width=\textwidth]{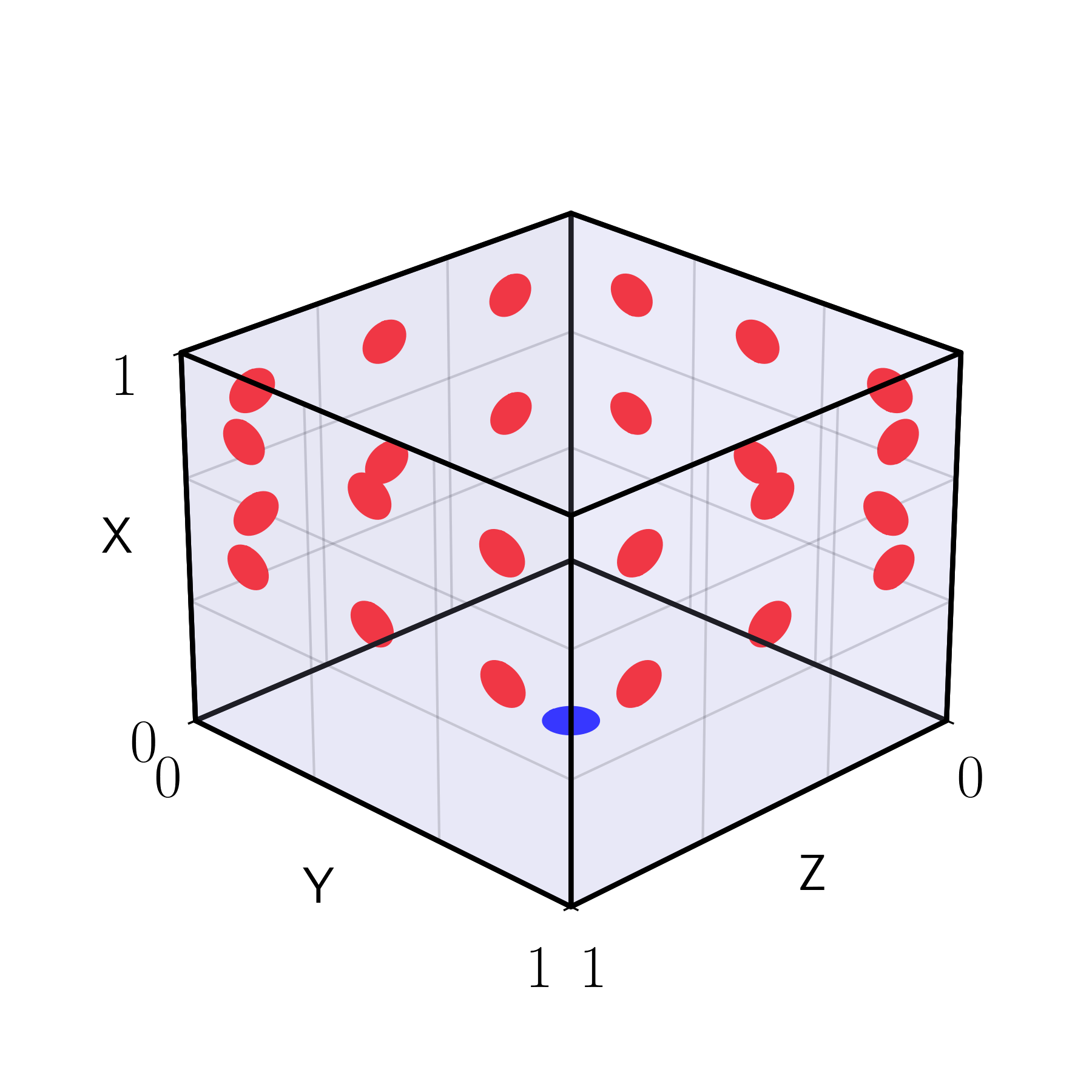}
        \caption{Setup}
        \label{fig:ex1_setup}
    \end{subfigure}
    \hfill
    \begin{subfigure}[b]{0.3\textwidth}
        \includegraphics[width=\textwidth]{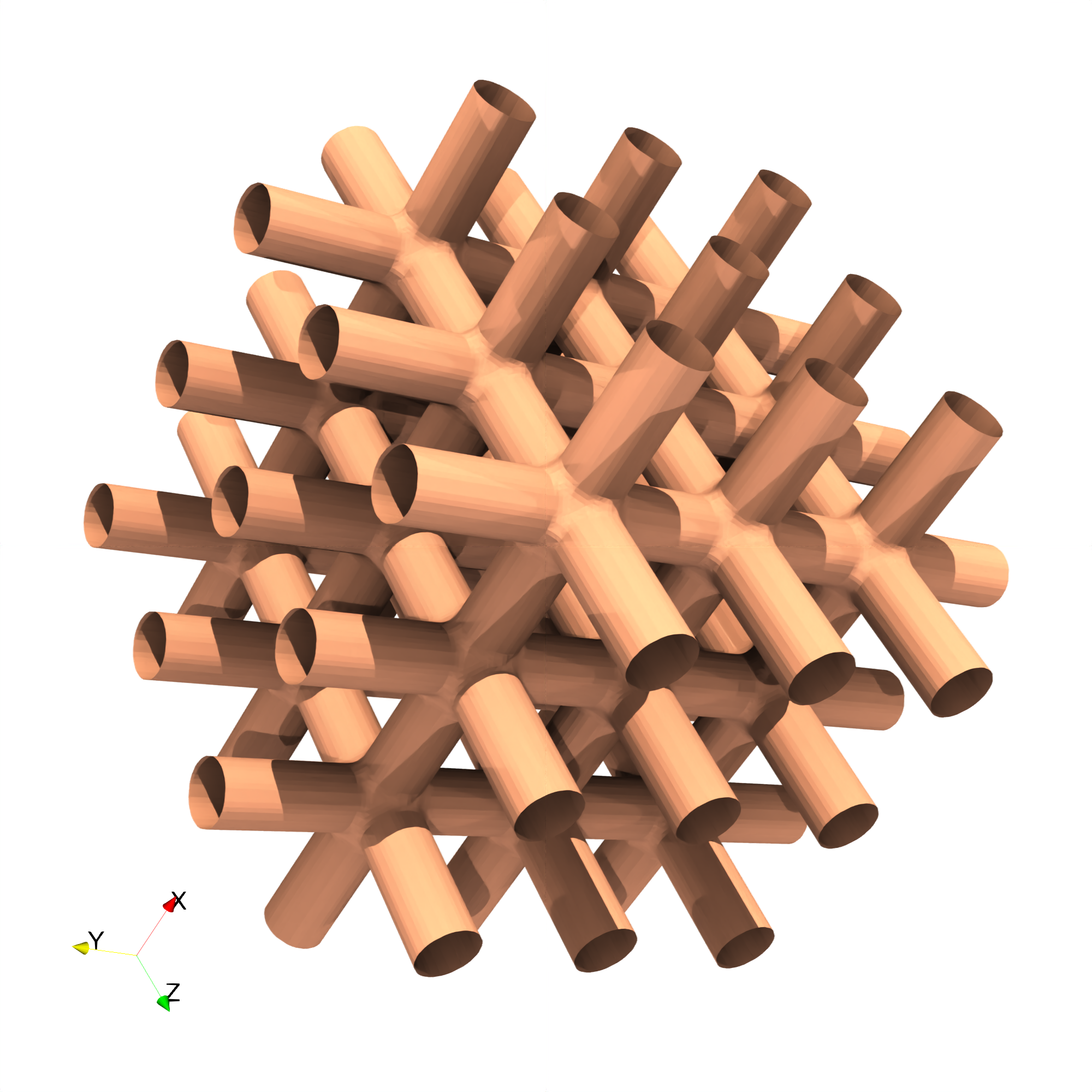}
        \caption{Initial design}
        \label{fig:ex1_initial}
    \end{subfigure}
    \hfill
    \begin{subfigure}[b]{0.3\textwidth}
        \includegraphics[width=\textwidth]{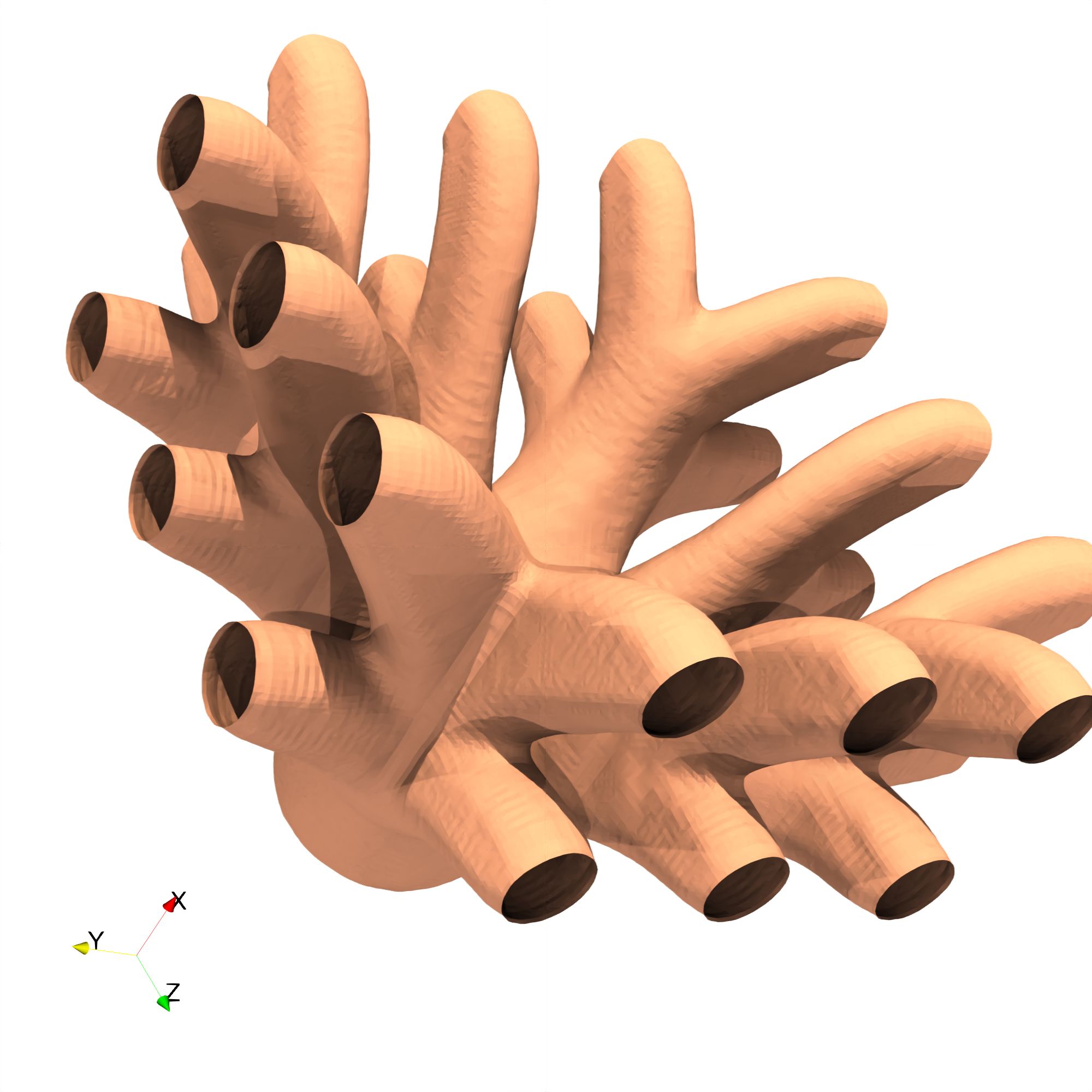}
        \caption{Final design}
        \label{fig:ex1_final}
    \end{subfigure}
    \caption{Minimization of the pressure drop for a fluid manifold with multiple outlets, volume constraint $V_0=0.15$.}
    \label{fig:fingers}
\end{figure}

In this subsection we revisit the pure-fluid manifold optimization problem with multiple outlets, as studied by Liu \textit{et al.} ~\cite{liu2022marker}.  
The objectives are twofold:  
\begin{enumerate}
  \item To compare the performance of the flow solver for the {full} Stokes–Brinkman problem with that of a {reduced} formulation in which isolated solid cells are excluded;
  \item To examine how varying the number of inner (SLP) iterations \(m\) in~\eqref{eq:inner_iterations} affects the convergence of the narrow-band optimization algorithm and the final design.
\end{enumerate}
Figure~\ref{fig:ex1_setup} sketches the flow configuration.  
A single circular inlet (blue) of radius \(r = 0.05\), denoted \(\partial\Omega_{\mathrm{in}}\), is located on the plane \(x = 0\).  
Twenty-four circular outlets (red) of the same radius, denoted \(\partial\Omega_{\mathrm{out}}\), are located on the planes \(y = 0\), \(y = 1\), \(z = 0\), and \(z = 1\).  

Dirichlet velocity boundary conditions are prescribed at the inlet and outlets:
\begin{equation}
\label{ex1:dirichlet_bc}
  u = u_{\mathrm{in}} \quad\text{on } \partial\Omega_{\mathrm{in}}, 
  \qquad
  u = u_{\mathrm{out}} \quad\text{on } \partial\Omega_{\mathrm{out}}, 
  \qquad
  u = 0 \quad\text{on } \partial\Omega_u^{D}\!\setminus\!\bigl(\partial\Omega_{\mathrm{in}}\cup\partial\Omega_{\mathrm{out}}\bigr),
\end{equation}
where \(u_{\mathrm{in}}\) and \(u_{\mathrm{out}}\) are fully developed parabolic profiles whose peak magnitudes satisfy  
\(\lVert u_{\mathrm{in}}\rVert_{\infty} = 24\,\lVert u_{\mathrm{out}}\rVert_{\infty} = 1\).  
Well-posedness of the Stokes–Brinkman problem requires:
\begin{equation}
\label{eq:mass_conservation}
  \int_{\partial\Omega_{\mathrm{in}}} u_{\mathrm{in}}\!\cdot\!\mathbf{n}\,\mathrm{d}S
  \;=\;
  \int_{\partial\Omega_{\mathrm{out}}} u_{\mathrm{out}}\!\cdot\!\mathbf{n}\,\mathrm{d}S.
\end{equation}

\begin{figure}[htbp]
    \centering
        \includegraphics[width=0.6\textwidth]{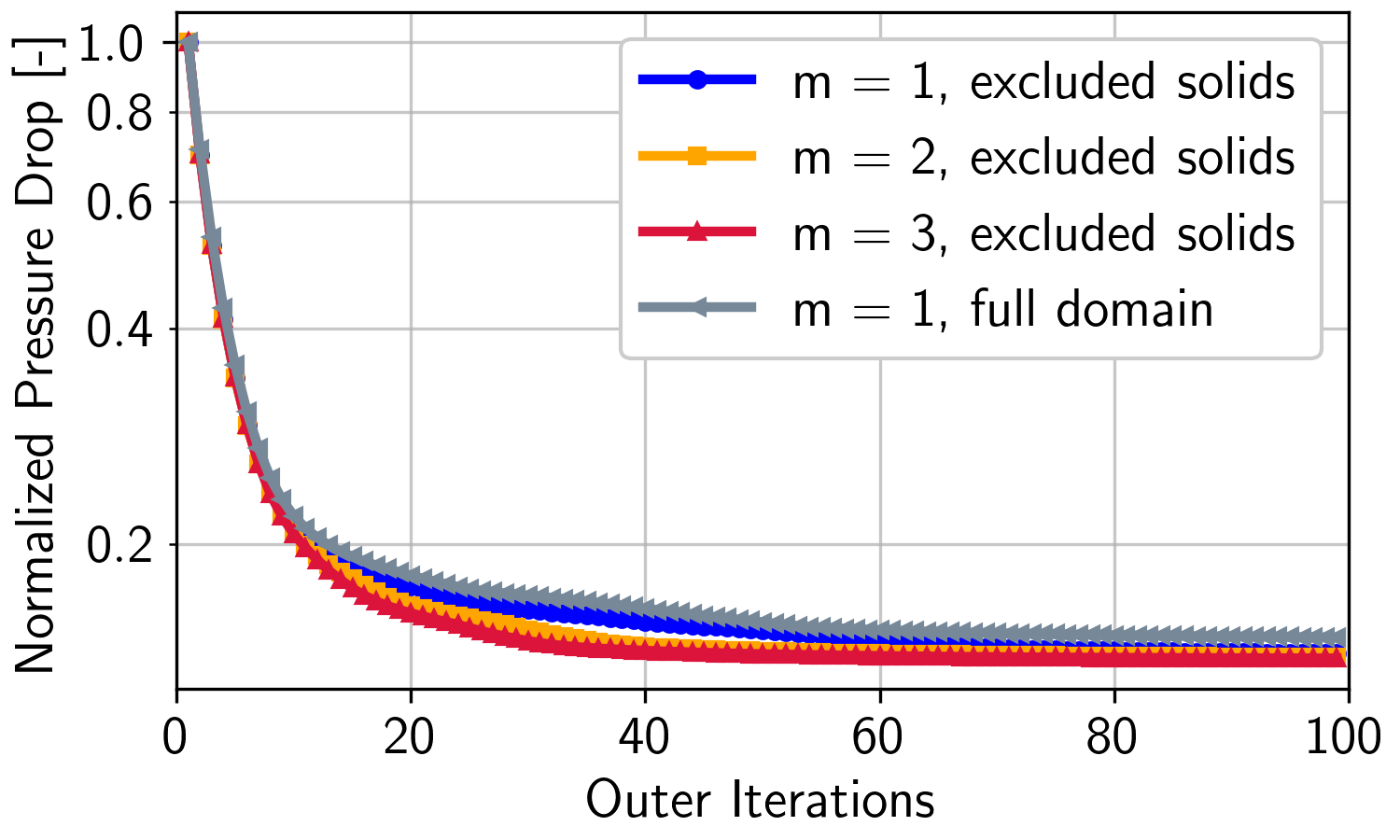}
    \caption{ Convergence History. For the first three scenarios isolated solid cells are excluded and the number of SLP iterations $m$ is varied. In the last scenario, the Stokes-Brinkman problem is solved in the full design domain for $m=1$ inner iterations.}
    \label{fig:ex1_convergence}
\end{figure}

Figure~\ref{fig:ex1_initial} shows the initial fluid–solid interface, a criss-cross network of circular tubes that satisfies the fluid-volume constraint \(V_{0}=0.15\).  
The final design, obtained by minimizing the pressure drop,
\begin{equation}
\label{ex1:pd_objective}
  \Phi_{(u,p)}(\gamma) 
  = 
  \int_{\partial\Omega_{\mathrm{in}}} p\,\mathrm{d}S 
  - 
  \int_{\partial\Omega_{\mathrm{out}}} p\,\mathrm{d}S,
\end{equation}
is depicted in Figure~\ref{fig:ex1_final}.
We discretize the domain with \(n = 180\) cells along each axis, yielding \(\approx 5.8\times10^{6}\) cells in total.  
Four scenarios are analyzed:  
\begin{itemize}
  \item[\textbf{(i)}] Isolated solid cells excluded, number of SLP iterations \(m = 1\);  
  \item[\textbf{(ii)}] Isolated solid cells excluded, number of SLP iterations \(m = 2\);  
  \item[\textbf{(iii)}] Isolated solid cells excluded, number of SLP iterations \(m = 3\);  
  \item[\textbf{(iv)}] Full-domain Stokes–Brinkman system, no cells excluded.
\end{itemize}
For cases (i)–(iii) we employ the CG–Uzawa algorithm described in Section~\ref{sec:cguzawa} with tolerance \(\varepsilon_S = 10^{-3}\).  
Case (iv) uses the CG–SIMPLE algorithm from ~\cite{pimanov2022workflow}.
Convergence histories are plotted in Figure~\ref{fig:ex1_convergence}, and Table~\ref{tab:ex1} reports the CPU time, RAM usage, and final objective values for all four scenarios. Since the final topologies are visually indistinguishable in all cases, only one representative design is shown (Fig. \ref{fig:ex1_final}). 

\begin{table}[h]
    \centering
    \caption{Comparison of CPU time, RAM usage, and final value of the objective value. Excluding isolated solids yields a $15\times$ speedup and $6.5\times$ reduction in memory.}
    \label{tab:ex1}
    \begin{tabular}{|c|c|c|c|}
        \hline
         & CPU time (s) & RAM (GB)  & Final objective, $\Phi_{(U,P)}$ \\ \hline
        $m=1$, excluded solids  & $1357$ & $4.0$ & $7.12\times10^8$ \\ \hline
        $m=2$, excluded solids  & $2391$ & $4.0$ & $7.04\times10^8$ \\ \hline
        $m=3$, excluded solids  & $2975$ & $4.0$ & $7.03\times10^8$ \\ \hline
        $m=1$, full domain      & $20075$ & $26.3$ & $7.11\times10^8$ \\ \hline
    \end{tabular}
\end{table}

Comparing scenarios~(i) and~(iv) shows that excluding isolated solid cells delivers a $15\times$ speed-up and a $6.5\times$ reduction in memory usage, while having no significant impact on the convergence or on the final objective value.
Increasing the number of inner SLP iterations does not qualitatively affect the outer-level convergence or the resulting design; we therefore fix \(m = 1\) in all subsequent examples.  

A further conclusion is that the CG–Uzawa solver—originally developed for pure Stokes flow—remains robust for the Stokes–Brinkman system when the Brinkman term is non-zero only in a single layer of solid cells.  This is consistent with the interpretation of the Brinkman term as a penalization that merely measures the proximity of the no-slip boundary.  
Finally, the integer optimization subproblem~\eqref{eq:subprob_pert_linear} exhibits an unexpectedly convenient property.  
Relaxing the binary constraint,
\begin{equation}
  (\Delta\Gamma_{a}^{0,k})_{j} \in 
  \bigl\{-\Gamma_{a}^{0,k}{}_j,\;1 - \Gamma_{a}^{0,k}{}_j\bigr\},
\end{equation}
to the interval constraint,
\begin{equation}
  (\Delta\Gamma_{a}^{0,k})_{j} \in 
  \bigl[-\Gamma_{a}^{0,k}{}_j,\;1 - \Gamma_{a}^{0,k}{}_j\bigr],
\end{equation}
and solving the relaxed problem with a Simplex algorithm yields a binary solution, removing the need for integer programming.

\subsection{Example 2: Optimization of a single‑fluid heat exchanger under a geometric constraint}
\noindent
In this section, we optimize a {single‑fluid} heat exchanger.  
The setting is similar to that studied by Kambampati~\textit{et al.}~\cite{kambampati2020level}, except that we model Stokes flow rather than Darcy flow and assume a larger fluid volume fraction.  
Our aim is to show how the proposed algorithm accommodates additional geometric constraints when only a subset of the domain~$\Omega$ is designable.  
The same treatment of non‑design regions is later used to impose a minimum‑thickness constraint.

\begin{figure}[htbp]
    \centering
    \begin{subfigure}[b]{0.35\textwidth}
        \includegraphics[width=\textwidth]{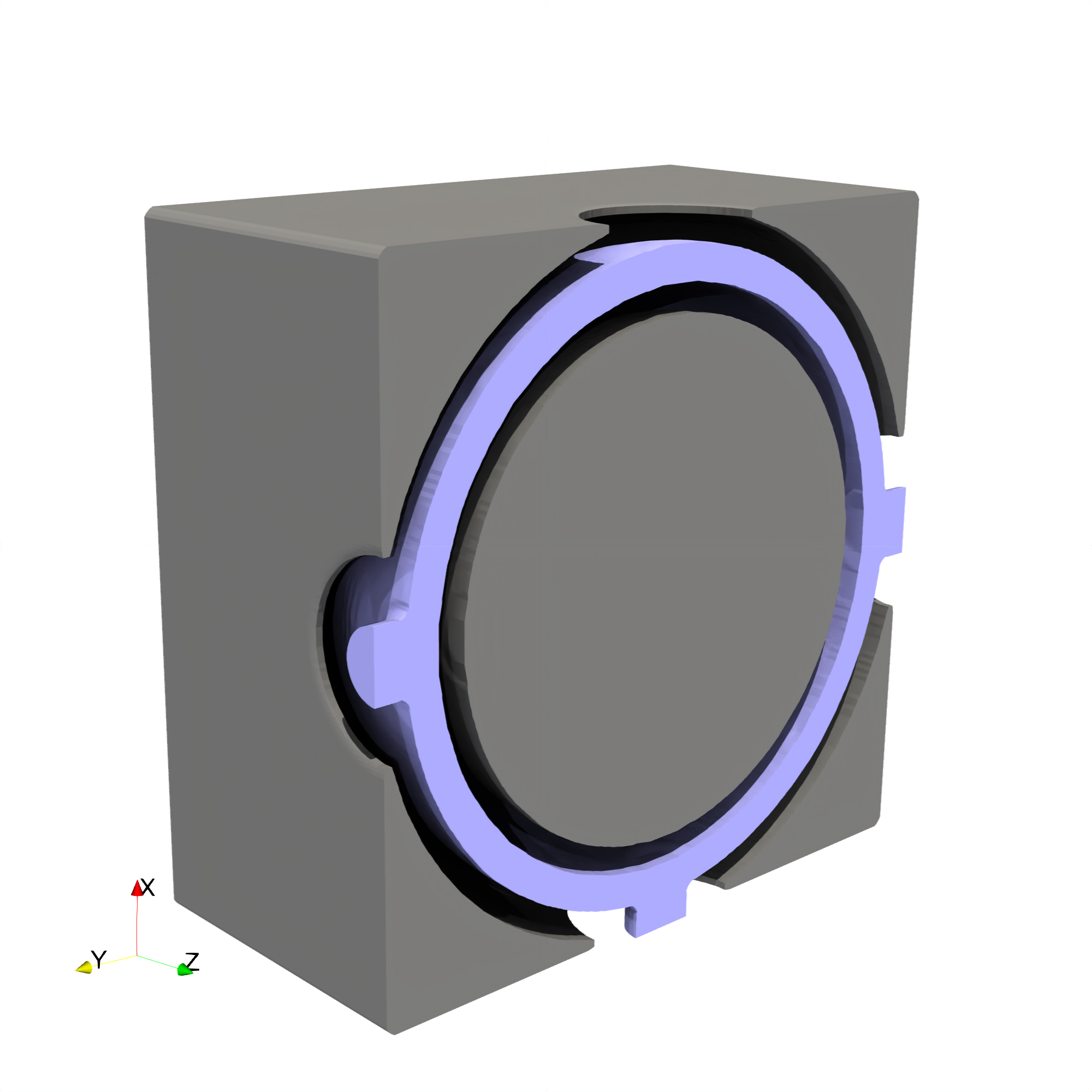}
        \caption{}
        \label{fig:ex2_initial_nondesign}
    \end{subfigure}
    \begin{subfigure}[b]{0.33\textwidth}
        \includegraphics[width=\textwidth]{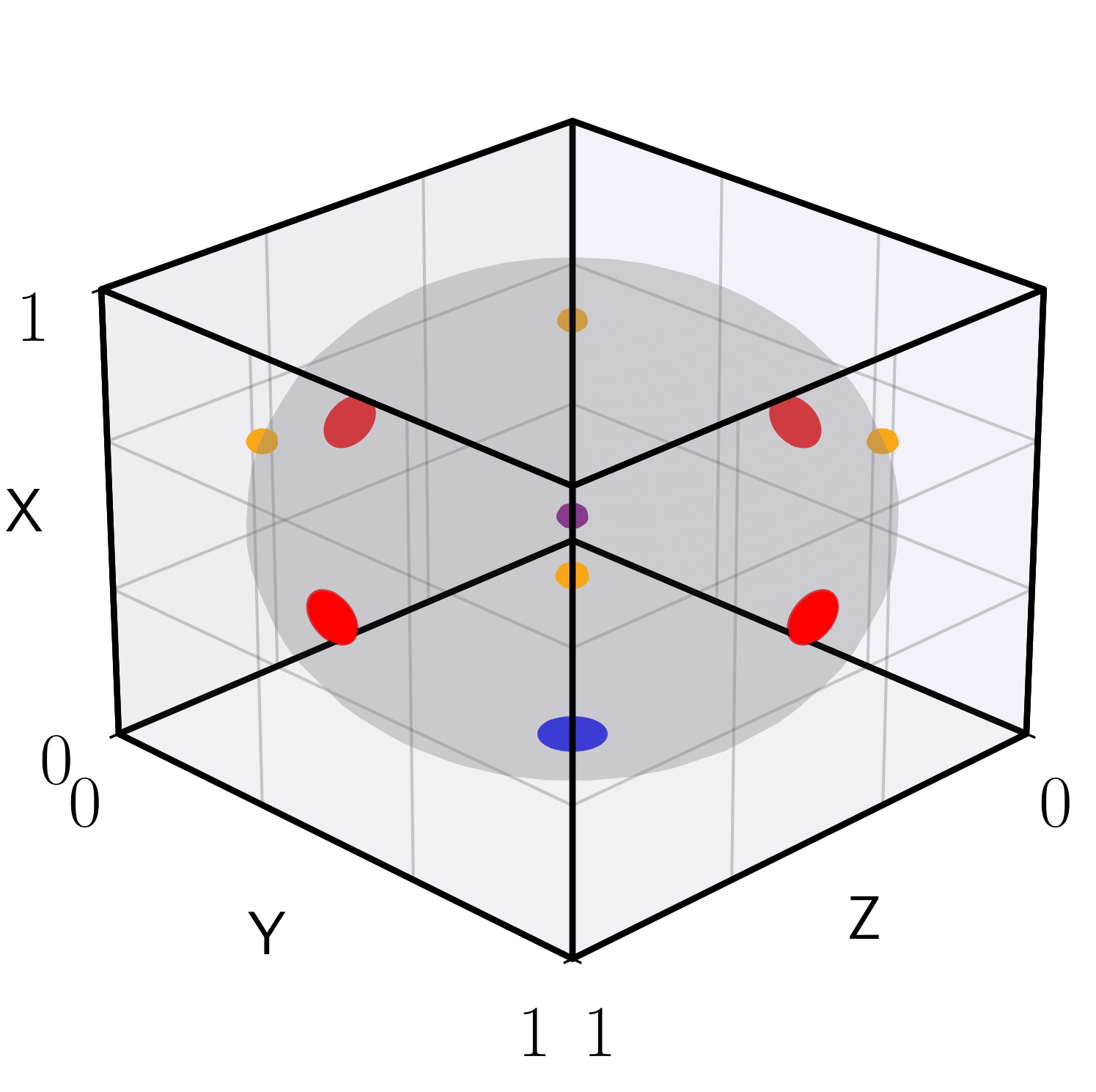}
        \caption{}
        \label{fig:ex2_setup}
    \end{subfigure}
    \caption{(a) Slice through the full domain: The non‑design solid region (grey) and the initial fluid region (purple).  
             (b) Problem set‑up: inlet (blue), four outlets (red), point heat source (violet), and four probe point locations (orange).}
    \label{fig:ex2_initial_setup}
\end{figure}

Figure~\ref{fig:ex2_initial_nondesign} shows a slice at $x = 0.5$.  
The non‑design solid occupies the region {inside} the inner sphere ($r < r_1$) and {outside} the outer sphere ($r > r_2$); thus the design domain is the spherical shell $r_1 \le r \le r_2$, where $r_1 = 0.36$ and $r_2 = 0.51$ are measured from the common center $\mathbf{x}_c = (0.5,0.5,0.5)$.
The initial fluid distribution (purple) satisfies the volume constraint $V_0 = 0.14$; the remaining designable solid is rendered transparent.  
The modification to the narrow‑band algorithm of Section~\ref{sec:narrow_band} is straightforward: Cells are first classified according to~\eqref{eq:voxel_types} without regard to the non‑design region, and any active cells located in that region are subsequently deactivated and removed from the optimization subproblem.
Figure~\ref{fig:ex2_setup} depicts the problem configuration.  
A single circular inlet \(\partial\Omega_{\mathrm{in}}\) (blue) of radius $r = 0.05$ is centred on the plane $x = 0$, and four outlets \(\partial\Omega_{\mathrm{out}}\) (red) of the same radius are centred on the planes $y = 0$, $y = 1$, $z = 0$, and $z = 1$.  
As in Example~1 we impose the Dirichlet velocity boundary condition~\eqref{ex1:dirichlet_bc} with parabolic profiles whose peak magnitudes satisfy $\lVert u_{\mathrm{in}}\rVert_{\infty} = 4\,\lVert u_{\mathrm{out}}\rVert_{\infty} = 1$.

For the heat equation~\eqref{eq:heat_equation} we prescribe a zero temperature at the inlet and place a point heat source at the domain center:
\begin{equation}
  \theta = \theta_{\mathrm{in}} = 0 \quad\text{on } \partial\Omega_\theta^{D}, 
  \qquad 
  \frac{\partial\theta}{\partial n}=0 \quad\text{on } \partial\Omega_\theta^{N}, 
  \qquad 
  q(\mathbf{x}) = Q\,\delta\bigl(\mathbf{x}-\mathbf{x}_c\bigr),
\end{equation}
where $\partial\Omega_\theta^{D} = \partial\Omega_{\mathrm{in}}$, 
$\partial\Omega_\theta^{N} = \partial\Omega\setminus\partial\Omega_{\mathrm{in}}$, and $Q = 1$.  
A constant thermal conductivity $k = 10^{-3}$ is assumed in both fluid and solid; this choice highlights the effect of convective transport.

The objective functional is a weighted combination:
\begin{equation}
\label{eq_weighted_objective}
  \Phi(\gamma) = \omega\,\Phi_{(u,p)}(\gamma) + \bigl(1-\omega\bigr)\,\Phi_{\theta}(\gamma),
\end{equation}
where $\Phi_{(u,p)}(\gamma)$ is the pressure drop defined in~\eqref{ex1:pd_objective} and,
\begin{equation}
  g_\theta = \sum_{i=1}^{4}\delta\bigl(\mathbf{x}-\mathbf{x}_i\bigr),
  \qquad 
  \Phi_\theta(\gamma) = (g_{\theta},\theta) = \sum_{i=1}^{4}\theta(\mathbf{x}_i).
\end{equation}
The four probe points $\mathbf{x}_i$ lie at the intersection of the plane $x = 0.75$ with the sphere of radius 
$r_3 = \tfrac{1}{2}(r_1 + r_2)$ centered at $\mathbf{x}_c$ (orange markers in Fig.~\ref{fig:ex2_setup}).
The domain is discretized with $n = 200$ cells per direction.  
To assess the influence of the thermal term we use two weights, $\omega = 1$ and $\omega = 0.01$: The former corresponds to the pure pressure‑drop minimization problem.

\begin{figure}[htbp]
    \centering
    \begin{subfigure}[b]{0.3\textwidth}
        \includegraphics[width=\textwidth]{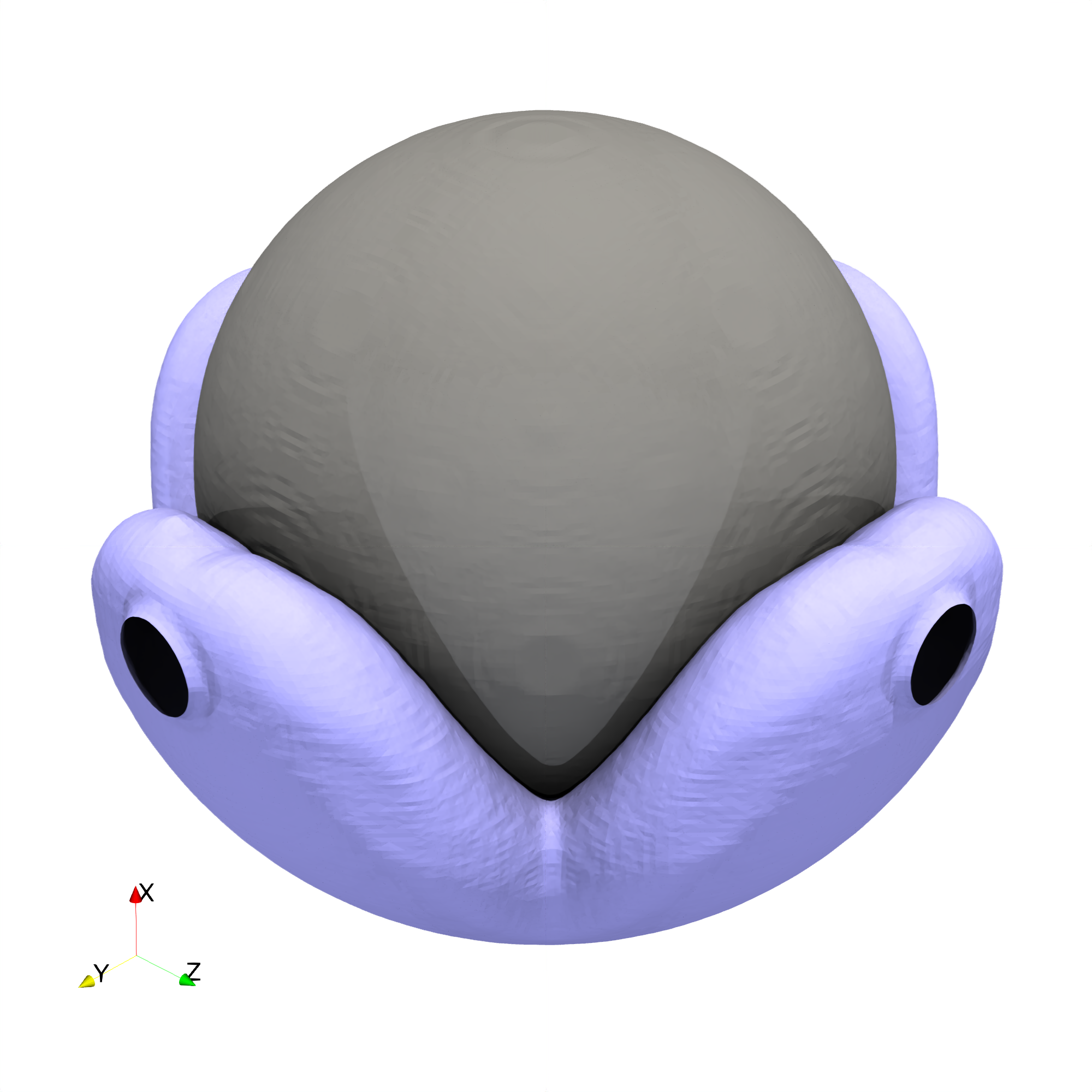}
        \caption{}
        \label{fig:ex2_final_1}
    \end{subfigure}
    \begin{subfigure}[b]{0.3\textwidth}
        \includegraphics[width=\textwidth]{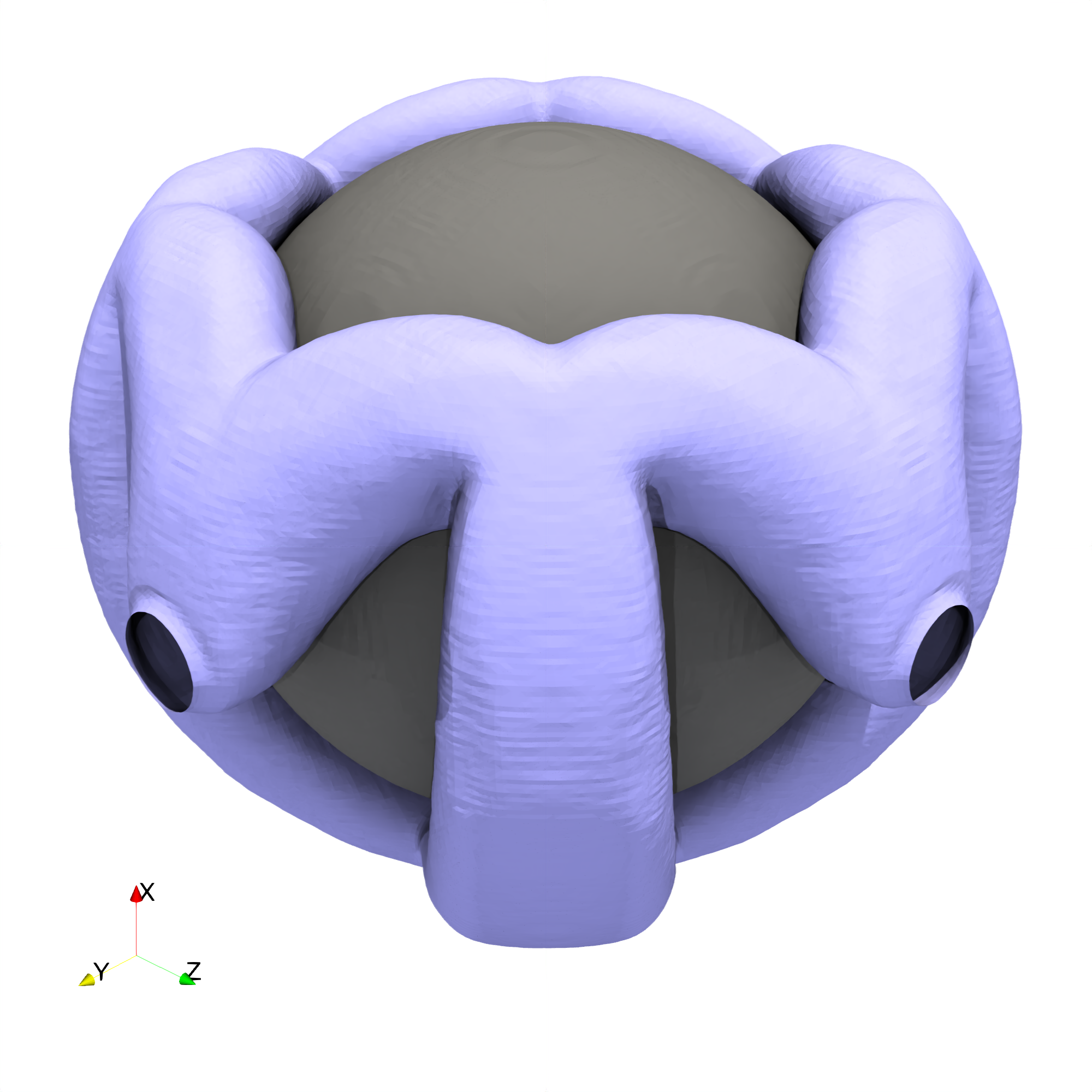}
        \caption{}
        \label{fig:ex2_final_001}
    \end{subfigure}
    \begin{subfigure}[b]{0.35\textwidth}
    \centering
    \includegraphics[width=\linewidth]{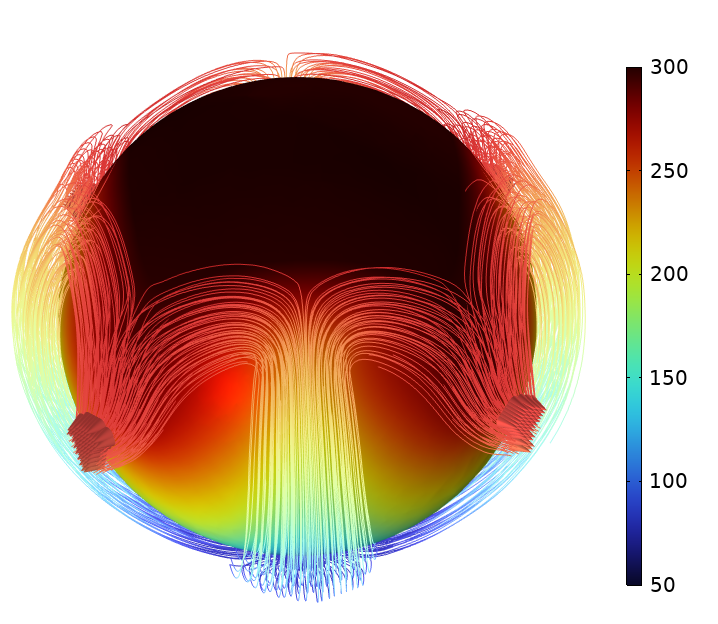}
    \caption{}
    \label{fig:ex_comsol}
\end{subfigure}
    \caption{Optimized heat‑exchanger designs.  
(a) Pressure‑drop minimization (\(\omega = 1\));  
(b) weighted objective~\eqref{eq_weighted_objective} with \(\omega = 0.01\);  
(c) temperature field and velocity streamlines for the design in~(b).}
    \label{fig:ex2_2}
\end{figure}

The final designs for both parameters are shown in Figs. \ref{fig:ex2_final_1} and \ref{fig:ex2_final_001}. When the pressure drop dominates, the optimizer eliminates almost every flow obstruction, producing four channels that connect the inlet with the outlets along the shortest paths. The result is efficient for pressure drop but not so effective for the thermal objective, because the flow bypasses the non‑design hot sphere and goes directly to the outlets, therefore picking up little heat. In contrast, when the thermal part of the objective is considered the design sacrifices pressure drop efficiency to improve thermal performance as seen in Table \ref{tab:ex2_final_objective}. The channels now “wrap” around the heated core, increasing the path length from the inlet to the outlets. This geometric re‑routing is also shown in Fig. \ref{fig:ex_comsol}, which shows the temperature distribution and velocity streamlines for the final design for $\omega=0.01$ where the streamlines first reach the probe points before going to the outlets.

\begin{figure}[htbp]
    \centering
    \begin{subfigure}[b]{0.49\textwidth}
        \includegraphics[width=\textwidth]{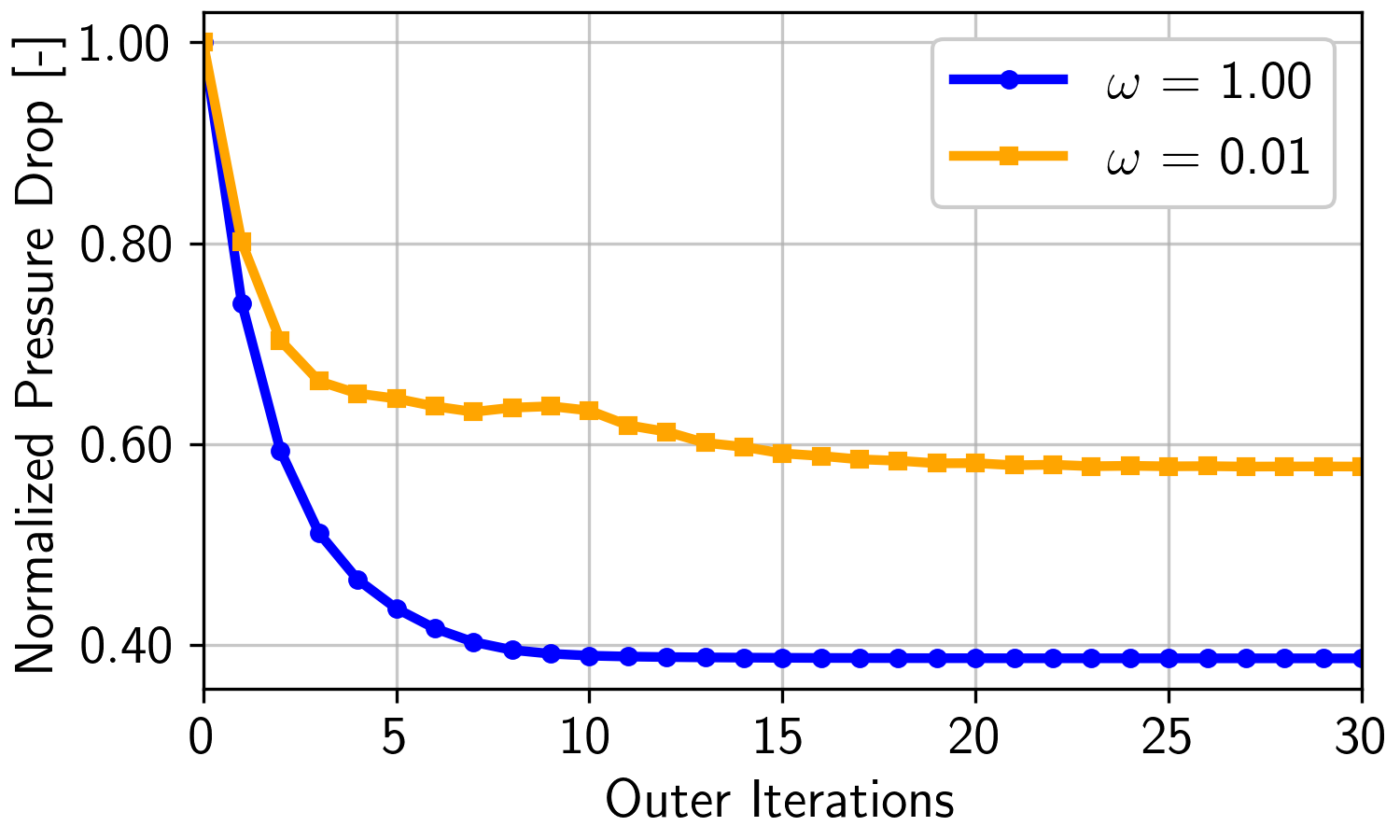}
        \label{fig:ex2_pd}
    \end{subfigure}
    \begin{subfigure}[b]{0.49\textwidth}
        \includegraphics[width=\textwidth]{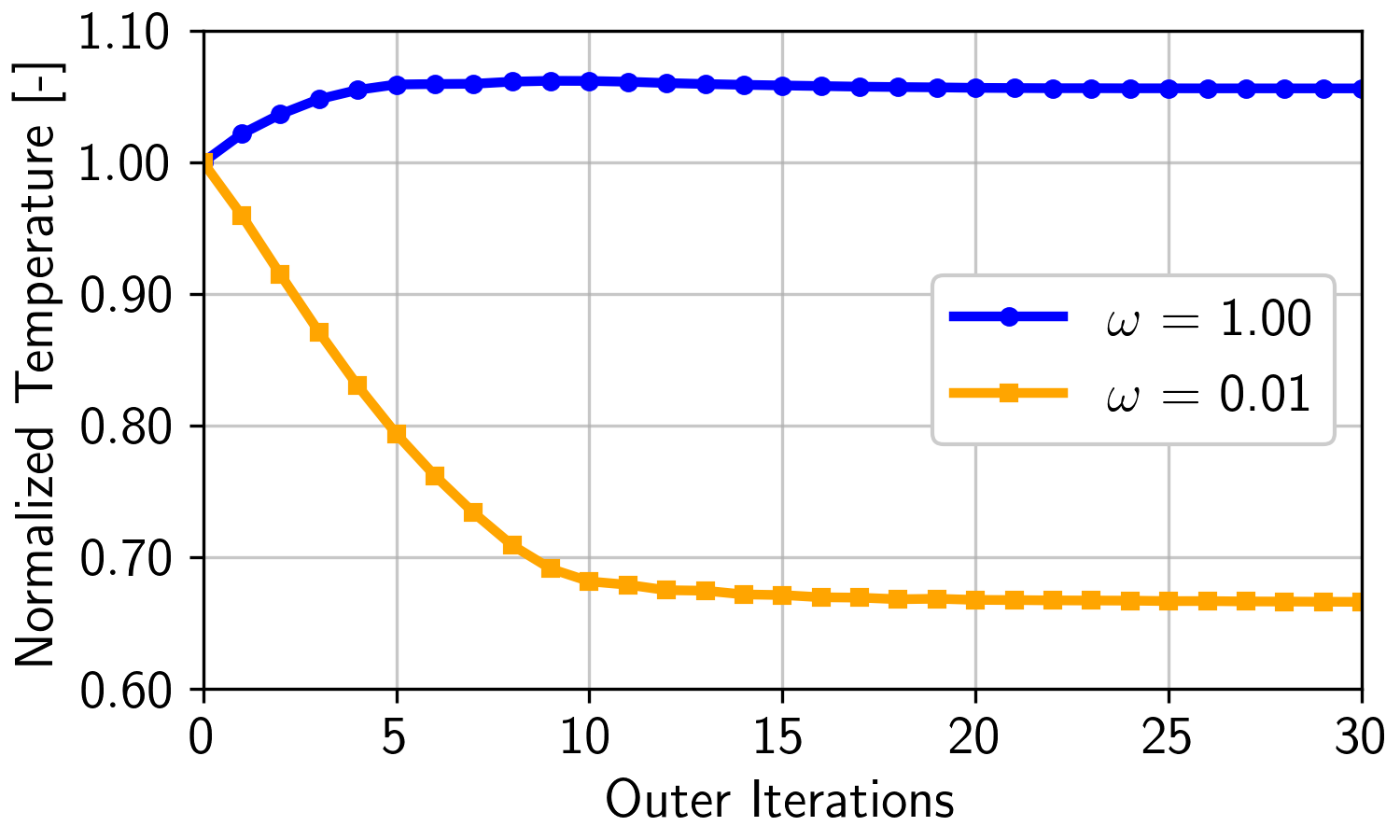}
        \label{fig:ex2_hx}
    \end{subfigure}
    \caption{Convergence history for the two selected weights. The objectives are normalized with initial values.}
    \label{fig:ex2_convergence}
\end{figure}

In Fig. \ref{fig:ex2_convergence}, we show the convergence histories for both objectives normalized by their respective initial values. The two panels of Fig. \ref{fig:ex2_convergence} demonstrate that the narrow–band SLP algorithm converges smoothly and monotonically for both objective weights. The objectives also converge rapidly. The pressure–drop objective settles after 10 to 20 iterations, while the normalized probe‑point temperature remains almost unchanged after 5 to 15 iterations, confirming stable convergence for both objectives.

The final values of the objective functionals are summarized in Table \ref{tab:ex2_final_objective}, which quantifies the pressure drop–thermal trade–off. We observe that decreasing the weight from $\omega=1$ to $\omega=0.01$ raises the pressure drop by approximately 33 \% but lowers the probe‑point temperature by nearly 38 \%. 

\begin{table}[h]
    \centering
    \caption{Final pressure drop and probe‐point temperature for two weights, together with their relative change
             $\Delta\Phi = 100\,(\Phi-\Phi_{\omega=1})/\Phi_{\omega=1}$ 
             with respect to the $\omega = 1$ baseline.}
    \label{tab:ex2_final_objective}
    \begin{tabular}{|c|c|c|c|c|}
        \hline
        Weight $\omega$ 
        & $\Phi_{(u,p)}$  
        & $\Phi_{\theta}$ 
        & $\Delta\Phi_{(u,p)}$ [\%] 
        & $\Delta\Phi_{\theta}$ [\%] \\ \hline
        1     & 0.52 & 7.20 & 0.0 & 0.0 \\ \hline
        0.01  & 0.69 & 4.48 & +32.7 & $-37.8$ \\ \hline
    \end{tabular}
\end{table}

Finally, the average computation times per iteration are shown in Table \ref{tab:ex2_avg_iter_time}. For a $200^{3}$ discretization, each outer iteration remains computationally tractable: The forward and adjoint Stokes–Brinkman solve averages 32 s while the forward and adjoint convection-diffusion solve averages 17 s. The entire optimization requires less than 17 GB of RAM, which is now commonly available in modest workstations, underscoring the practicality of the sparse discretization scheme.

\begin{table}[h]
    \centering
    \caption{Average computation time per iteration. The RAM memory usage is $16.8$GB.}
    \label{tab:ex2_avg_iter_time}
    \begin{tabular}{|l|c|}
        \hline
         & Time (s) \\ \hline
        Stokes–Brinkman + adjoint & 32 \\ \hline
        Convection–diffusion + adjoint  & 17 \\ \hline
    \end{tabular}
\end{table}

\subsection{Example 3: Optimization of a Two‑Fluid Heat Exchanger}
\label{subsec:two_fluid_hx}

In this example, we revisit the two‑fluid heat‑exchanger problem from Feppon \textit{et al.}~\cite{feppon2021body}.  
The domain consists of two separate channels, $\Omega_f^{\mathrm{hot}}$ and $\Omega_f^{\mathrm{cold}}$, that carry different fluids. 
The solid phase conducts heat between the channels while keeping them apart, ensuring that the two fluids do not interpenetrate.

\begin{figure}[htbp]
    \centering
    \begin{subfigure}[b]{0.4\textwidth}
        \includegraphics[width=\textwidth]{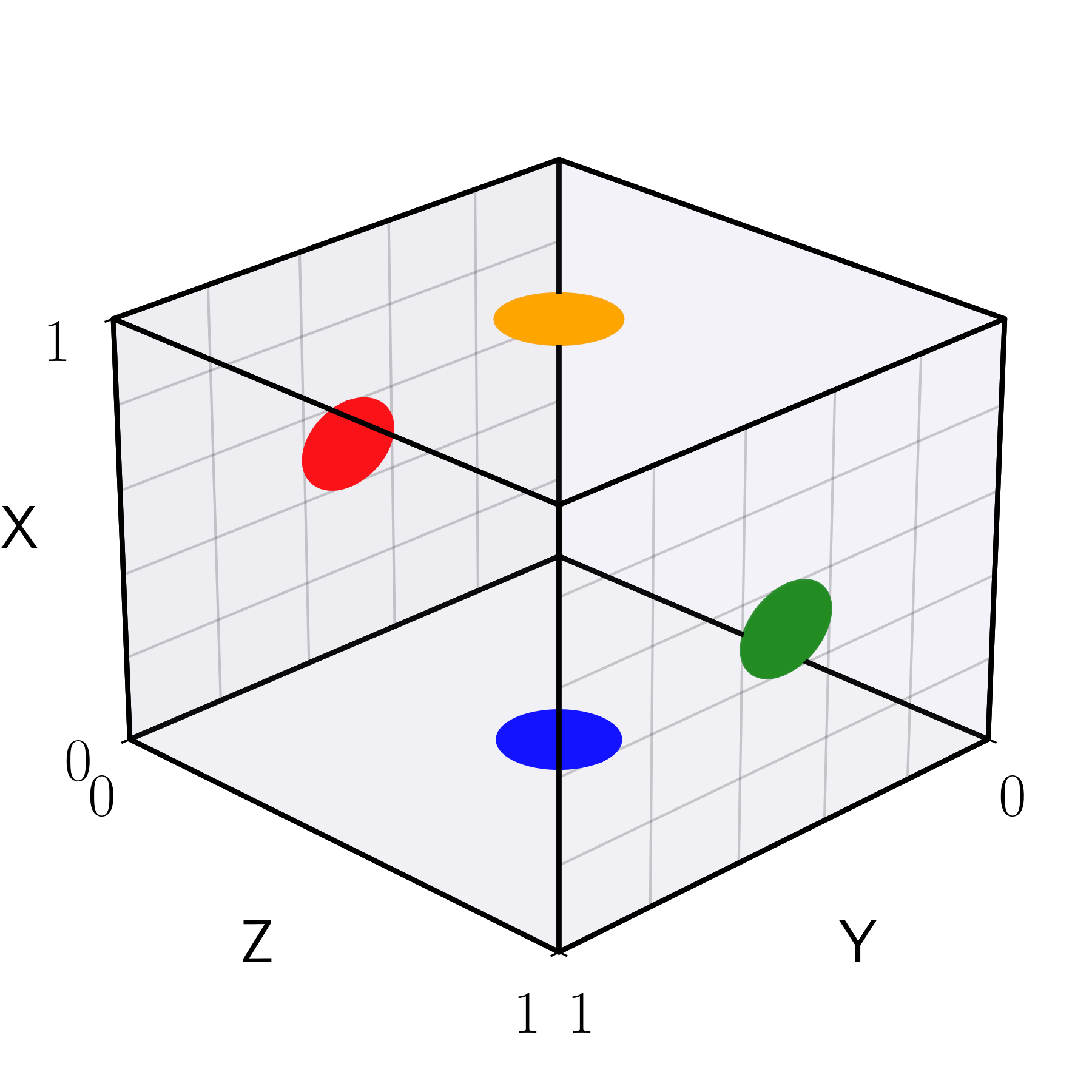}
        \caption{Problem setup}
        \label{fig:ex3_setup}
    \end{subfigure}
    \begin{subfigure}[b]{0.4\textwidth}
        \includegraphics[width=\textwidth]{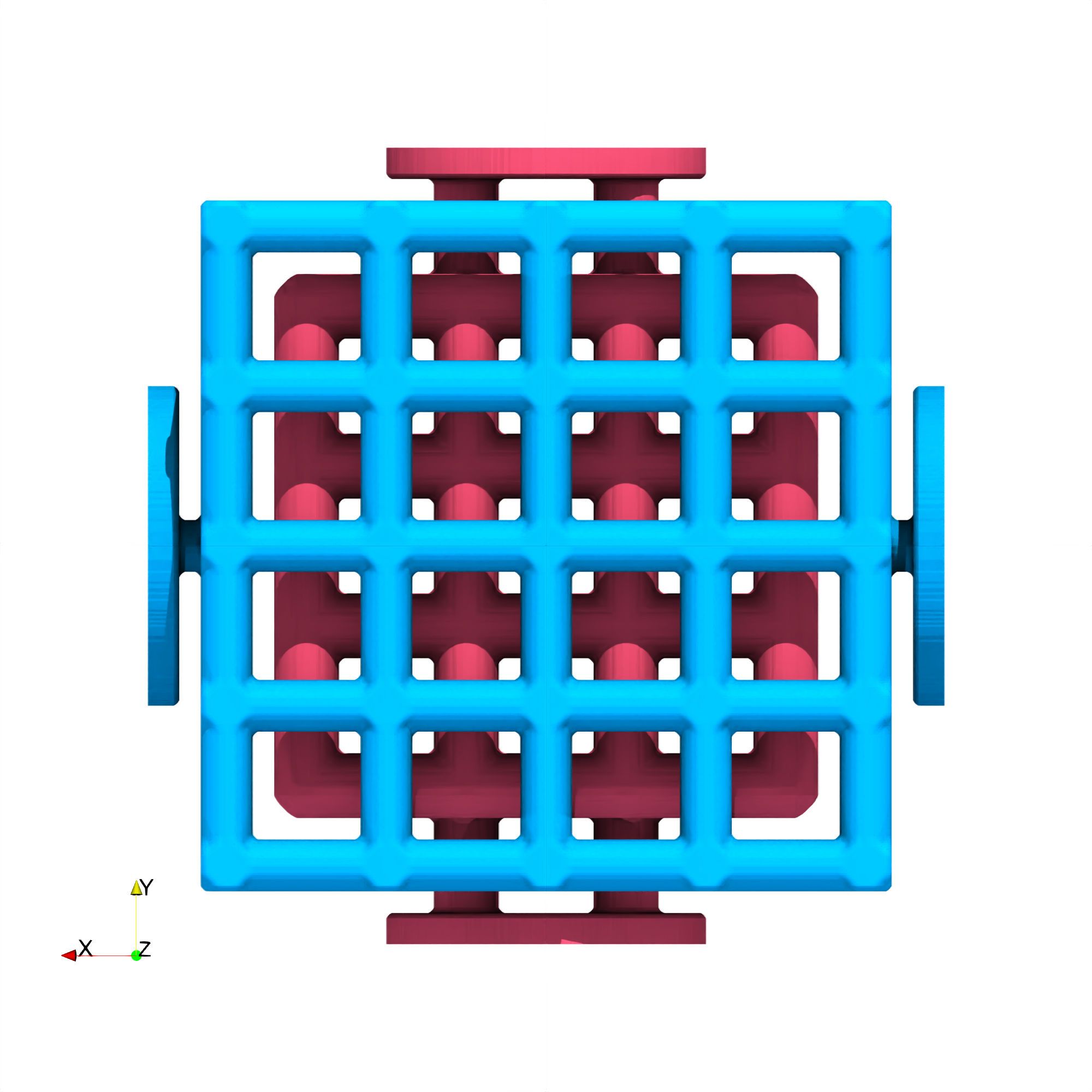}
        \caption{Initial design}
        \label{fig:ex3_init}
    \end{subfigure}
    \caption{Boundary conditions for velocity and temperature; inlet/outlet radius $r = 0.1$.}
    \label{fig:ex3}
\end{figure}

Figure~\ref{fig:ex3_setup} shows the boundary conditions \cite{feppon2021body}.  
The hot inlet, $\partial\Omega_{\mathrm{in}}^{\mathrm{hot}}$ (red), is a circle of radius $r = 0.1$ on the plane $y = 0$, and the hot outlet, $\partial\Omega_{\mathrm{out}}^{\mathrm{hot}}$ (green), lies on $y = 1$.  
The cold inlet, $\partial\Omega_{\mathrm{in}}^{\mathrm{cold}}$ (blue), is located on $x = 0$, and the cold outlet, $\partial\Omega_{\mathrm{out}}^{\mathrm{cold}}$ (orange), is on $x = 1$.

\paragraph{Flow boundary conditions}
The Dirichlet velocity profiles are prescribed at every inlet and outlet:
\begin{equation}
\label{ex3:dirichlet_bc}
\begin{aligned}
  u &= u_{\mathrm{in}}^{\mathrm{hot}}
       &&\text{on } \partial\Omega_{\mathrm{in}}^{\mathrm{hot}},
  &
  u &= u_{\mathrm{out}}^{\mathrm{hot}}
       &&\text{on } \partial\Omega_{\mathrm{out}}^{\mathrm{hot}}, \\[4pt]
  u &= u_{\mathrm{in}}^{\mathrm{cold}}
       &&\text{on } \partial\Omega_{\mathrm{in}}^{\mathrm{cold}},
  &
  u &= u_{\mathrm{out}}^{\mathrm{cold}}
       &&\text{on } \partial\Omega_{\mathrm{out}}^{\mathrm{cold}}, \\[4pt]
  u &= 0
       &&\text{on } \partial\Omega_u^{D}\setminus\bigl(
         \partial\Omega_{\mathrm{in}}^{\mathrm{hot}}\cup
         \partial\Omega_{\mathrm{out}}^{\mathrm{hot}}\cup
         \partial\Omega_{\mathrm{in}}^{\mathrm{cold}}\cup
         \partial\Omega_{\mathrm{out}}^{\mathrm{cold}}\bigr).
\end{aligned}
\end{equation}
Each profile is fully developed and parabolic, with peak magnitudes satisfying
\[
  \lVert u_{\mathrm{in}}^{\mathrm{hot}}\rVert_{\infty}
  \;=\;
  \lVert u_{\mathrm{out}}^{\mathrm{hot}}\rVert_{\infty}
  \;=\;
  \lVert u_{\mathrm{in}}^{\mathrm{cold}}\rVert_{\infty}
  \;=\;
  \lVert u_{\mathrm{out}}^{\mathrm{cold}}\rVert_{\infty}
  \;=\; 1,
\]
so that the mass‑conservation condition~\eqref{eq:mass_conservation} is met for both fluids.

\paragraph{Thermal boundary conditions}

Heat transfer is driven by imposing a high temperature at the hot inlet and a zero temperature at the cold inlet:
\begin{equation}
  \theta = \theta_{\mathrm{in}}^{\mathrm{hot}} = 100
  \quad\text{on } \partial\Omega_{\mathrm{in}}^{\mathrm{hot}}, 
  \qquad
  \theta = \theta_{\mathrm{in}}^{\mathrm{cold}} = 0
  \quad\text{on } \partial\Omega_{\mathrm{in}}^{\mathrm{cold}}.
\end{equation}
A homogeneous Neumann boundary condition is prescribed on all remaining boundaries:
\begin{equation}
  \frac{\partial\theta}{\partial n} = 0
  \quad\text{on } 
  \partial\Omega_\theta^{N}
  := \partial\Omega \setminus
  \bigl(
    \partial\Omega_{\mathrm{in}}^{\mathrm{hot}}
    \cup
    \partial\Omega_{\mathrm{in}}^{\mathrm{cold}}
  \bigr).
\end{equation}

\paragraph{Objective functionals}
For the fluid objective we {minimize} the sum of the pressure drops in the cold and hot channels:
\begin{equation}
  \Phi_{(u,p)}(\gamma)
  \;=\;
  \Phi_{(u,p)}^{\mathrm{cold}}
  +
  \Phi_{(u,p)}^{\mathrm{hot}}
  \;=\;
  \Bigl(
    \int_{\partial\Omega_{\mathrm{in}}^{\mathrm{cold}}} p\,\mathrm{d}s
    -
    \int_{\partial\Omega_{\mathrm{out}}^{\mathrm{cold}}} p\,\mathrm{d}s
  \Bigr)
  +
  \Bigl(
    \int_{\partial\Omega_{\mathrm{in}}^{\mathrm{hot}}}  p\,\mathrm{d}s
    -
    \int_{\partial\Omega_{\mathrm{out}}^{\mathrm{hot}}} p\,\mathrm{d}s
  \Bigr).
\end{equation}
For the thermal objective we maximize the heat carried away by the cold fluid minus the heat carried away by the hot fluid:
\begin{equation}
\label{ex3_objective}
  \Phi_{\theta}(\gamma)
  \;=\;
  \Phi_{\theta}^{\mathrm{cold}} - \Phi_{\theta}^{\mathrm{hot}},
  \qquad
  \Phi_{\theta}^{\mathrm{cold}}
  = \int_{\partial\Omega_{\mathrm{out}}^{\mathrm{cold}}}
      \theta\,u\!\cdot\! n \,\mathrm{d}s,
  \quad
  \Phi_{\theta}^{\mathrm{hot}}
  = \int_{\partial\Omega_{\mathrm{out}}^{\mathrm{hot}}}
      \theta\,u\!\cdot\! n \,\mathrm{d}s,
\end{equation}
where the unit normal \(n\) points outward from \(\Omega\).  
This objective promotes efficient heat exchange: It rewards heat removed by the cold stream and penalizes heat that leaves with the hot stream.
The weighted objective to be {minimized} is therefore:
\begin{equation}
\label{eq:ex3_weighted}
  \omega\,\Phi_{(u,p)}(\gamma)
  \;-\;
  (1-\omega)\,\Phi_{\theta}(\gamma).
\end{equation}

\begin{remark}
Because the velocity vanishes on the fluid–solid interface and both velocity and temperature are prescribed at the inlets, maximizing \(\Phi_{\theta}(\gamma)\) is equivalent to maximizing the objective used by Feppon \textit{et al.}~\cite{feppon2021body}:
\begin{equation}
\label{objective_feppon}
  \Phi_{\theta}^{\ast}(\gamma)
  \;=\;
  \int_{\partial\Omega_{f}^{\mathrm{cold}}} \theta\,u\!\cdot\! n \,\mathrm{d}s
  \;-\;
  \int_{\partial\Omega_{f}^{\mathrm{hot}}}  \theta\,u\!\cdot\! n \,\mathrm{d}s
  \;=\;
  \int_{\Omega_{f}^{\mathrm{cold}}}  u\!\cdot\!\nabla\theta \,\mathrm{d}x
  \;-\;
  \int_{\Omega_{f}^{\mathrm{hot}}}   u\!\cdot\!\nabla\theta \,\mathrm{d}x.
\end{equation}
For \(\Phi_{\theta}^{\ast}(\gamma)\) the following equilibrium relation holds:
\begin{equation}
\label{equilibrium_eq}
  \int_{\partial\Omega_{f}^{\mathrm{hot}}} \theta\,u\!\cdot\! n \,\mathrm{d}s
  \;+\;
  \,\int_{\partial\Omega_{f}^{\mathrm{cold}}} \theta\,u\!\cdot\! n \,\mathrm{d}s
  \;=\; 0.
\end{equation}
For our objective \(\Phi_{\theta}(\gamma)\) this relation becomes:
\begin{equation}
  \Phi_{\theta}^{\mathrm{hot}}
  \;+\;
  \,\Phi_{\theta}^{\mathrm{cold}}
  \;=\;
  -\int_{\partial\Omega_{\mathrm{in}}^{\mathrm{hot}}} \theta\,u\!\cdot\! n \,\mathrm{d}s \;=\; \mathrm{const},
\end{equation}
where the inlet term is constant during optimization and can therefore be omitted.

We also minimize the weighted multi‑objective functional in Eq.~\eqref{eq:ex3_weighted}, whereas Feppon \textit{et al.} maximize the exchanged heat and treat the pressure‑drop terms as separate optimization constraints (one per phase). 
\end{remark}
\begin{remark}
Compared with the original formulation in \cite{feppon2021body}, we impose the {Dirichlet} velocity conditions at both the inlets and outlets instead of prescribing outlet pressure.  
Consequently, the functional \(\Phi_{\theta}(\gamma)\) is linear in the state variable~\(\theta\), which simplifies the sensitivity analysis.
\end{remark}

\begin{remark}
Finally, we model Stokes flow rather than the full Navier–Stokes equations, which greatly simplifies computation of sensitivities.  
For the Reynolds numbers considered in \cite{feppon2021body}, this simplification introduces a difference in the objectives of less than \(1\%\).
\end{remark}

\paragraph{Initial design} The fluid-solid interface of the initial design is shown in Fig. \ref{fig:ex3_init} (red color for hot and blue color for cold). The volume constraint is initially satisfied for the two phases: $V_0^{cold} = V_0^{hot} = 0.15$.

\paragraph{Implementation of the minimum‑thickness constraint}
In the previous example we showed how the proposed algorithm can handle {non‑design} constraints, i.e., regions of the domain that remain fixed during optimization. The same mechanism is used to impose a minimum separation constraint between the two fluid phases: a solid wall at least \(d\) voxels thick must always lie between the hot‑ and cold‑fluid channels.
Each outer iteration is divided into two sub‑iterations:

1. \textbf{Hot‑phase step.}  
   The current cold channel \(\Omega_f^{\mathrm{cold}}\) together with a solid halo of width \(d\) is marked as non‑design.  The optimizer then updates only the hot phase, deforming \(\Omega_f^{\mathrm{hot}}\) while respecting the non-design mask.

2. \textbf{Cold‑phase step.}  
   The roles are reversed: the updated \(\Omega_f^{\mathrm{hot}}\) and its
   \(d\)-voxel halo are frozen, and the cold phase is optimized around the hot phase.

The halo is generated by scanning all neighbors of each voxel within a sphere of radius \(d\) and if any neighbor belongs to the opposite fluid phase, the voxel is flagged as non‑design. Fluid voxels may still be converted to solid, so the interface is free to retreat but not to advance toward the opposite phase.

The initial design must already satisfy this \(d\)-voxel separation.  If it does not, a preprocessing step—e.g., eroding the fluid regions (or,
equivalently, dilating the solid halo) by \(d\) voxels—can be applied to
produce a feasible starting point.

\paragraph{Study scenarios and parameters}
In what follows, two scenarios are considered.  
Section~\ref{sec:omega} imposes a non‑mixing constraint, i.e., the minimum‑thickness constraint with a one‑voxel separation.  
The thermal conductivity is set to \(k = 5\times 10^{-4}\), which yields a Peclet number \(Pe = 2000\), matching the value used by Feppon \emph{et al.} \cite{feppon2021body}.  
The domain is discretized with \(n = 300\) grid points per direction, corresponding to \(27\times10^{6}\) design variables.  
Within this scenario we examine (i) the dependence on the weight \(\omega\) in the objective~\eqref{eq:ex3_weighted} (Section~\ref{sec:omega}), and {(ii) the effect of the initial total fluid volume while fixing the weight \(\omega\) (Section~\ref{sec:volume}).}

Section~\ref{sec:large} uses the same problem setup but at a higher resolution, \(n = 370\), corresponding to about \(51\times10^{6}\) design variables.  
A single weight \(\omega\) is employed.  
Here the thermal conductivity is reduced to \(k = 10^{-4}\), resulting in \(Pe = 10\,000\).  
The separation distance is set to nine voxels, which corresponds to \(0.025\) in physical units. 
This larger minimum‑distance constraint allows the final design to be converted to a body‑fitted mesh and validated in COMSOL Multiphysics.

\subsubsection{Effect of the weight \(\omega\)}
\label{sec:omega}

\begin{figure}[htbp]
    \centering
    \begin{subfigure}[b]{0.4\textwidth}
        \includegraphics[width=\textwidth]{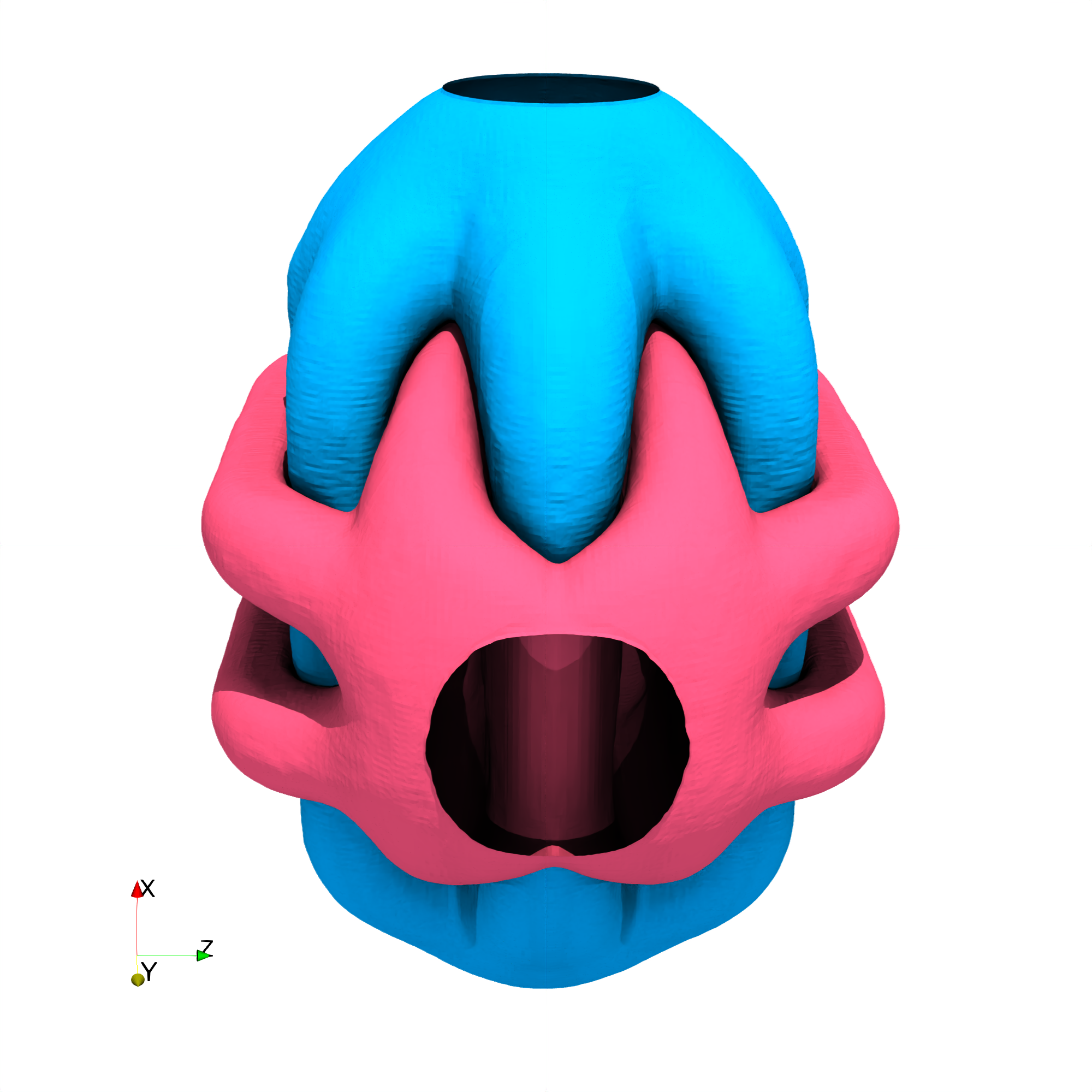}
        \caption{\(\omega = 0.9\)}
    \end{subfigure}
    \begin{subfigure}[b]{0.4\textwidth}
        \includegraphics[width=\textwidth]{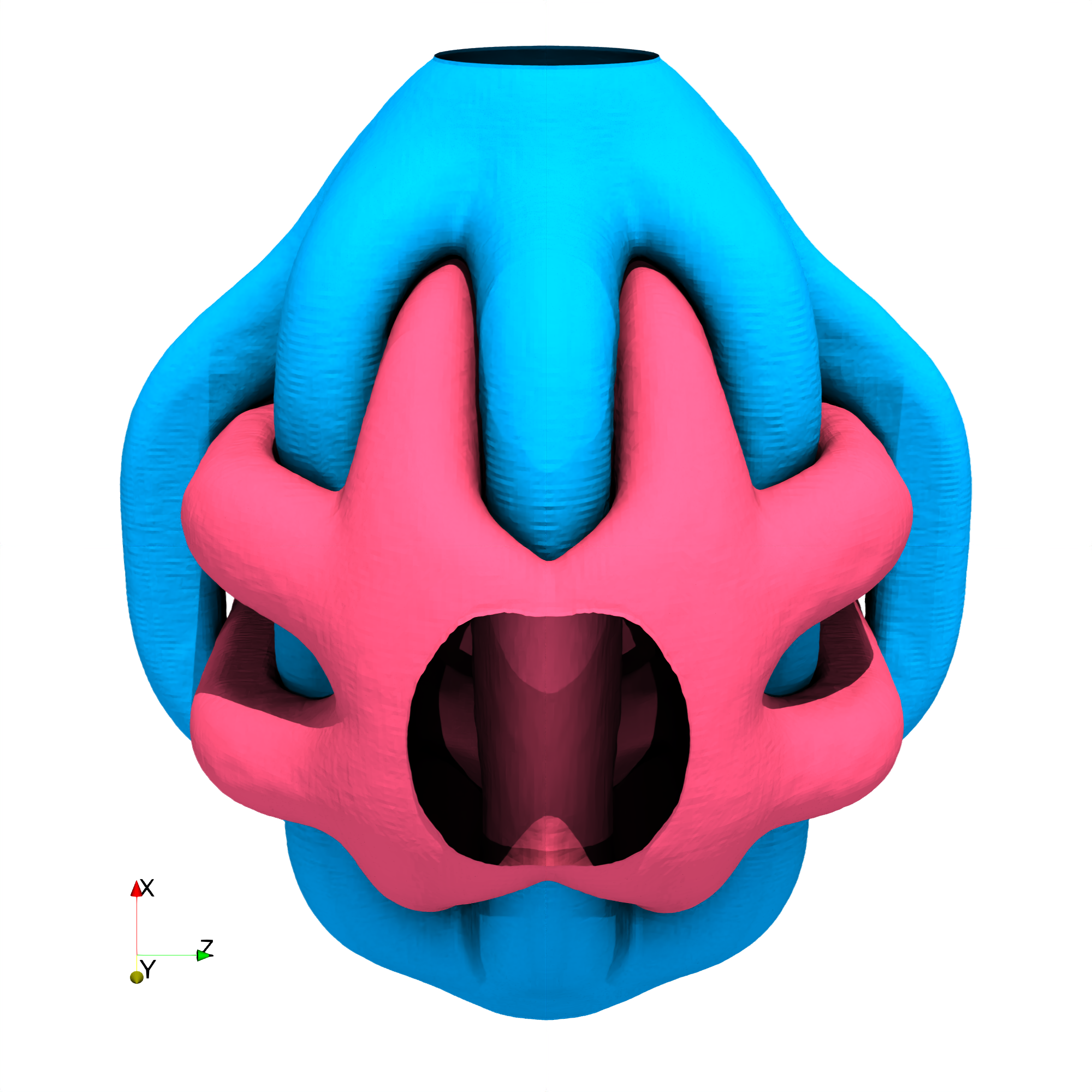}
        \caption{\(\omega = 0.8\)}
    \end{subfigure}
    \begin{subfigure}[b]{0.4\textwidth}
        \includegraphics[width=\textwidth]{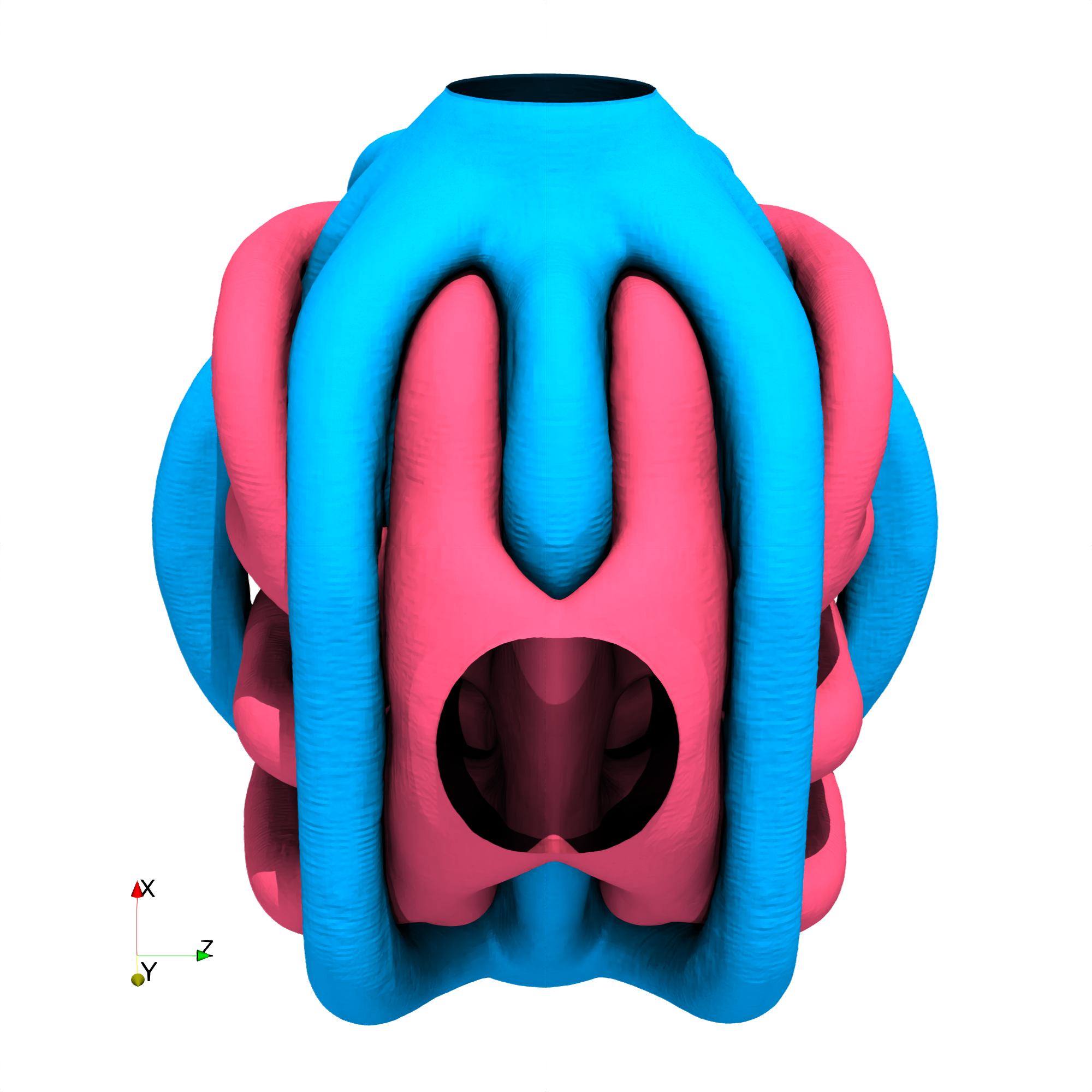}
        \caption{\(\omega = 0.6\)}
    \end{subfigure}
    \begin{subfigure}[b]{0.4\textwidth}
        \includegraphics[width=\textwidth]{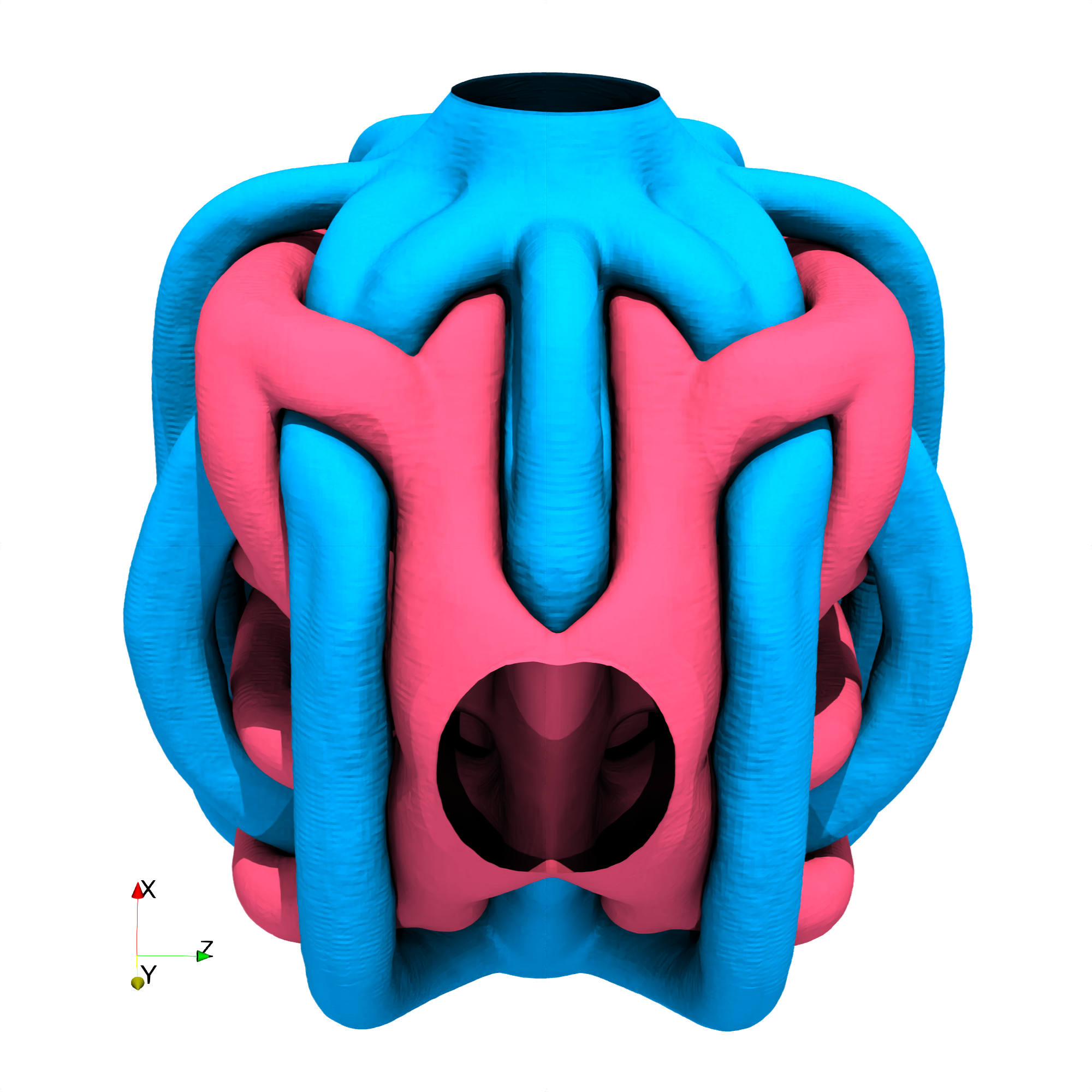}
        \caption{\(\omega = 0.4\)}
    \end{subfigure}
    \caption{Final designs for different weights \(\omega\).  
             The non‑mixing constraint between cold and hot phases is enforced.}
    \label{fig:comparison}
\end{figure}

We investigate four weights, \(\omega = 0.9, 0.8, 0.6,\) and \(0.4\). The corresponding optimized geometries are displayed in Fig. \ref{fig:comparison}. As the optimization shifts from pressure drop‑dominated ($\omega=0.9$) to heat‑exchange‑dominated ($\omega=0.4$) regimes the channels systematically elongate, and interdigitate, enlarging the fluid–solid contact area at the expense of longer flow paths. This trend is mirrored quantitatively as shown in Table~\ref{tab:comparison_phi}, which compares the final objective values. Indeed, the combined pressure drop rises from $10.1$ ($\omega=0.9$) to $33.0$ ($\omega=0.4$) while the net heat removed by the cold stream changes sign and grows from $-0.07$ ($\omega=0.9$) to $0.67$. ($\omega=0.4$). 

\begin{table}[h]
    \centering
    \caption{Final objective values for the two‑fluid heat exchanger at different weights \(\omega\).}
    \label{tab:comparison_phi}
    \begin{tabular}{|c|c|c|c|c|c|c|}
        \hline
        &
        \(\Phi_{(u,p)}^{\text{cold}}\) &
        \(\Phi_{(u,p)}^{\text{hot}}\) &
        \(\Phi_{(u,p)}^{\text{cold}}+\Phi_{(u,p)}^{\text{hot}}\) &
        \(\Phi_{\theta}^{\text{cold}}\) &
        \(\Phi_{\theta}^{\text{hot}}\) &
        \(\Phi_{\theta}^{\text{cold}}-\Phi_{\theta}^{\text{hot}}\) \\ \hline
        \(\omega = 0.9\) & 3.1 & 7.0 & 10.1 & 0.75 & 0.82 & \(-0.07\) \\ \hline
        \(\omega = 0.8\) & 4.7 & 9.9 & 14.6 & 0.86 & 0.71 & \phantom{\(-\)}0.15 \\ \hline
        \(\omega = 0.6\) & 8.6 & 14.7 & 23.3 & 1.02 & 0.54 & \phantom{\(-\)}0.48 \\ \hline
        \(\omega = 0.4\) & 13.3 & 19.7 & 33.0 & 1.12 & 0.45 & \phantom{\(-\)}0.67 \\ \hline
    \end{tabular}
\end{table}

\begin{figure}[htbp]
    \centering
    \begin{subfigure}[b]{0.49\textwidth}
        \includegraphics[width=\textwidth]{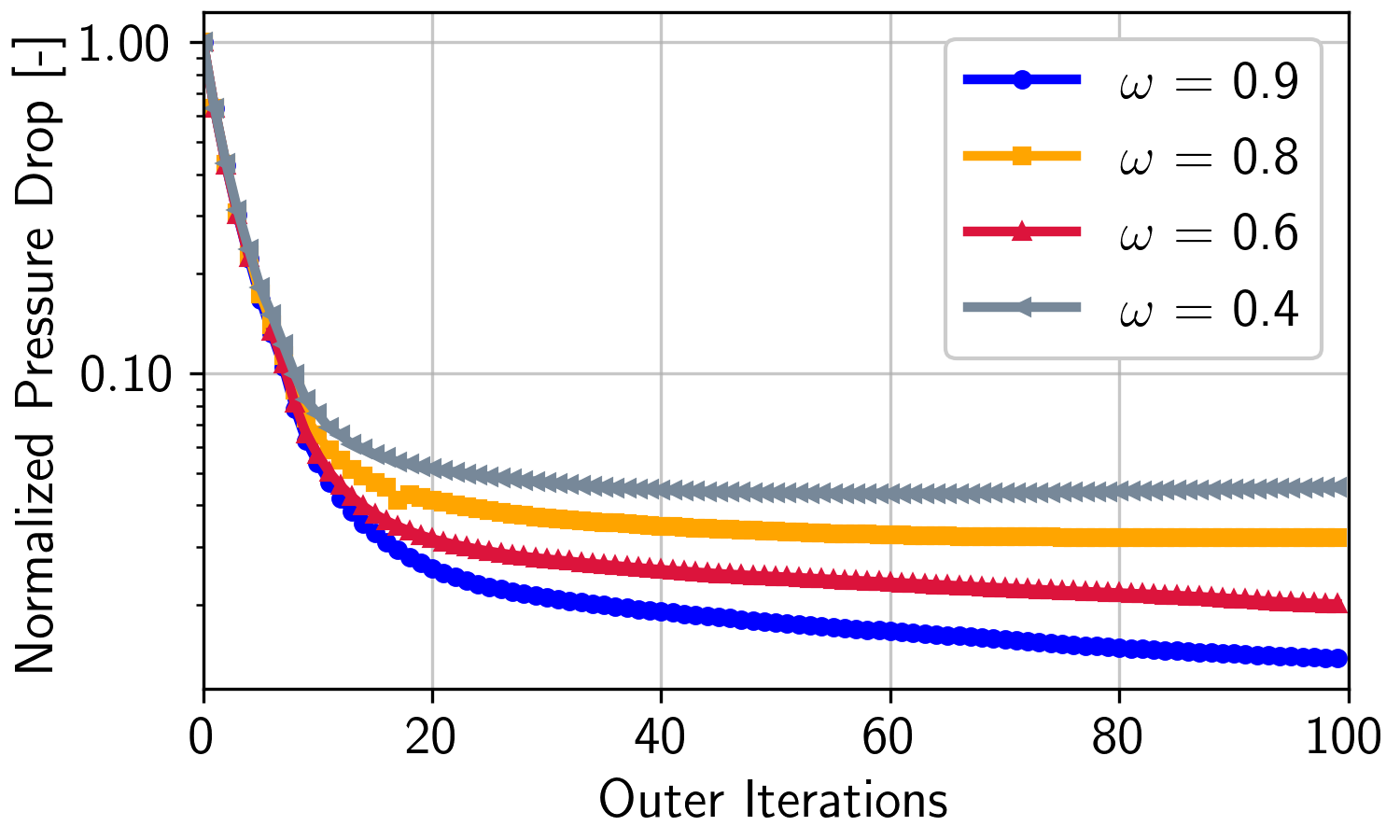}
    \end{subfigure}
    \hfill
    \begin{subfigure}[b]{0.49\textwidth}
        \includegraphics[width=\textwidth]{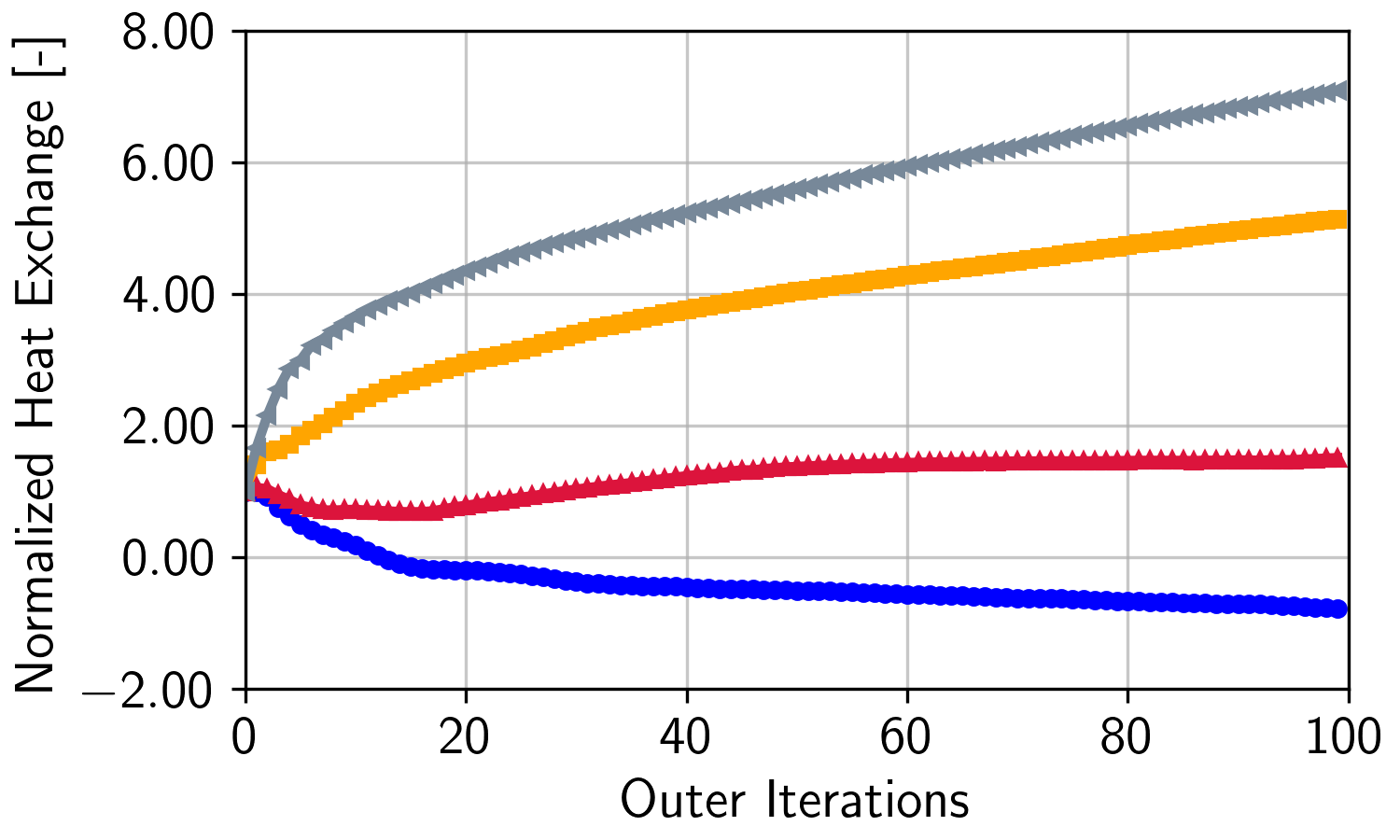}
    \end{subfigure}
    \caption{Convergence histories of the normalized objectives \(\Phi_{(u,p)}\) (left) and \(\Phi_{\theta}\) (right) for the four weights \(\omega\).}
    \label{fig:ex3_convergence}
\end{figure}

The convergence remains smooth for all weights as shown in Fig. \ref{fig:ex3_convergence} as the interface area develops. The solver timings are shown in Table \ref{tab:ex_1_timing} where total time for the Stokes–Brinkman includes system assembly, AMG setup, and forward and adjoint solves with CG–Uzawa and the total time for convection–diffusion includes the corresponding steps for the convection-diffusion equation. The timings show that the Stokes–Brinkman cost increases significantly, with up to 91\% increase, with the geometric complexity induced by smaller $\omega$, which promotes more heat exchange. On the other hand, the convection–diffusion solve stays nearly constant, keeping the total wall time per iteration below 8 minutes even for the most tortuous design.

{
    Although the choice \(V_0^{\mathrm{cold}}=V_0^{\mathrm{hot}}=0.15\) (used in part to enable direct comparison with Feppon \emph{et al.})  implies that a large fraction of the {computational bounding box} is solid, most of this solid remains outside of the area where conduction and convection are significant. The amount of solid {between} the phases—the part that matters for conduction—is governed by the minimum‑separation constraint. Starting from well‑separated initial channels, the optimizer systematically moves the phases closer to each other (as close as the prescribed gap allows): The intervening wall becomes thinner and is often perforated by conductive bridges that enhance exchange, rather than forming bulky barriers.}

\begin{table}[h!]
    \centering
    \caption{Wall‑clock time for the two‑fluid heat‑exchanger optimization at different weights \(\omega\).  
             Stokes–Brinkman time includes system assembly, AMG construction, and forward/adjoint solves with CG–Uzawa.  
             Convection–diffusion time includes analogous steps for the thermal subproblems.  Peak memory usage was 58GB.}
    \label{tab:ex_1_timing}
    \begin{tabular}{|c|c|c|}
        \hline
        Weight \(\omega\) &
        Stokes–Brinkman (s) &
        Convection–diffusion (s) \\ \hline
        0.9 & 197 & 72 \\ \hline
        0.8 & 265 & 75 \\ \hline
        0.6 & 304 & 72 \\ \hline
        0.4 & 378 & 75 \\ \hline
    \end{tabular}
\end{table}

\begin{figure}[htbp]
    \centering
    \begin{subfigure}[b]{0.45\textwidth}
        \includegraphics[width=\textwidth]{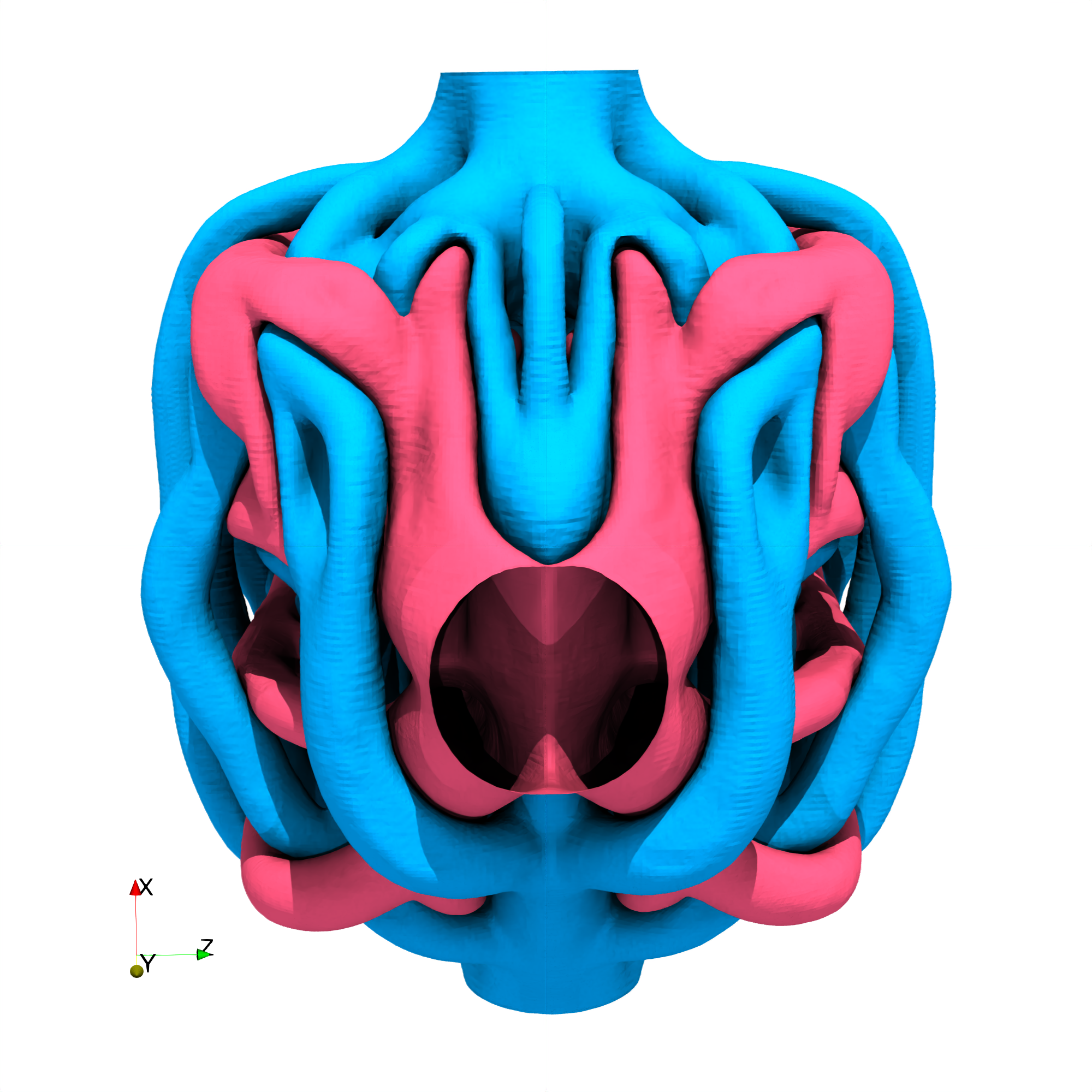}
        \caption{{\(V_0^{\mathrm{cold}} + V_0^{\mathrm{hot}} = 0.2\)}}
    \end{subfigure}\hfill
    \begin{subfigure}[b]{0.45\textwidth}
        \includegraphics[width=\textwidth]{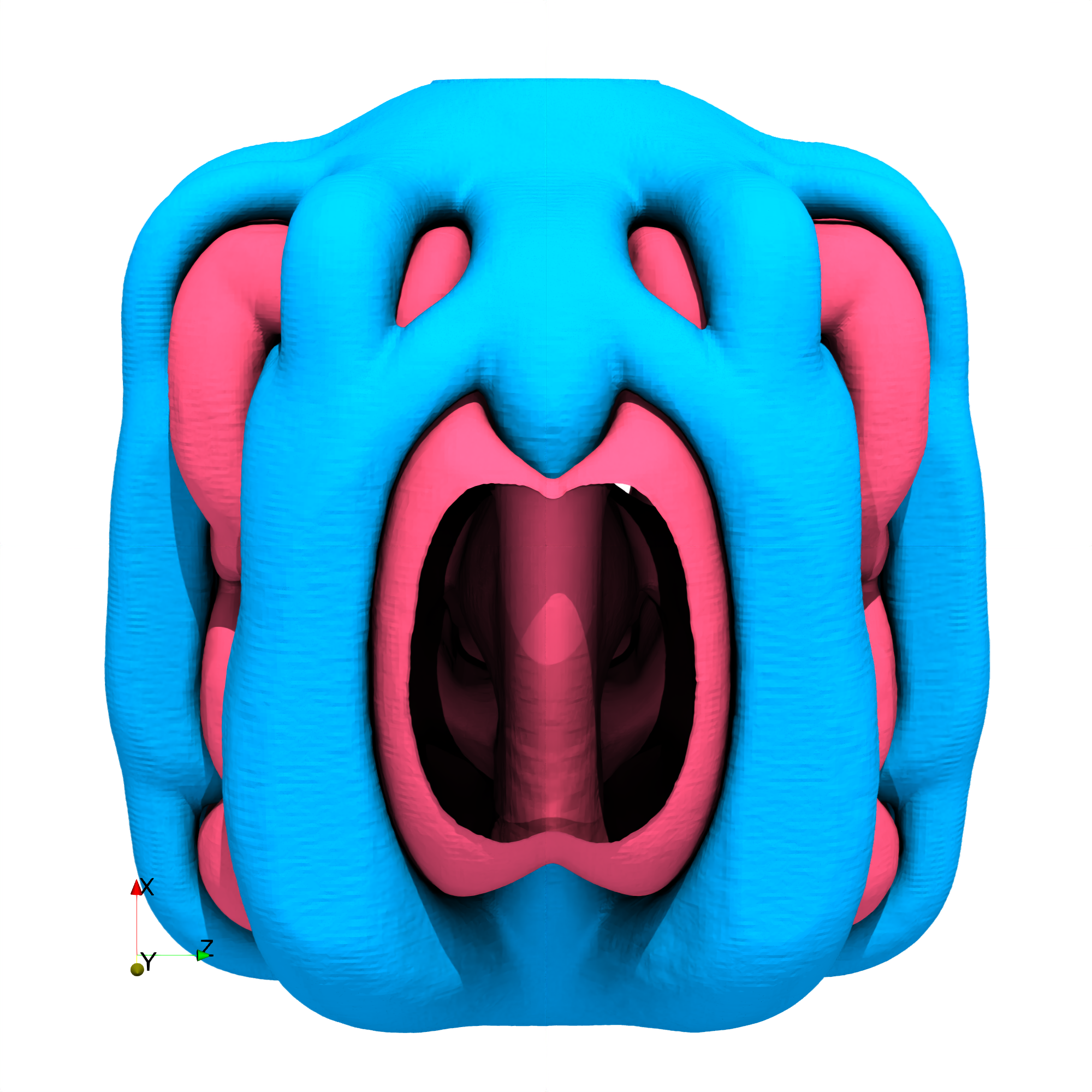}
        \caption{{\(V_0^{\mathrm{cold}} + V_0^{\mathrm{hot}} = 0.5\)}}
    \end{subfigure}
    \caption{{Optimized designs at \(\omega=0.4\) for different total initial fluid volumes.  
             The non‑mixing constraint and boundary conditions are identical to those in Section~\ref{sec:omega}.}}
    \label{fig:comparison_vf}
\end{figure}

{
\subsubsection{Effect of the initial fluid volume}
\label{sec:volume}
Since the optimization algorithm preserves the initial volume of each phase, we additionally investigate how the {initial} total fluid volume influences the outcome.    
Using the same boundary conditions as in Fig.~\ref{fig:ex3_setup}, the same Dirichlet velocity profiles~\eqref{ex3:dirichlet_bc}, and the same non‑mixing constraint as in Section~\ref{sec:omega}, we fix \(\omega=0.4\) and vary the total fluid volume \(V_0^{\mathrm{cold}}+V_0^{\mathrm{hot}}\in\{0.2,\,0.3,\,0.5\}\).  
Different volumes are realized by adjusting the initial channel thicknesses while keeping inlet/outlet radii fixed.  
Optimized designs for the extreme cases \(0.2\) and \(0.5\) are shown in Fig.~\ref{fig:comparison_vf}; the \(\,0.3\) case corresponds to the design shown in Fig.~\ref{fig:comparison}(d).}

{Table~\ref{tab:comparison_vf} summarizes the objective values.  
Since the Dirichlet inlet/outlet velocities are identical in both cases, the net exchanged heat \(\Phi_{\theta}^{\text{cold}}-\Phi_{\theta}^{\text{hot}}\) varies only modestly with the total fluid volume.
In contrast, the combined pressure drop \(\Phi_{(u,p)}^{\text{cold}}+\Phi_{(u,p)}^{\text{hot}}\) increases sharply as the available fluid volume decreases, reflecting the higher viscous losses in thinner channels.}

\begin{table}[h]
    \centering
    {
    \caption{{Final objective values for \(\omega=0.4\) for different total initial fluid volumes.}}
    \label{tab:comparison_vf}
    \begin{tabular}{|c|c|c|c|c|c|c|}
        \hline
        \(V_0^{\mathrm{cold}} + V_0^{\mathrm{hot}}\)
        &
        \(\Phi_{(u,p)}^{\text{cold}}\) &
        \(\Phi_{(u,p)}^{\text{hot}}\) &
        \(\Phi_{(u,p)}^{\text{cold}}+\Phi_{(u,p)}^{\text{hot}}\) &
        \(\Phi_{\theta}^{\text{cold}}\) &
        \(\Phi_{\theta}^{\text{hot}}\) &
        \(\Phi_{\theta}^{\text{cold}}-\Phi_{\theta}^{\text{hot}}\) \\ \hline
        0.2 & 35.4 & 65.2 & 100.6 & 1.13 & 0.43 & 0.69 \\ \hline
        0.3 & 13.3 & 19.7 & 33.0  & 1.12 & 0.45 & 0.67 \\ \hline
        0.5 & 4.1  & 9.3  & 13.4  & 1.07 & 0.50 & 0.57 \\ \hline
    \end{tabular}
    }
\end{table}

{Overall, reducing the initial fluid volume produces thinner and more tortuous channels that preserve heat‑exchange performance but substantially amplify the required pressure drop.  
Conversely, increasing the fluid volume reduces the hydrodynamic cost with only a modest decrease in net heat removed by the cold fluid.
}

\subsubsection{{Dependence on the initial design}}
\begin{figure}[htbp]
    \centering
    \begin{subfigure}[b]{0.45\textwidth}
        \includegraphics[width=\textwidth]{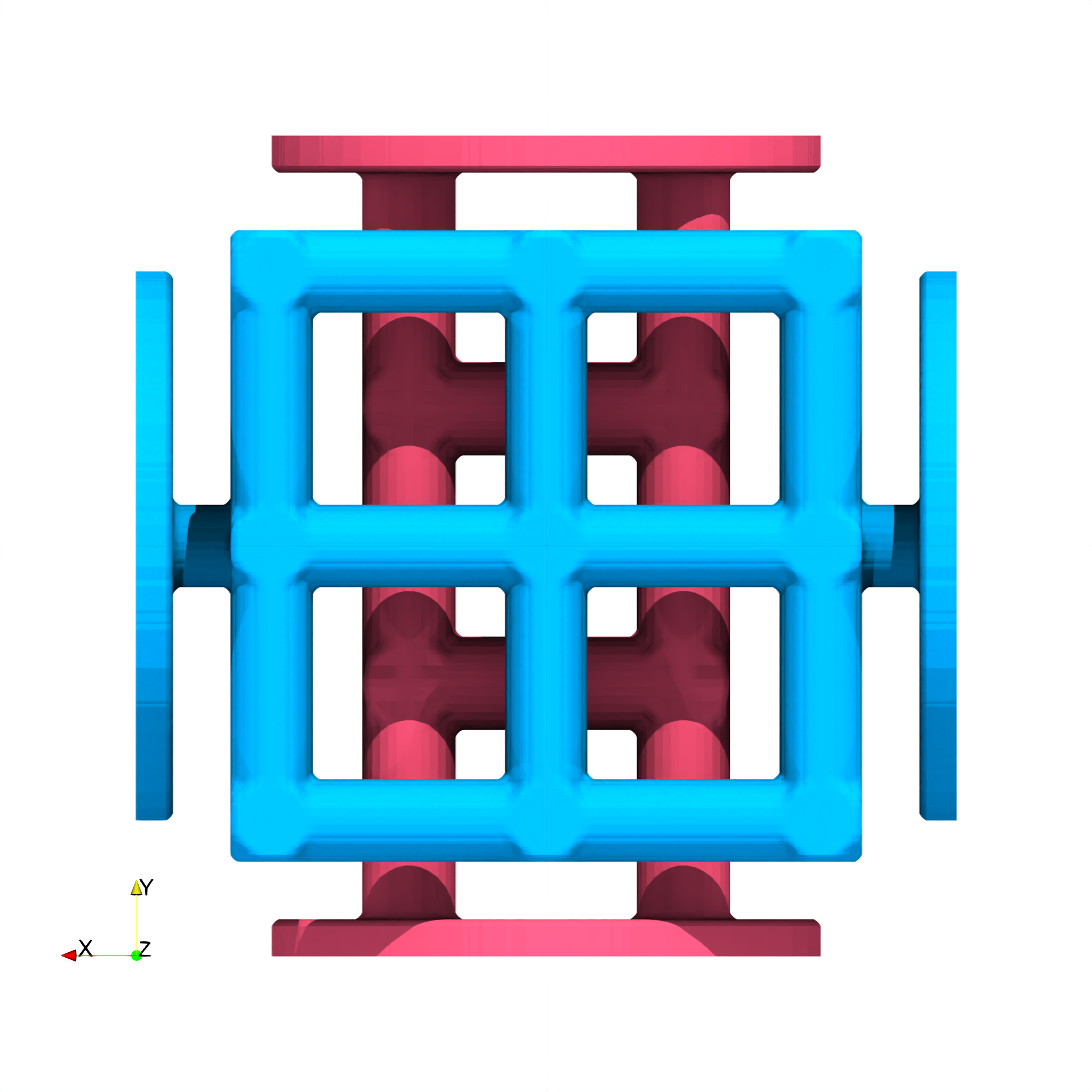}
        \caption{{Initial: \(3\times 3\) (cold), \(2\times 2\) (hot)}}
    \end{subfigure}\hfill
    \begin{subfigure}[b]{0.45\textwidth}
        \includegraphics[width=\textwidth]{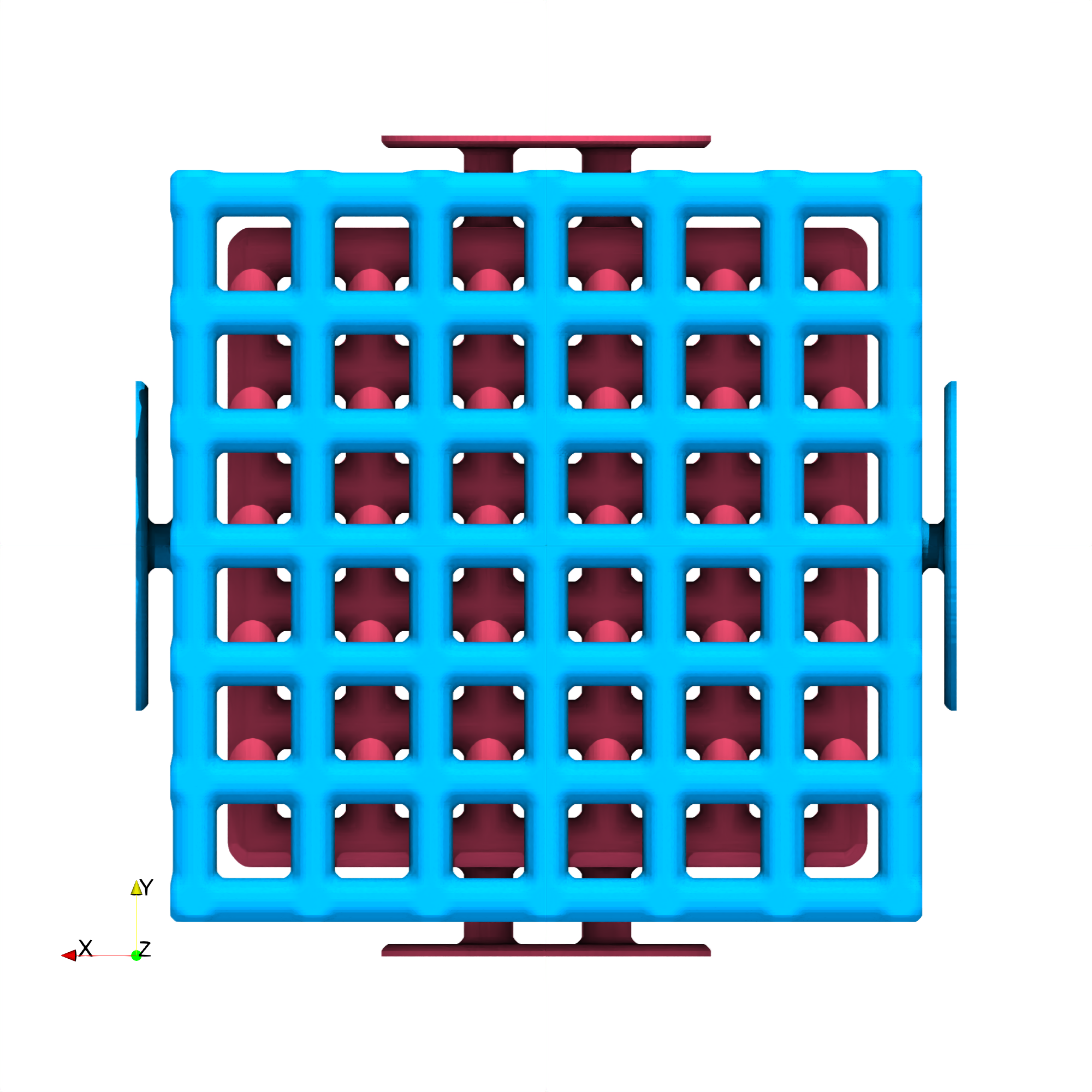}
        \caption{{Initial: \(7\times 7\) (cold), \(6\times 6\) (hot)}}
    \end{subfigure}

    \vspace{0.75em}

    \begin{subfigure}[b]{0.45\textwidth}
        \includegraphics[width=\textwidth]{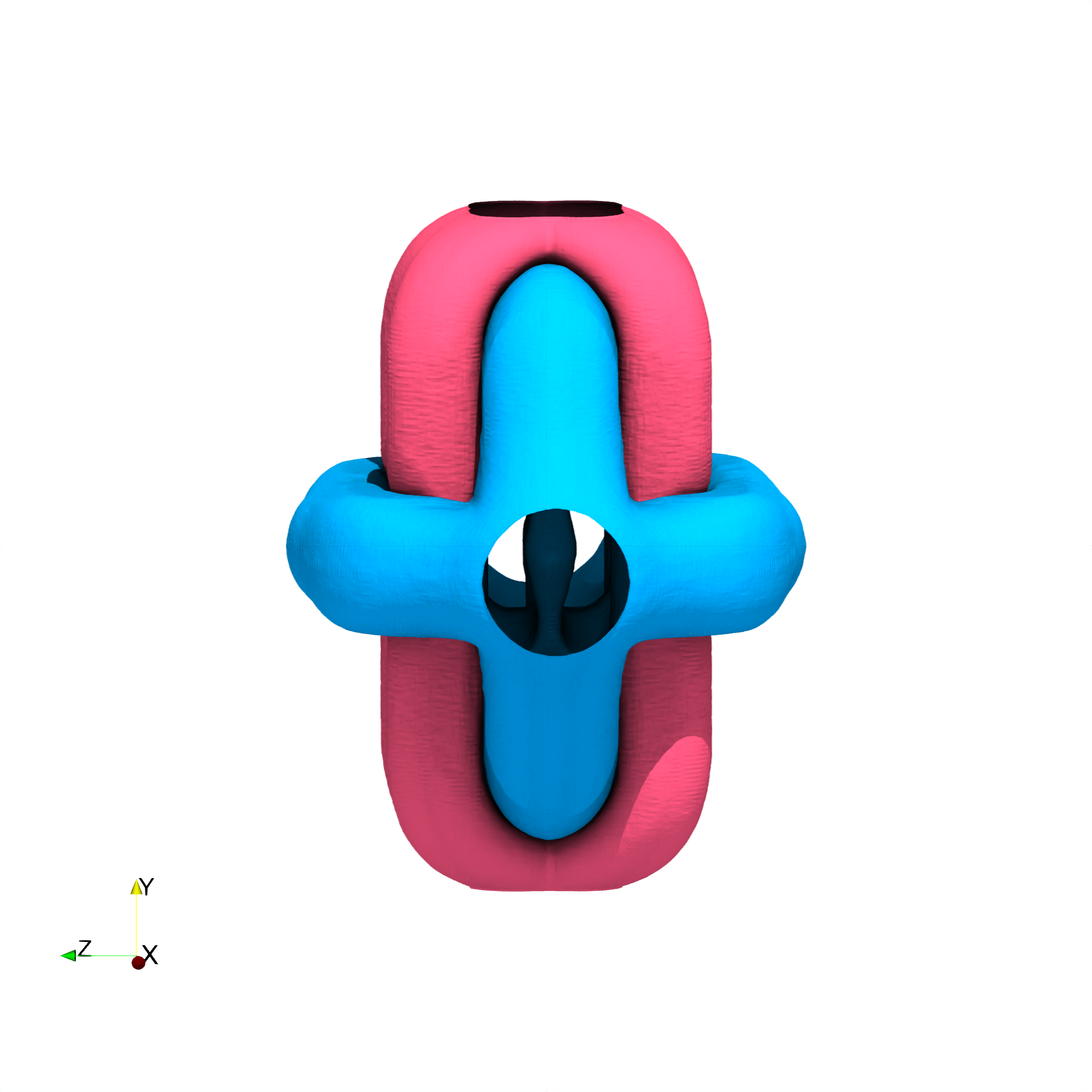}
        \caption{{Final (from \(3\times 3\) / \(2\times 2\))}}
    \end{subfigure}\hfill
    \begin{subfigure}[b]{0.45\textwidth}
        \includegraphics[width=\textwidth]{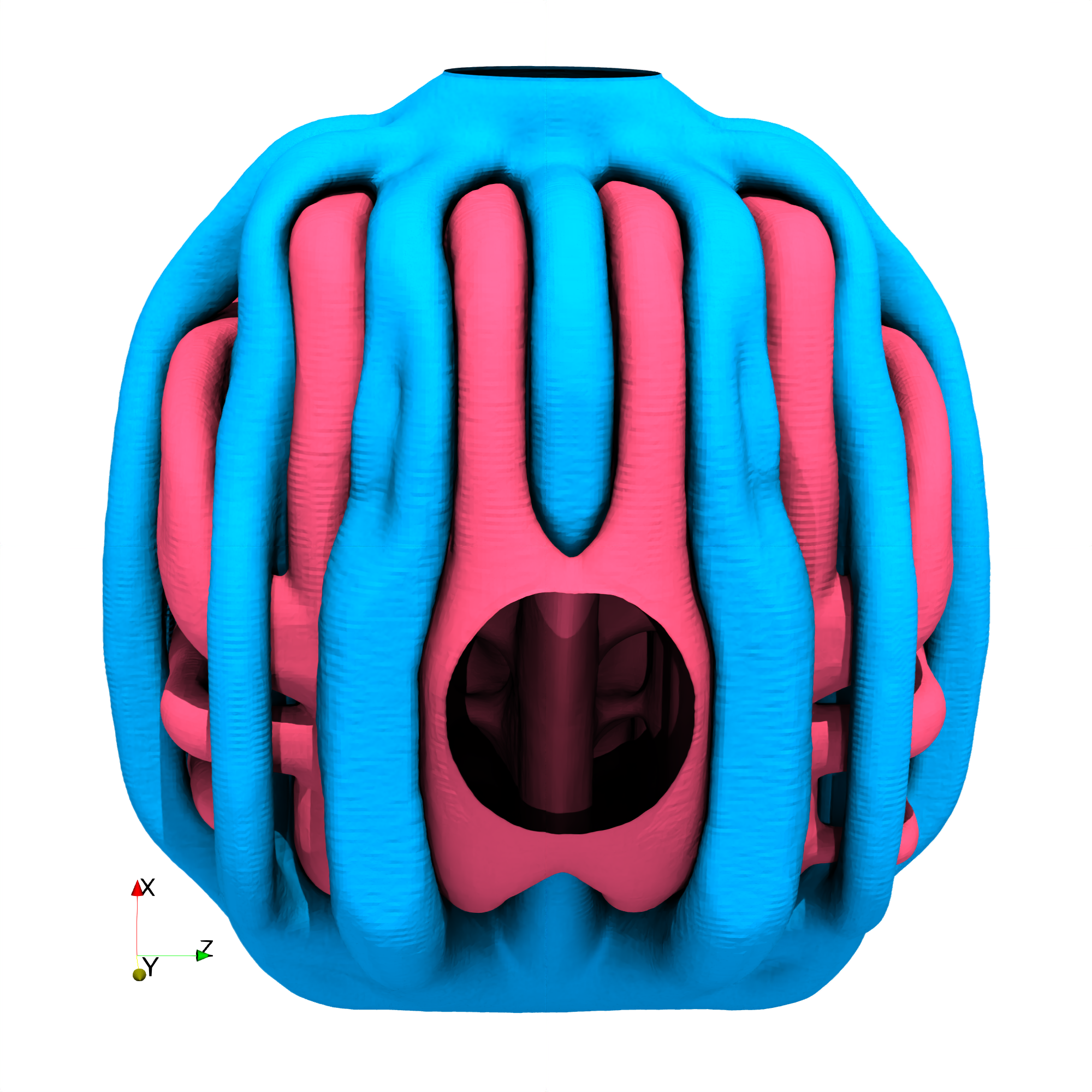}
        \caption{{Final (from \(7\times 7\) / \(6\times 6\))}}
    \end{subfigure}
    \caption{{Initial (top) and optimized (bottom) designs for two different initial topologies.  
             The non‑mixing constraint between cold and hot phases is enforced.}}
    \label{fig:dependence_initial}
\end{figure}

{For completeness, we also perform optimizations starting from different initial guesses. 
Following \cite{feppon2021body}, our baseline initialization consists of two nested arrays of pipes (see Fig.~\ref{fig:ex3_init}): a \(5\times 5\) array for the cold phase and a \(4\times 4\) array for the hot phase. 
Here we consider two additional initial designs: (i) a \(3\times 3\) array for the cold phase and a \(2\times 2\) array for the hot phase, and (ii) a \(7\times 7\) array for the cold phase and a \(6\times 6\) array for the hot phase. 
The corresponding initial total fluid volumes are \(V_0^{\mathrm{cold}} + V_0^{\mathrm{hot}} = 0.25\) and \(V_0^{\mathrm{cold}} + V_0^{\mathrm{hot}} = 0.40\), respectively. 
We perform the optimizations with weights \(\omega=0.1\) and \(\omega=0.2\), respectively. 
The initial and final designs are shown in Fig.~\ref{fig:dependence_initial}.
}

{From these cases we observe a pronounced dependence on the initial design.
This is a result of the stiff optimization problem primarily arising from the non-mixing constraint.
Accordingly, a reasonable choice of the starting topology can influence both the convergence behavior and the quality of the final design when optimizing two-fluid heat exchangers with non-mixing constraint. In practice, additional manufacturing and space constraints would indicate a good starting topology.}

\subsubsection{Large-scale validation with COMSOL Multiphysics}
\label{sec:large}
\begin{figure}[htbp]
    \centering
    \begin{subfigure}[b]{0.45    \textwidth}
        \includegraphics[width=\textwidth]
        {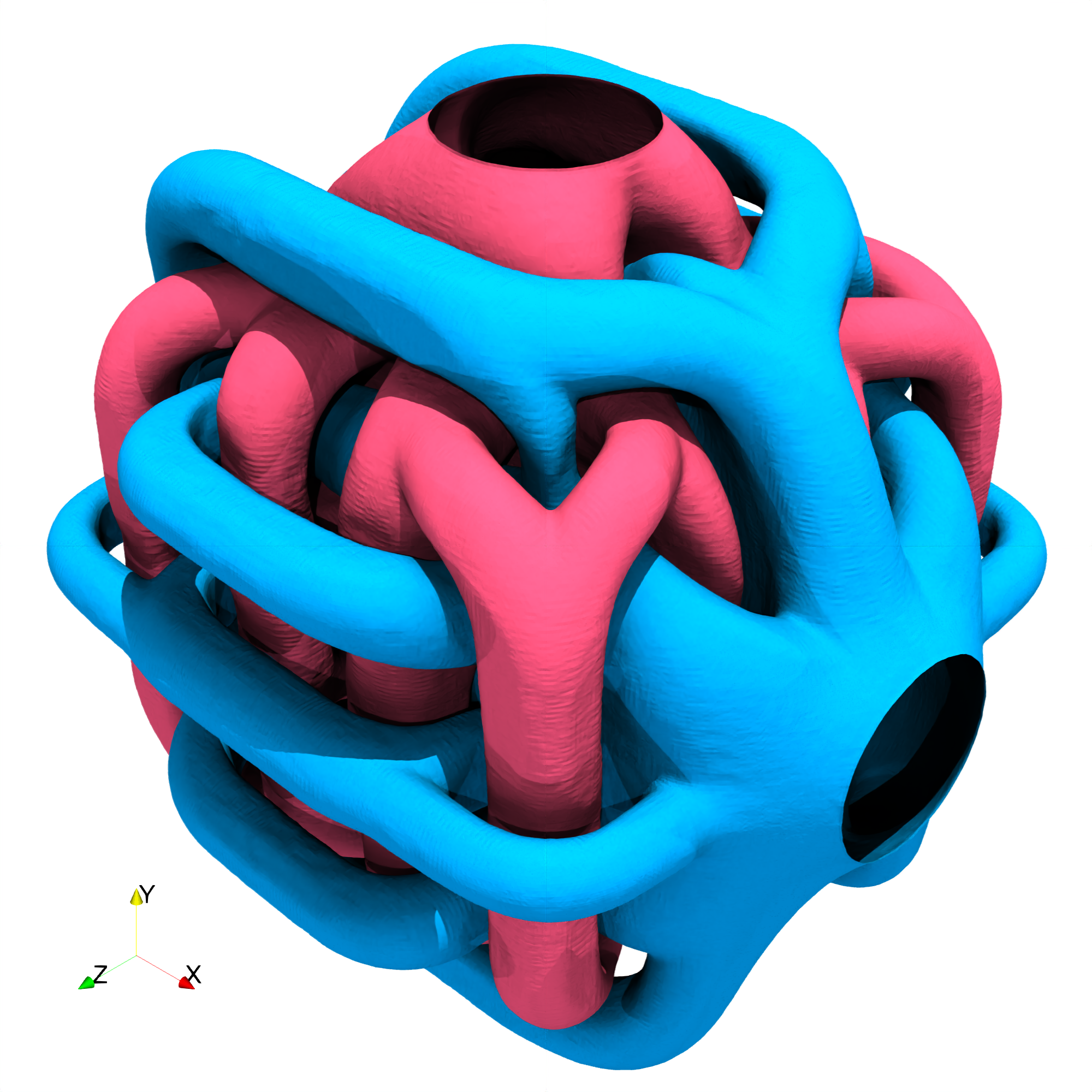}
        \caption{}
        \centering
        \label{fig:ex3_minthick_design}
    \end{subfigure}
    \begin{subfigure}[b]{0.45\textwidth}
        \includegraphics[width=\textwidth]{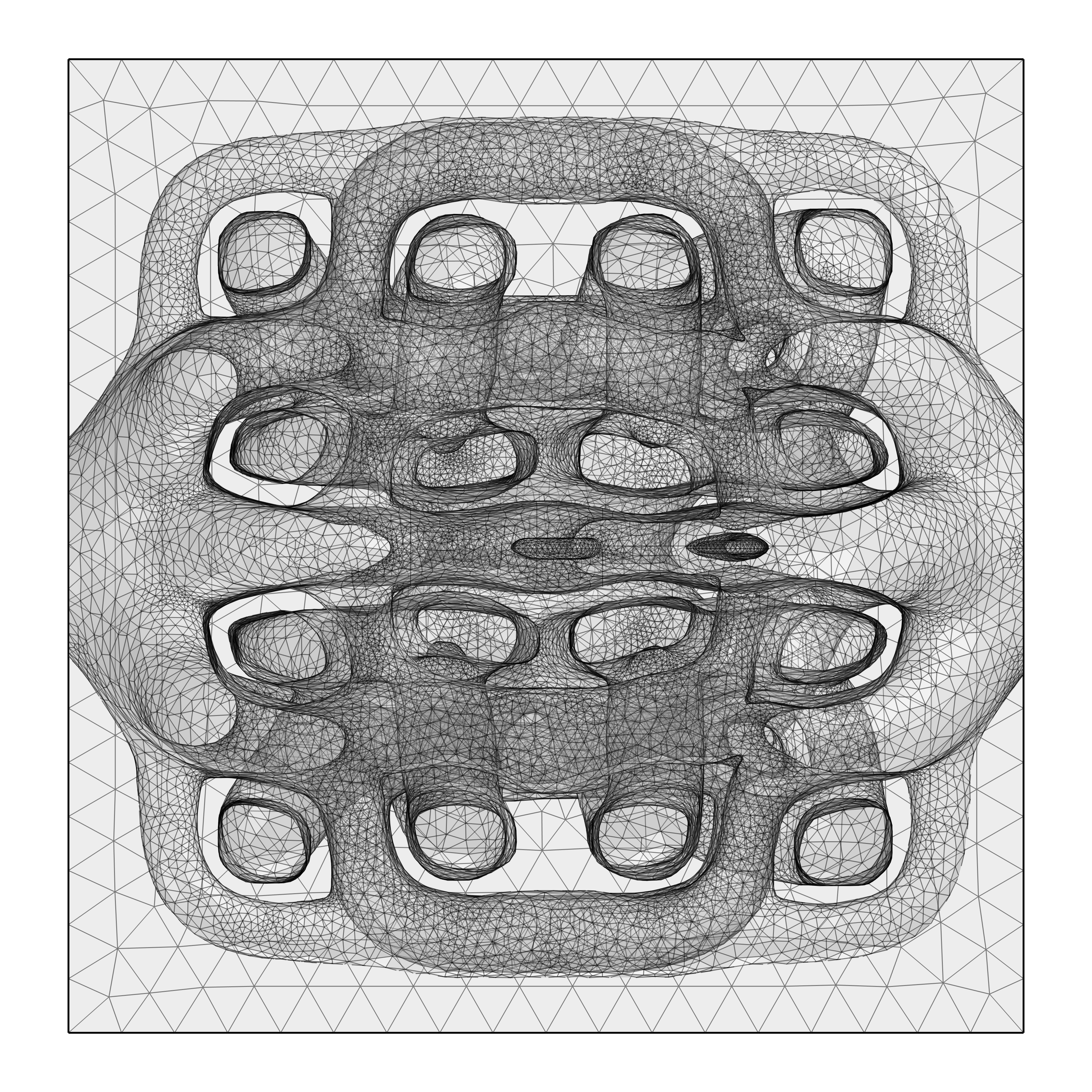}
        \caption{}
        \centering
        \label{fig:ex3_minthick_bf}
    \end{subfigure}
    \caption{a) Final large-scale design with minimal thickness constraint imposed; the resolution is $n = 370$, b) Body-fitted mesh.}
    \label{fig:ex3_minthick}
\end{figure}

Fig. \ref{fig:ex3_minthick_design} shows the $370^{3}$‑voxel optimized topology obtained with a nine‑voxel non‑mixing constraint, which corresponds to \(0.025\) in physical units. The geometry was converted to a body‑fitted tetrahedral mesh (Fig. \ref{fig:ex3_minthick_bf}) and re-simulated in COMSOL Multiphysics for independent verification. Table \ref{tab:comparison_phi_2} summarizes the comparison of the pressure‑drop and heat‑exchanged objectives. We observe that the objectives obtained with our solver differ by at most $4.3\%$ from the COMSOL prediction, which can be explained by errors introduced due to conversion from a structured grid to a body-fitted mesh and different discretization methods. 

\begin{table}[h]
    \centering
    \caption{Comparison of flow and thermal objective values for cold and hot phases.}
    \label{tab:comparison_phi_2}
    \begin{tabular}{|c|c|c|c|c|c|c|}
        \hline
         & $\Phi_{(u,p)}^{\mathrm{cold}}$ & $\Phi_{(u,p)}^{\mathrm{hot}}$ &$\Phi_{(u,p)}^{\mathrm{cold}}$ + $\Phi_{(u,p)}^{\mathrm{hot}}$ & $\Phi_{\theta}^{\mathrm{cold}}$ & $\Phi_{\theta}^{\mathrm{hot}}$ & $\Phi_{\theta}^{\mathrm{cold}} - \Phi_{\theta}^{\mathrm{hot}}$ \\ \hline
         Our method & 14.0 & 17.9 & 31.9 & 1.36 & 0.47 & 0.89 \\ \hline
         COMSOL P1/P1-P2 & 14.2 & 18.5 & 32.7 & 1.39 & 0.46 & 0.93 \\ \hline
    \end{tabular}
\end{table}

\begin{figure}[htbp]
    \centering
    \begin{subfigure}[b]{0.45\textwidth}
        \centering
        \includegraphics[width=\linewidth]{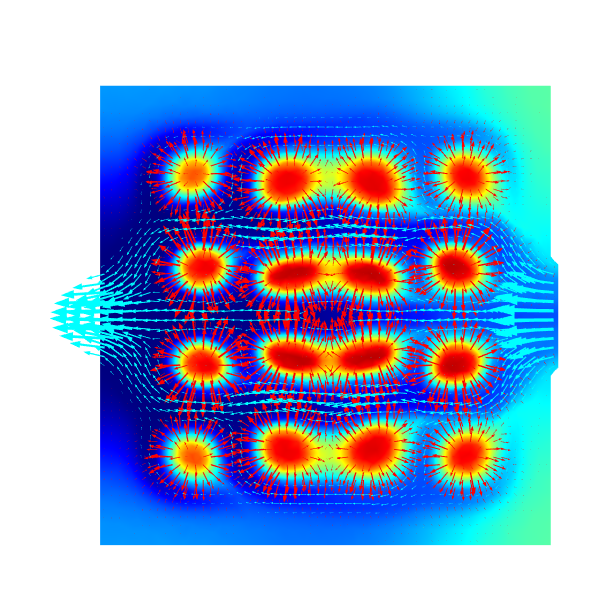}
        \label{fig:ls_flux}
    \end{subfigure}
    \begin{subfigure}[b]{0.45\textwidth}
        \centering
        \includegraphics[width=\linewidth]{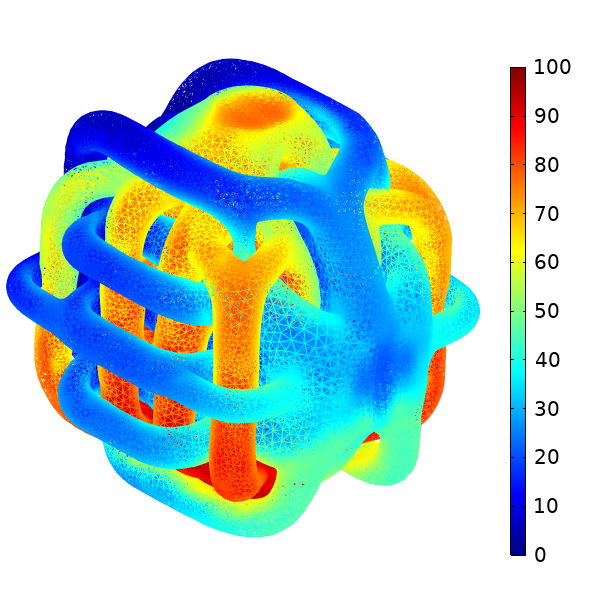}
        \label{fig:ls_temperature}
    \end{subfigure}
    \caption{Final large-scale design with minimal thickness constraint imposed; the resolution is $n = 370$. Left: Convective (cyan) and conductive (red) heat fluxes for a two-dimensional slice. Right: Temperature field on the surface.}
    \label{fig:ex3_states}
\end{figure}

Fig. \ref{fig:ex3_states} provides additional physical insight into the optimized heat exchanger. The cyan arrows show intra-channel convective transport $\rho c_p \theta \mathbf{u}$ due to fluid flow, while the red arrows reveal inter-channel conduction $-k \nabla \theta$ across solid bridges where temperature gradients are the steepest. The hot fluid enters at $100$ and leaves, on average, at $69.4$. Conversely, the cold fluid warms from $0$ to $30.1$.

The workflow remains computationally practical at this scale as summarized in Table \ref{tab:stokes_brinkman_timings} and Table \ref{tab:convection_diffusion_timings}. A forward and adjoint solves of the Stokes–Brinkman problem take $\approx 9$ min, while the convection-diffusion solves add another $\approx 2.5$ min. The peak memory is below $130$ GB on a workstation with 12 cores. 

\begin{table}[h]
    \centering
    \caption{Wall–clock times for the Stokes–Brinkman problem (12 proc) with the CG-Uzawa method. Total number of fluid cells (active + isolated fluids) is approx. $16.5\times 10^6$ . Tolerance for outer CG-Uzawa iteration is $10^{-3}$, tolerance for inner iterations is $10^{-6}$.}
    \label{tab:stokes_brinkman_timings}
    \begin{tabular}{|l|c|}
        \hline
        \textbf{Step}             & \textbf{Time (s)} \\ \hline
        Assemble                  & 15   \\ \hline
        Build AMG            & 26   \\ \hline
        Solve forward (66 CG-Uzawa iterations)             & 306  \\ \hline
        Solve adjoint (38 CG-Uzawa iterations)            & 194  \\ \hline
        \textbf{Total}            & \textbf{541} \\ \hline
    \end{tabular}
\end{table}

\begin{table}[h]
    \centering
    \caption{Wall-clock times for the Convection–diffusion problem (12 proc). Total number of cells is approx. $50.6\times 10^6$ (full domain). Tolerance is $10^{-6}$.}
    \label{tab:convection_diffusion_timings}
    \begin{tabular}{|l|c|}
        \hline
        \textbf{Step}                         & \textbf{Time (s)} \\ \hline
        Assemble                              & 44  \\ \hline
        Forward problem (build AMG + solve)    & 32 + 10 = 42 \\ \hline
        Adjoint problem (build AMG + solve)    & 31 + 36 = 67 \\ \hline
        \textbf{Total}                        & \textbf{153} \\ \hline
    \end{tabular}
\end{table}





\newpage
\section{Conclusions and future work}

{
We have introduced a sparse, narrow‑band topology optimization methodology for conjugate thermal–fluid problems. We have demonstrated that fictitious solid voxels can be eliminated from the forward and adjoint flow analyses. In doing so, we retain the flexibility of density methods while approaching the computational efficiency of the advanced level‑set techniques. Through multiple large‑scale examples, we have shown that the optimization converges to binary designs within the classical Brinkman penalization framework, without introducing filters or projections.}

To demonstrate the method’s efficiency, we optimized a two‑fluid exchanger with up to $51\times10^{6}$ design variables (Example 3) subject to a minimum‑thickness constraint. A complete forward–adjoint cycle for the final design took $9$ min for the Stokes–Brinkman system and $2.5$ min for the convection–diffusion system, while using less than $130$ GB of memory on $12$ CPU cores. In addition, the performance predictions deviate by less than $4.3\%$ from an independently meshed COMSOL model, confirming the solver’s accuracy. These results demonstrate that large‑scale, multiphysics topology optimization—previously confined to distributed‑memory computers—can be executed on high‑end workstations with turnaround times compatible with industrial design loops. {As future work, we plan to extend the present Stokes–Brinkman formulation to the Navier– Stokes flow.}

\appendix
\section*{Acknowledgments}
\label{app1}
The authors would like to acknowledge the U.S. Department of Energy, Advanced Research Projects Agency–Energy, for recognizing the importance of thermal-fluid topology optimization for aerospace applications and for funding portions of this work under award number DE-AR0001477. The content of this paper is solely the responsibility of the authors and does not necessarily reflect the official views of the U.S. Department of Energy.

\section*{Declaration of Interest Statement}
The authors declare that they have no known competing financial interests or personal relationships that could have appeared to influence the work reported in this paper.




\bibliographystyle{elsarticle-num} 
\bibliography{refs.bib}

\end{document}